\documentclass[pmlr]{jmlr}% new name PMLR (Proceedings of Machine Learning)

\usepackage{xcolor}

\setcitestyle{authoryear,open={(},close={)}}

\usepackage[utf8]{inputenc} % allow utf-8 input
\usepackage[T1]{fontenc}    % use 8-bit T1 fonts
\usepackage{hyperref}       % hyperlinks
\usepackage{url}            % simple URL typesetting
\usepackage{microtype}      % microtypography
\usepackage{nicefrac}       % compact symbols for 1/2, etc.
\usepackage{enumitem}

\usepackage{wrapfig}
\usepackage{float}
\usepackage{caption}
\usepackage{chngcntr} % make figure numbering in appendix work

\makeatletter
\def\set@curr@file#1{\def\@curr@file{#1}} %temp workaround for 2019 latex release
\makeatother

\usepackage{booktabs}

\usepackage{enumitem}

\jmlrvolume{219}
\jmlryear{2023}
\jmlrworkshop{Machine Learning for Healthcare}

\firstpageno{1}
\usepackage{titletoc}
\usepackage{titlesec}

\usepackage{subcaption}
\title[Detecting Heart Disease from Multi-View Ultrasound via SAMIL]{Detecting Heart Disease from Multi-View Ultrasound Images via Supervised Attention Multiple Instance Learning}
\author{\Name{Zhe Huang}$^{1}$
		\Email{\textsc{zhe.huang@tufts.edu}}
\AND
        \Name{Benjamin S. Wessler}$^{2}$
		\Email{\textsc{bwessler@tuftsmedicalcenter.org}}
\AND       
        \Name{Michael C. Hughes}$^1$
        \Email{\textsc{michael.hughes@tufts.edu}}
        \\
        \addr $^1$ Dept. of Computer Science, Tufts University, Medford, MA, USA
        \\
        \addr $^2$ Division of Cardiology, Tufts Medical Center, Boston, MA, USA
}%endauthortag

\begin{document}
%%% MCH SPACE-REDUCING HACK 3/3
% Reset vertical space for equations 
% (must be after \begin{document})
\setlength{\abovedisplayskip}{2pt plus 3pt}
\setlength{\belowdisplayskip}{2pt plus 3pt}

\maketitle

\begin{abstract}
  %Incoporating Ben's feedback
Aortic stenosis (AS) is a degenerative valve condition that causes substantial morbidity and mortality. This condition is under-diagnosed and under-treated. In clinical practice, AS is diagnosed with expert review of transthoracic echocardiography, which produces dozens of ultrasound images of the heart. Only some of these views show the aortic valve. To automate screening for AS, deep networks must learn to mimic a human expert’s ability to identify views of the aortic valve then aggregate across these relevant images to produce a study-level diagnosis. We find previous approaches to AS detection yield insufficient accuracy due to relying on inflexible averages across images. We further find that off-the-shelf attention-based multiple instance learning (MIL) performs poorly. We contribute a new end-to-end MIL approach with two key methodological innovations. First, a supervised attention technique guides the learned attention mechanism to favor relevant views. Second, a novel self-supervised pretraining strategy applies contrastive learning on the representation of the whole study instead of individual images as commonly done in prior literature. Experiments on an open-access dataset and a temporally-external heldout set show that our approach yields higher accuracy while reducing model size.
\end{abstract}
\let\thefootnote\relax\footnotetext{Open-source Code for our Supervised Attention MIL (SAMIL): \url{https://github.com/tufts-ml/SAMIL}}

\startcontents[sections]

\section{Introduction}
\label{sec:Introduction}
%incorporating Ben's feedback
Aortic stenosis (AS) is a progressive degenerative valve condition that is the result of fibrotic and calcific changes to the heart valve. These structural changes occur over years, eventually leading to obstruction of blood flow and can be fatal if not treated. AS is common and affects over 12.6 million adults and causes an estimated 102,700 deaths annually. AS can be effectively treated when it is identified in a timely manner, though diagnosis remains challenging~\citep{yadgir2020global}. One promising route to improving AS detection is to consider automatic screening of patients at risk using cardiac ultrasound. Automatic screening could provide a systematic, reproducible process and augment current approaches that rely on cardiac auscultation and miss a significant number of cases~\citep{gardezi2018cardiac}.

% Aortic stenosis is an important valve disease.
% \todo{Explain it impacts million people over age 65}
% \todo{Explain it is easily treatable, but difficult to detect.}

% One promising route to improving detection of AS is to consider automatic screening of all routinely-collected transthoracic echocardiograms (TTEs). 
% Automatic screening could provide a systematic, reproducible process and augment current manual screening known to have inter-rater reliability issues.

\begin{figure}[!t]
\begin{tabular}{c c c}
\begin{minipage}{.45\textwidth}
(a) Human expert approach
\end{minipage}
& &
\begin{minipage}{.45\textwidth}
(b) Filter then Average {\scriptsize \citep{holste2022automated}}
\end{minipage}
\\
\includegraphics[width=0.45\textwidth]{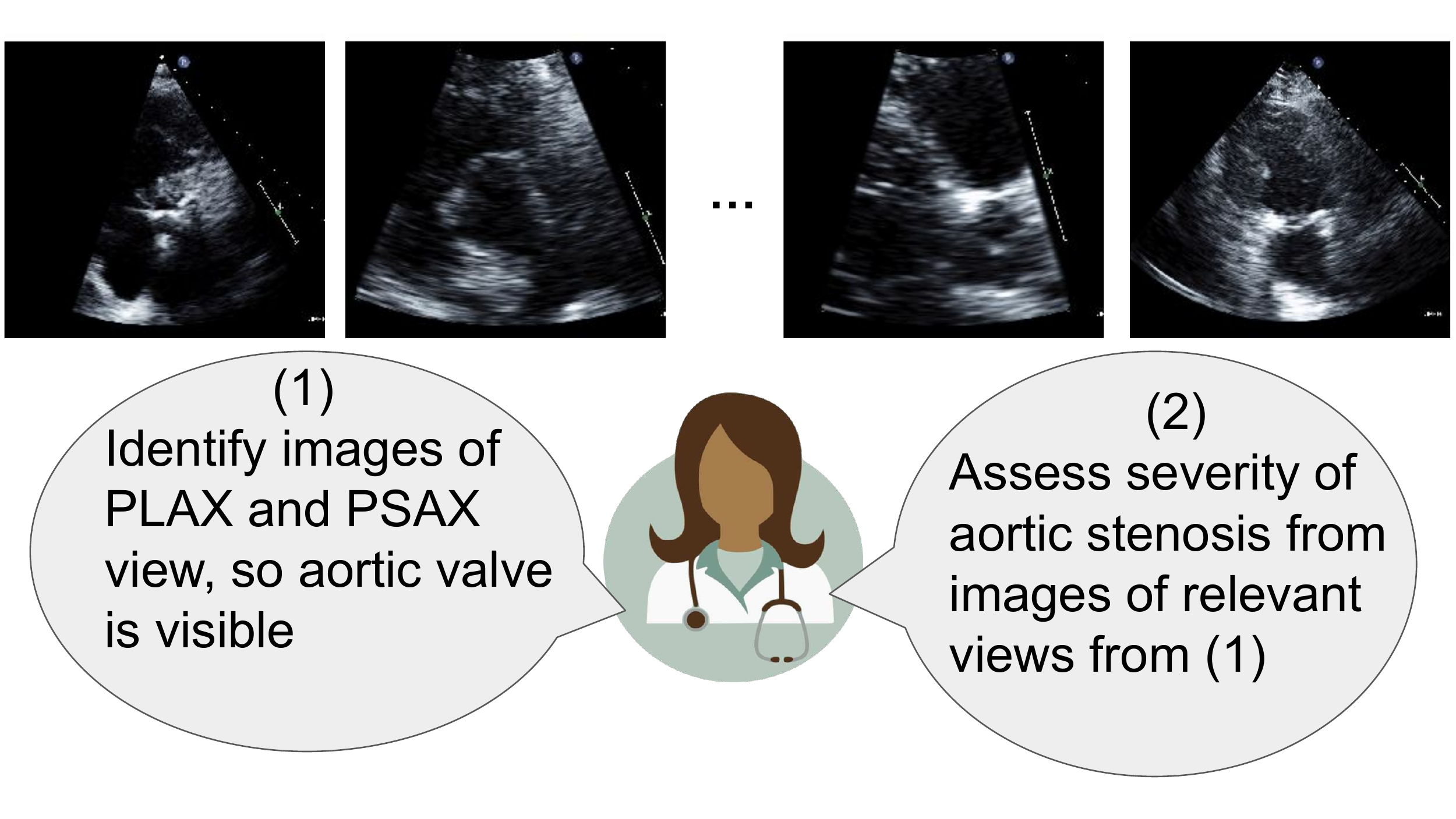}
& &
\includegraphics[width=0.45\textwidth]{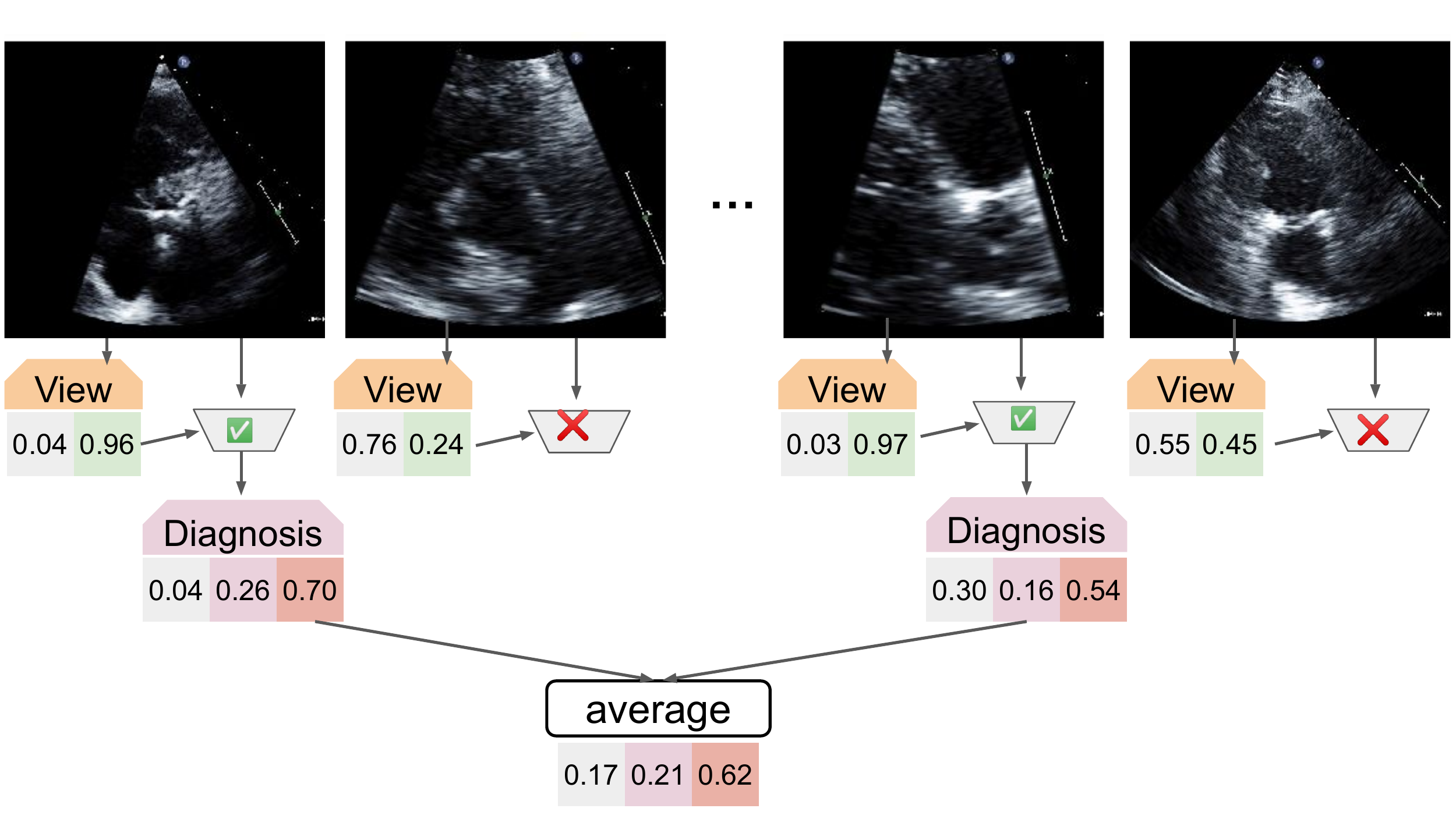}
\\
\includegraphics[width=0.45\textwidth]{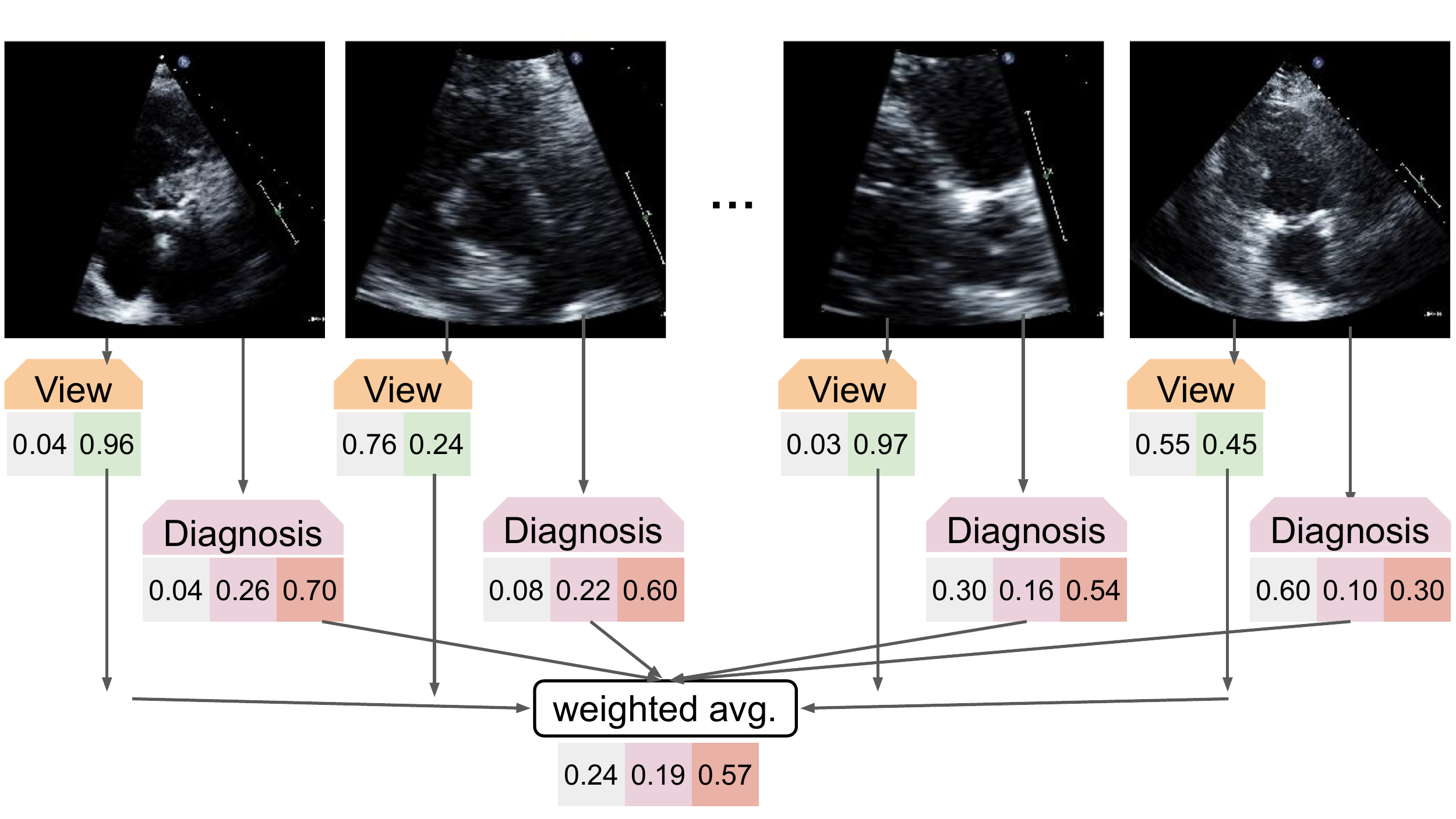}
& &
\includegraphics[width=0.45\textwidth]{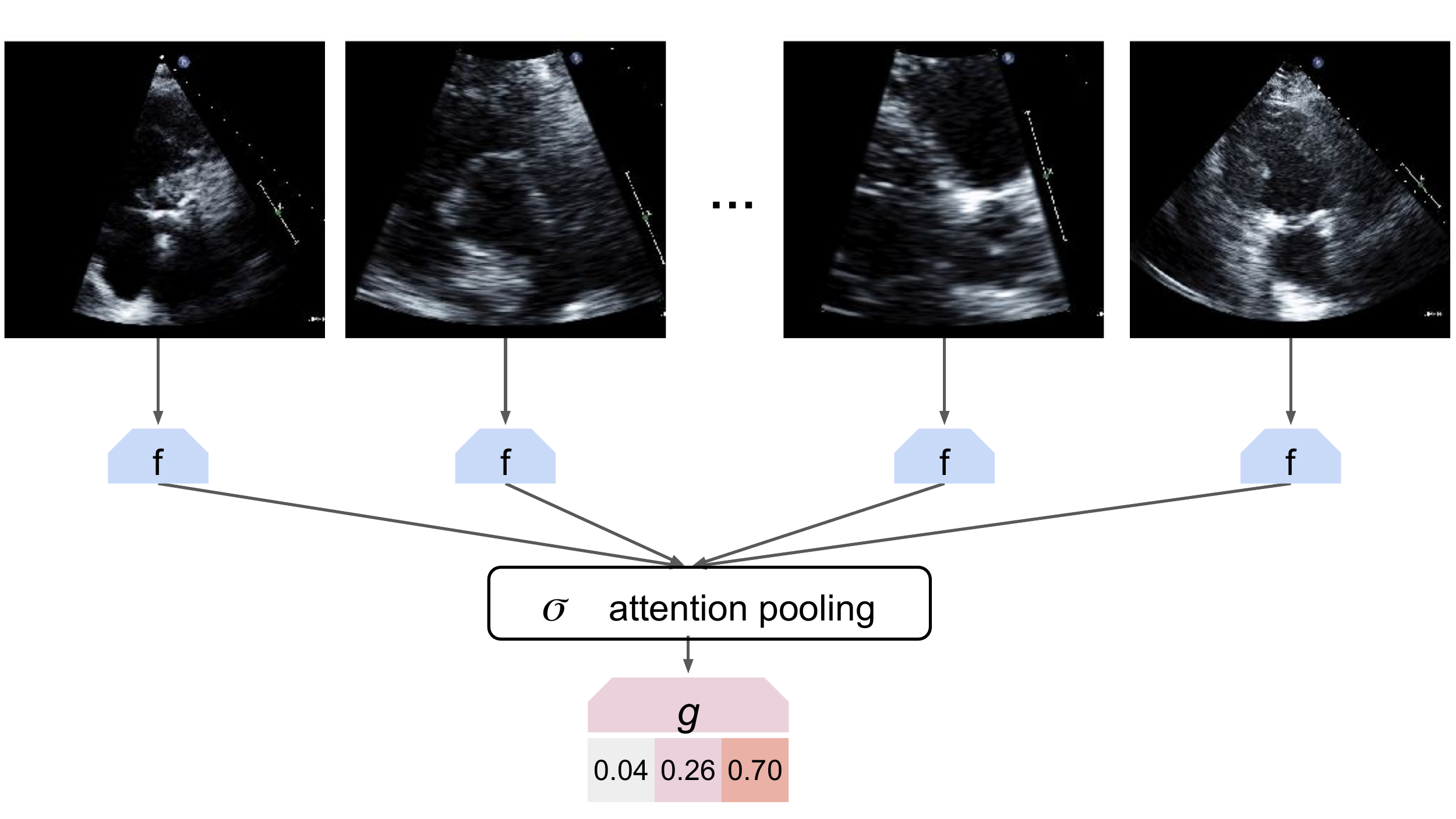}
\\
\begin{minipage}{.45\textwidth}
(c) Weighted Average by View Relevance {\scriptsize \citep{wessler2023automated,huang2021new}}
\end{minipage}
& &
\begin{minipage}{.45\textwidth}
(d) Attention-based MIL ~\\
\end{minipage}
\end{tabular}
\caption{\textbf{Overview of methods for diagnosing aortic valve disease from multiple images of the heart.}
In our chosen diagnostic problem, the input is multiple ultrasound images representing different canonical view types of the heart's complex anatomy (e.g. PLAX, PSAX, A2C, A4C, and more, see \citet{mitchell2019guidelines} for a taxonomy).
The output is a probabilistic prediction of the severity of Aortic Stenosis (AS), on a 3-level scale of no / early / significant disease.
We wish to develop deep learning methods that can solve this problem like expert cardiologists (panel a).
Two recent efforts (panel b by others, panel c by our group) made progress using a separately-trained view type classifier and per-image diagnosis classifier, but rely on combining diagnosis probabilities across images via average pooling that cannot learn how to distribute attention non-uniformly among images of relevant views.
In this work, we develop flexible attention-based multiple instance learning (MIL, panel d), with crucial contributions of supervised attention (Sec.~\ref{sec:methods_SA}) and improved pretraining strategies (Sec.~\ref{sec:methods_CL}) that substantially improve performance at our task.
}%endcaption
\label{fig:diagrams}
\end{figure}

The challenge in developing an automated system for diagnosing AS is that each echocardiogram study consists of \emph{dozens} of images or videos (typically 27-97 in our data) that show the heart’s complex anatomy from different acquisition angles. As illustrated in Fig.~\ref{fig:diagrams}(a), clinical readers are trained to look across many images to identify those that show the aortic valve at sufficient quality and then use these ``relevant'' images to assess the valve’s health. Training an algorithm to mimic this expert diagnostic process is difficult. Standard deep learning classifiers are designed to consume only one image and produce one prediction. Automatic screening of echocardiograms requires the ability to make one coherent prediction from \emph{many} images representing diverse view types. To make matters more difficult, each image’s view type is not typically recorded in digital health records during routine collection.

% The challenge is that an echocardiogram study consists of dozens of images or videos that show the heart's complex anatomy from many different acquisition angles.
% Clinical readers are trained to look across many images to identify those that show the aortic valve at sufficient quality and then use these ``relevant'' images to assess the valve's health. Training an algorithm to mimic this expert human diagnostic process is difficult.
% Most standard deep learning classifiers are designed to consume only one image and produce one prediction. Automatic screening of echocardiograms requires the ability to process \emph{many} images representing diverse view types. Each image's view type is not stored in the EHR during routine collection.

\emph{Multiple-instance learning} (MIL) is a branch of weakly supervised learning in which classifiers can consume a variable-sized set of images to make one prediction.
Recent impressive advances in deep attention-based MIL have been published~\citep{ilse2018attention, lee2019set, sharma2021cluster, shao2021transmil}. 
However, their success at medical diagnostic tasks, especially those with ultrasound images from many possible view types, has not been previously evaluated.

% Our approach does not require the separately-trained filtering step (Fig.~\ref{fig:diagrams} (b)) for selecting relevant views for diagnosis needed by some previous AS screening methods~\citep{holste2022automated}.
\subsection*{Contributions to Clinical Translation and MIL Methodology}
This study's contribution to applied clinical research is the development and validation of a new deep MIL approach for automatic diagnosis of heart valve disease from multiple ultrasound images produced by a  routine trans-thoracic echocardiogram (TTE) study. 
Our end-to-end approach can take as input any number of images from various view types, eliminating the need for a separately-trained filtering step (Fig.~\ref{fig:diagrams}(b)) to select relevant views for diagnosis required by some prior AS screening methods~\citep{holste2022automated}.
Our approach is also more flexible and data-driven than the weighted average (Fig.~\ref{fig:diagrams}(c)) of our team's previous efforts for AS screening~\citep{huang2021new,wessler2023automated}.
Head-to-head evaluation in Sec.~\ref{sec:Results} demonstrates that our approach can yield superior balanced accuracy for assigning AS severity grades to new studies, while keeping model size over 4x smaller than previous efforts like \citep{holste2022automated}.
Small model sizes enable faster predictions and ease portability to new hospital systems.

Our approach's success is made possible by two methodological contributions. % MIL research. 
First, we propose a supervised attention mechanism (Sec.~\ref{sec:methods_SA}) that steers focus toward images of relevant views, mimicking a human expert.
On our AS diagnosis task, supervised attention yields notable gains -- balanced accuracy jumps from 60\% to over 70\% -- over previous off-the-shelf attention-based MIL~\citep{ilse2018attention}.
Second, we introduce a self-supervised pretraining strategy (Sec.~\ref{sec:methods_CL}) that focuses contrastive learning on the embedding of an entire study (a.k.a. the embedding of the ``bag'', using MIL vocabulary). In contrast, most previous pretraining focuses on representations of individual images.
Both innovations are broadly applicable to other MIL problems involving imaging data of multiple view types.
%we introduce a novel self-supervised pretraining strategy that contrasts the bag-level embedding instead of image-level embedding, which are shown to be more effective for our problem (might also be suitable for other MIL problem).

%This study makes three contributions towards the automatic diagnosis of Aortic Stenosis using machine learning. Firstly, we propose an end-to-end multi-instance-learning approach that can diagnose Aortic Stenosis in realistic clinical scenarios, consuming various numbers of images from different view types in a realistic TTE study, without requiring pre-filtering. Secondly, we leverage clinical insights and introduce a supervised attention mechanism that uses predicted view relevance to guide the model's attention, significantly improving the performance of off-the-shelf MIL algorithms. Thirdly, we propose a novel self-supervised pre-training strategy that contrasts the bag-level embeddings, instead of the image-level embeddings, which we demonstrate to be more effective for our problem and may also be suitable for other MIL problems.

\subsection*{Generalizable Insights about Machine Learning in the Context of Healthcare}

This study offers critical insight into how multiple-instance learning can be applied to routine echocardiography studies.
We show that recent MIL architectures are insufficient to achieve competitive performance because they attend to irrelevant instances and thus lack the ability to make \textbf{clinically plausible} decisions.
Our two innovations -- supervised attention (Sec. \ref{sec:methods_SA}) and bag-level self-supervised pretraining (Sec.~\ref{sec:methods_CL}) can be broadly applicable to many clinical image analysis problems that require non-trivial aggregation over multiple images from multiple acquisition angles (views) to make one diagnosis. 
Beyond echocardiography, these insights could be useful for lung ultrasound, fetal ultrasound, head CT, and more.
%all of which require the clinical expert to aggregate information across multiple images from multiple acquisition angles (views).

\section{Related Work}
\label{sec:RelatedWork}
\subsection{Multiple-Instance Learning.}
Multiple-instance learning~\citep{dietterich1997solving,maron1997framework} describes a type of supervised learning problem where an unordered bag of instances and a corresponding bag label are provided as input for model training, and the goal is to predict the bag label for unseen bags. 
This type of problem appears in many medical applications, including whole-slide image (WSI) analysis in pathology~\citep{cosatto2013automated,shao2021transmil,li2021dual},
diabetic retinopathy screening~\citep{quellec2012multiple, li2021deep}, and cancer diagnosis~\citep{ding2012breast,campanella2019clinical}.
See App.~\ref{app:relatedworks} for a broader summary of classic MIL techniques and more medical applications. For extensive reviews of the MIL literature, see~\citet{zhou2004multi, quellec2017multiple,carbonneau2018multiple}.
%% MCH trimmed: kandemir2015computer was used twice in same paragraph

Two primary ways for modeling multiple instance learning problems are the instance-based approach and the embedding-based approach. In the instance-based approach, an instance classifier is used to score each instance, and a pooling operator is then used to aggregate the instance scores to produce a bag score. In the embedding-based approach, a feature extractor generates an embedding for each instance, which is then aggregated into a bag-level embedding. A bag-level model is subsequently employed to compute a bag score based on the embedding. The embedding-based approach is argued to deliver better performance than the instance-based approach~\citep{wang2018revisiting}, but at the same time harder to determine the key instances that trigger the classifier~\citep{liu2017detecting}. 
% MCH: moved dual-stream to the next parag

% \paragraph{Classic approaches.}
% % Numerous studies have been conducted on MIL even before the  ``deep learning era'' \footnote[1]{It is usually refered to 2012 when AlexNet achieved a significant improvement in image recognition accuracy over previous methods in the ImageNet Large Scale Visual Recognition Challenge}. 
% % Numerous studies on MIL have been conducted even prior to the advent of deep learning. 
% Examples of classic MIL methods includes iARP~\citep{dietterich1997solving}, 
% Diverse Density~\citep{maron1997framework}, 
% Citation-kNN~\citep{wang2000solving}, MI-Kernels~\citep{zhang2001dd}, MI/mi-SVM~\citep{andrews2002support}, mi-Graph ~\citep{zhou2009multi}, MILBoost~\citep{zhang2005multiple}, among others. 

% Classic approaches toward multiple-instance learning are limited in the choice of $f$. For example, using pre-computed features as the representation of each instance~\citep{dietterich1997solving}or using only fully-connected layers as $f$~\citep{wang2018revisiting} for feature extraction. The choice of $\sigma$ (using restricted and very often non-trainable pooling functions such as mean pooling or max pooling ~\citep{feng2017deep, pinheiro2015image, zhu2017deep}.) 

% ~\cite{pappas2014explaining} proposed an attention-based MIL method where the attention weights are learned from an auxiliary linear regression model.

\paragraph{When input is one image from each desired view type.}
Some recent medical imaging work assumes that instead of an unordered ``bag'' of instances of arbitrary size, the provided input will contain exactly one image for each of a few known view types (usually 2 or 4).
Examples include work on 2-view chest x-rays \citep{rubinLargeScaleAutomated2018,hashirQuantifyingValueLateral2020} as well as work on breast cancer screening using 2 views~\citep{carneiroUnregisteredMultiviewMammogram2015,vantulderMultiviewAnalysisUnregistered2021} or 4 views~\citep{wuDeepNeuralNetworks2020,nasirkhanMultiViewFeatureFusion2019}.
Methods differ in whether they fuse view-specific branches early or late, with latest innovations transferring information across views via transformers~\citep{vantulderMultiviewAnalysisUnregistered2021}.
In contrast to such work, the MIL methods we develop consume dozens of images for which a view type is not known in advance, reflecting the lack of recorded view annotations in typical echocardiograms.

\paragraph{Deep attention-based MIL.} 
Our proposed method builds upon recent works advancing attention-based deep neural networks for MIL. 
ABMIL~\citep{ilse2018attention} is an embedding approach where a two-layer neural network computes attention weights for each instance, with the final representation formed by averaging over instance embeddings weighted by attention.
Set Transformer~\citep{lee2019set} proposed to model the interactions among instances by using self-attention with multi-head attention~\citep{vaswani2017attention}. TransMIL ~\citep{shao2021transmil} uses a Transformer-based architecture to capture correlations among patches for whole-slide image classification. C2C~\citep{sharma2021cluster} divides patches from a whole-slide image into clusters, and sample multiple patches from each cluster for training. 
C2C then tries to guide attention weights to be similar to a predefined uniform distribution, aiming to minimize intra-cluster variance for patches from the same cluster. A recent method called DSMIL~\citep{li2021dual} attempts to benefit from instance-based and embedding-based approaches via a dual-stream architecture. That work pretrains an \emph{instance-level} feature extractor using self-supervised contrastive learning.  
%It also uses KL divergence to guide the attention weights, but different from ours, the KL-divergence is computed between the attention weights and the predefined uniform distribution, aiming to minimize intra-cluster variance for patches from the same cluster. 

%DSMIL~\citep{li2021dual} solves WSI classification with a two-stream architecture and multi-scale fusion. 

% However, such an idea is not directly applicable to our problem of diagnosing AS from multi-view ultrasound images, \todo{TODO, HOW TO EXPLAIN THIS CLEARER. for example, cardiologists can make AS diagnosis decision from a few key images showing the aortic valves, and does not need to look at irrelevant images such as images from other view types other than PLAX and PSAX.}

%DSMIL: where the first stream uses the standard max-pooling to identify the critical instance in the bag, and the second stream computes an attention score for each instance by measuring its distance to the critical instance.

 % predicting the color values in an image~\citep{zhang2016colorful},
\subsection{Self-supervised Learning and Pretraining of MIL}
Self-supervised learning (SSL) has demonstrated success in learning visual representations ~\citep{ chen2020simple, he2020momentum, chen2020improved, grill2020bootstrap, caron2020unsupervised, chen2021exploring,huang2023accuracy}. %Self-supervised learning methods generally 
SSL requires defining a pretext task such as
predicting the future in latent space~\citep{oord2018representation},
predicting the rotation of an image ~\citep{gidaris2018unsupervised},
or solving a jigsaw puzzle~\citep{noroozi2016unsupervised}.  %2. loss functions, such as logistic loss ~\citep{mikolov2013efficient}, margin loss~\citep{schroff2015facenet}, triplet loss~\citep{weinberger2009distance}. 
The term ``pretext'' suggests that the task being solved is not of primary downstream interest, but rather serves as a means to learn a better data representation. 
After selecting a pretext task, an appropriate loss function must also be selected.
%and loss functions can often be examined separately. 
Here, we focus on the instance discrimination task~\citep{wu2018unsupervised} and  InfoNCE loss~\citep{oord2018representation} following the success of \emph{momentum contrastive learning} (MoCo)~\citep{he2020momentum,chen2020improved}. 

% Loss functions can often be investigated independently of pretext tasks and various loss functions have been tried in prior works ~\citep{hadsell2006dimensionality, doersch2015unsupervised, zhang2016colorful, wang2015unsupervised, wu2018unsupervised, hjelm2018learning}.

Recently, self-supervision has been successfully applied to pretrain MIL models \citep{holste2022self, holste2022automated, liu2022multiple, lu2019semi, li2021dual, saillard2021self, dehaene2020self, rymarczyk2023protomil}. However, these studies all apply self-supervised contrastive learning to representations of individual images.
For example, \citet{liDomainGeneralizationMammography2021} encourage the embeddings of different views of the same patient to be similar, while \citet{chengContrastiveLearningEchocardiographic2022} specifically develop contrastive learning strategies for images from echocardiograms when view labels are known.
In our experiments, we observe image-level pretraining is not beneficial and sometimes \textbf{slightly harmful} for our AS diagnosis task.
This may be because the pretraining task's objective (learning good image level representations) being too distant from (or even contradict) the downstream task's objective (learning good bag-level representations for AS diagnosis). This could relate to an issue prior literature calls \emph{class collision}~\citep{arora2019theoretical, chuang2020debiased, khosla2020supervised, dwibedi2021little, zheng2021weakly, ash2021investigating, li2021comatch}. 

% This could be due to the pretraining objective, which is task-agnositic (pulling together an image with the augmentation of itself, and pushing away all the other) is too different from the objective of our downstream task which is task-specific (diagnosis AS using all images from a study with multiple instance learning), or more broadly the class collision issue that has been studied in prior literatures\citep{arora2019theoretical, chuang2020debiased, khosla2020supervised, dwibedi2021little, zheng2021weakly, ash2021investigating, li2021comatch} 

%\subsection{Connection to Knowledge Distillation}

% where the pretrained view classifier teaches the MIL model concept of 'relevant views'. 

% used its softmax predictions on each instance to teach the MIL model.

\subsection{Automated Screening of Aortic Stenosis.}

Work on automatic screening for aortic stenosis from echocardiograms has accelerated in the past few years~\citep{ginsbergDeepVideoNetworks2021,dai2023identifying,holste2022automated,wessler2023automated}, including recent work contemporaneous with this paper~\citep{vaseliProtoASNetDynamicPrototypes2023}.
Very recent work by \citet{krishna2023fully} demonstrated that a commercial deep learning system can closely emulate human performance on most of the elementary echocardiogram-derived measures for AS assessment, such as aortic valve area, peak velocity of blood through the valve, and mean pressure gradients. However, the inability to assign a study-level AS severity rating limits its usefulness as a screening tool.

Among previous efforts that can assign study-level AS grades, there are key differences in how they overcome the challenge of multi-view images available in each patient scan or \emph{study}.
Some groups have taken the \emph{Filter then Average} approach diagrammed in Fig.~\ref{fig:diagrams}(b). 
\citet{dai2023identifying} used a single video of the PLAX view to screen for AS.
\citet{holste2022automated} similarly filters to several PLAX videos, then uses a deep learning architecture specialized to video.
%This latter study reports strong external validation performance. 
Our team has previously pursued the \emph{Weighted Avg. by View Relevance} strategy in Fig.~\ref{fig:diagrams}(c),
combining separately-trained image-level view classifiers and image-level diagnostic classifiers via weighted averaging~\citep{huang2021new}.
This weighted averaging method 
was later refined for a clinical audience with external validation in \citet{wessler2023automated}.
A limitation of both filtering and weighting strategies is that by construction they treat images of relevant views equally; they cannot attend to some relevant views more than others.

Other work has pursued automated AS screening beyond echo images.
Some have created classifiers based on time-varying electrocardiogram signals~\citep{cohen2021electrocardiogram, elias2022deep}. Others have used wearable sensors~\citep{yangClassificationAorticStenosis2020}.
We argue that 2D echocardiograms remain the gold-standard information source for diagnosis.

The use of video, rather than still frames, is an advantage of some prior work~\citep{dai2023identifying,holste2022automated} over our approach.
%Our current study uses still frames available in an open-access dataset.
However, these video efforts evaluate on proprietary data, while our work emphasizes reproducibility by using still images from the open-access TMED dataset described below.
The MIL architecture proposed here could be extended to video by a straightforward adaptation of the instance representation layer.

\section{Dataset}
\label{sec:Dataset}

% \paragraph{TMED2 Dataset.}
% We evaluate our approach on an open-access echocardiogram dataset TMED2~\citep{huang2021new,huang2022tmed}. TMED2-2 is a dataset of 2D echocardiogram images. It provides a labeled set of 599 echocardiogram studies and an authentic unlabeled set of 5486 studies. Each study represents a routine TTE scan of a patient that contains images from various views of the heart. A typical study in the dataset contains around 68 images (median=68, 10-90th percentile range=27-97). For each study in the labeled set, a diagnosis label is provided on the study level and a subset of the images in this study (around 40\%)are provided with view labels while the others remain unlabeled. For each study in the unlabeled set, neither the diagnosis label nor any view labels are provided. According to \cite{mitchell2019guidelines}, at least 9 canonical view types frequently appear in routine TTEs. However, for our purpose of diagnosing AS, only some of the view types (PLAX and PSAX) are clinically relevant, as they show the aortic valve. We use the predefined train/val/test splits from TMED2, which has 360/119/120 studies respectively. 

%Trans-thoracic echocardiography (TTE) is a gold-standard way to non-invasively capture the heart's anatomy for measurement and diagnosis. 
In this work, for model training and primary evaluation we use an open-access dataset 
that our team created.
The Tufts Medical Echocardiogram Dataset (TMED)~\citep{huang2021new},
now in its latest version known as TMED-2~\citep{huang2022tmed},  is a collection of 2D echocardiogram images gathered during routine care at Tufts Medical Center in Boston, MA, USA from 2016-2021. 
Our research study of these  \emph{fully deidentified} images has been approved by 
%our Institutional Review Board (details withheld to preserve anonymity).
the Tufts Medical Center institutional review board.

Each study in the dataset represents a routine transthoracic echocardiogram (TTE) scan of one patient and includes \emph{all} collected 2D ultrasound images of the heart, with a median of 68 images per study (10-90th percentile range = 27-97). No filtering to specific views was applied except removal of Doppler images via metadata inspection. 
Each study's available images are exactly the 2D TTE images available to cardiologists in the health records system.

TMED-2 contains a labeled set of 599 studies. 
Every study in the labeled set has a standard 5-level rating of aortic stenosis (AS) severity assigned by a board-certified expert during routine reading. 
To focus on automated screening use cases, we followed our previous clinical work~\citep{wessler2023automated} and mapped each rating to one of 3 diagnostic classes: ``no AS'', ``early AS'' (combining mild and mild-to-moderate), and ``significant AS'' (combining moderate and severe). 
See App.~\ref{tab:TMED2_label_mapping} for further details on this label mapping.
Experts who assign these labels have access to more information than our algorithms (see Sec.~\ref{sec:Discussion}).

% in addition to the 2D images, clinician readers also see videos over time, Doppler images of blood flow, and other clinical variables not available in TMED-2.

\textbf{Splits.}
To make the most of the available data, we follow \citet{huang2022tmed} and average over 3 predefined training/validation/test splits. Each split divides the labeled set into 360/119/120 studies, each with similar proportions of no, early, and signficant AS.

\textbf{View labels for view classifiers.}
Roughly 40\% of images in TMED-2's labeled set are labeled with \emph{view type}, using 5 possible view labels: PLAX, PSAX, A2C, A4C, or Other. Only PLAX and PSAX views show the aortic valve and thus are relevant for AS assessment.
As per \citet{mitchell2019guidelines}, there are at least 9 canonical view types in routine TTEs, so many images in TMED-2 depict views that are ``irrelevant'' for AS diagnosis.
%View type labels are useful for training view classifiers.
Our later MIL approach does not need view labels at training or test time.
It does rely on a view classifier during training (Sec.~\ref{sec:methods_SA}), which we pretrain  using view labels in TMED-2's train set.
% which we fit using TMED-2's view labels.

\textbf{Unlabeled set for pretraining.}
TMED-2 additionally makes available a large \emph{unlabeled set} of 5486 studies from distinct patients. 
Studies in the unlabeled set have no diagnosis label nor view label.
We use this unlabeled set for pretraining representations (Sec.~\ref{sec:methods_CL}), but cannot use them for the supervised training of our MIL due to the lack of labels.

\textbf{2022-Validation dataset.}
For further evaluation, we obtained with IRB approval additional deidentified images from routine TTEs of 323 patients at our institution, collected during 2022 and thus temporally-external to the TMED-2 data.
Each study was again assigned an AS severity grading by a clinical expert during routine care.
We call this data \emph{2022-Validation}. 
It contains 225/48/50 examples of no/early/significant AS.

%We use the recommended predefined train/val/test splits from TMED2, which each consist of 360/119/120 studies. By averaging over 3 

% of 5486 studies (353,500 images).

\section{Methods}
\begin{figure}[!t]
% \includesvg[width=1\textwidth]{figures/SAMIL_diagram_draft1.svg}
\includegraphics[width=1\textwidth]{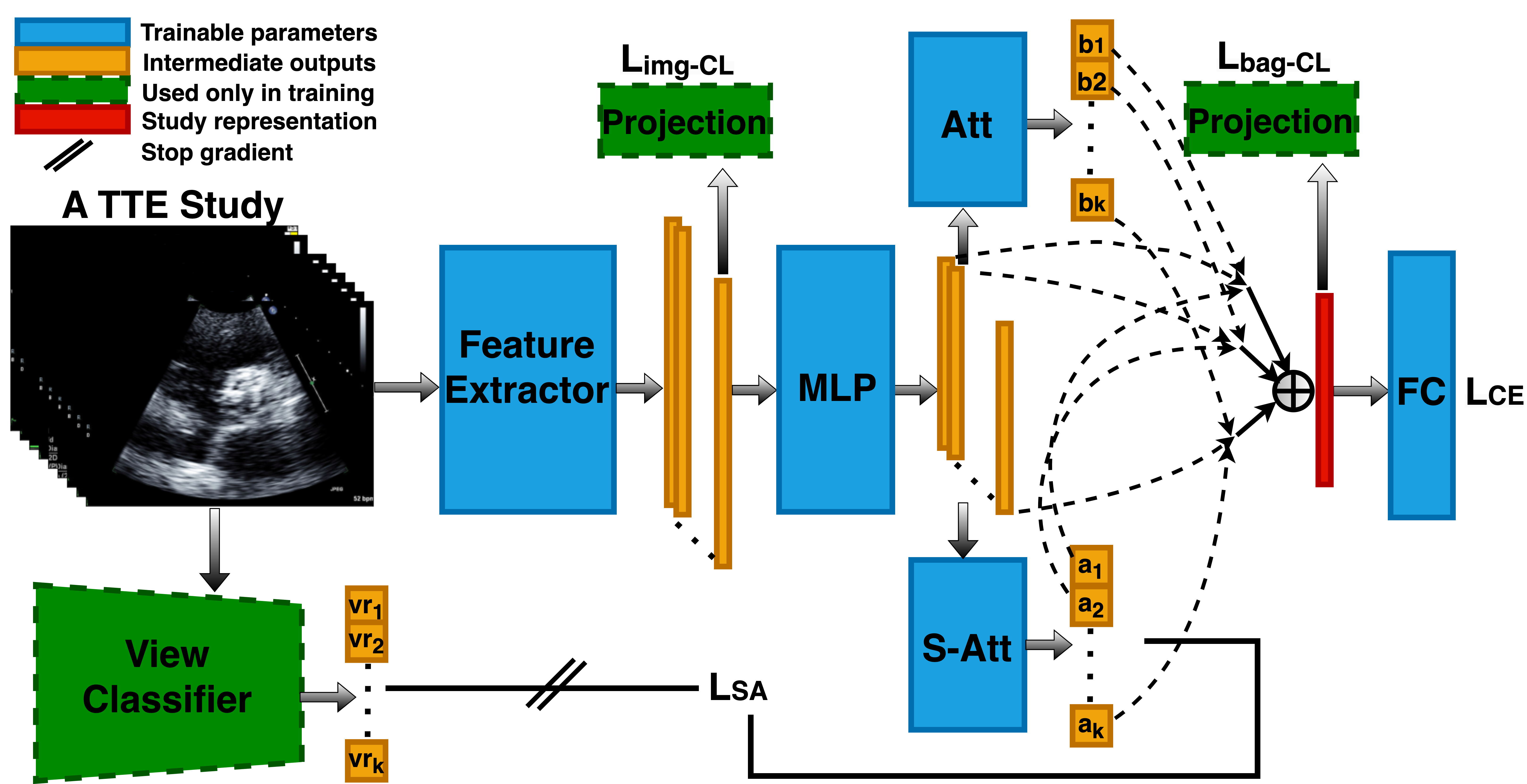}
\caption{\textbf{Overview of proposed method: Supervised Attention Multiple Instance Learning (SAMIL)}.
Given a study or ``bag'' with many images of diverse views of unknown type, a feature extractor processes each image individually into an embedding vector. Two attention modules (one supervised by a view classifier and one without) produce attention weights for each instance. The final study representation averages the image embeddings by combining the two attentions (Eq.~\eqref{eq:patient_embedding_samil}). A fully-connected (FC) layer maps the study representation to a 3-class diagnosis (no/early/significant AS). 
\emph{Pretraining:} SAMIL can be pretrained using bag-level (recommended, Sec.~\ref{sec:methods_CL}) or image-level contrastive learning. In either case, a projection head maps representations to a latent space where the contrastive loss is applied \citep{chen2020simple, chen2020improved}. The projection head is discarded after pretraining.
}%endcaption	
\label{fig:workflow_diagram}
\end{figure}

We now introduce our formulation of AS diagnosis as an MIL problem in Sec~\ref{sec:methods_formulation} and
discuss a general architecture for MIL (Sec.~\ref{sec:base_arch}).
We then present the two key innovations of our proposed method, which we call \emph{Supervised Attention Multiple Instance Learning} or SAMIL.
First, Sec.~\ref{sec:methods_SA} presents our supervised attention module that improves the MIL pooling layer to better attend to clinically relevant views.
Second, Sec.~\ref{sec:methods_CL} presents our study-level contrastive learning strategy to improve representation of entire studies (rather than individual images). Fig~\ref{fig:workflow_diagram} gives an overview of SAMIL.
%%Our proposed method is most similar to ABMIL, with two key components: A view regularization module that regularized the attention weights to align with the predicted view relevance, and a novel pretraining strategy that performs 'study-discrimination' instead of 'instance-discrimination'. 

\label{sec:Methods}
\subsection{Problem Formulation}
\label{sec:methods_formulation}

Let $D=\{(X_1, Y_1), \ldots, (X_N, Y_N)\}$ be a training dataset containing $N$ TTE studies. Each study, indexed by $i$, consists of a bag of images $X_i$ and an (optional) diagnostic label $Y_i$.

\textbf{Prediction task.} Given a training set of size $N$, our goal is to build a classifier that can consume a new echo study $X_*$ and assign the appropriate label $Y_*$.

\textbf{Input.}
Each ``bag'' $X_i$ contains $K_i$ distinct images: $\{x_{i1}, x_{i2}, \ldots, x_{iK_i}\}$, which are all 2D TTE images gathered during a routine echocardiogram. 
The number of images $K_i$ varies across studies (TMED-2's typical range 27-97).  
Each $x_{ik}$ is a grayscale 112x112 pixel image.

\textbf{Output.}
Each study's diagnostic label $Y_i \in \{0, 1, 2\}$ indicates the assessed severity level of aortic stenosis (0 = no AS, 1 =  early AS, 2 = significant AS). These labels are assigned by a cardiologist with specialty training in echocardiography during a routine clinical interpretation of the entire study. Diagnosis labels for individual images are unavailable. 
%Our goal is to build a model that given a new study $S_{new}$ predicts the diagnosis of Aortic Stenosis of this study.   

\textbf{Image preprocessing.}
We used the released dataset without additional preprocessing. 
As documented in \citet{huang2022tmed}, the images are extracted from DICOM files in the health record by taking the first frame of the corresponding cineloop, removing identifying information, padding the shorter axis to a square aspect ratio, and resizing to 112x112.

\subsection{General MIL architecture}
\label{sec:base_arch}

Following past work on deep neural network approaches to MIL~\citep{ilse2018attention, li2021dual}, a typical architecture has 3 components, as illustrated in Fig.~\ref{fig:diagrams}(d). First, an instance representation layer $f$ transforms each instance into a feature representation. Second, a pooling layer $\sigma$ aggregates across instances to form a bag-level representation in permutation-invariant fashion. Finally, an output layer $g$ 
maps the bag-level representation to a prediction.
%A general procedure for modeling multiple-instance learning problems consists of 3 steps :

%A mapping function $g$ that maps the aggregated feature representation to the bag label. 

We now describe the forward prediction process for one study or ``bag'' $X$ under a 3-component architecture specialized to our AS prediction problem.
Let $X = \{x_1, \ldots, x_K\}$ be the input bag of K instances, with individual instances indexed by integer $k$.
(We use $X$ interchangably with $X_i$, dropping study-specific index $i$ to reduce notational clutter.)

% Our work is built most directly upon ABMIL~\citep{ilse2018attention}, so when needed, we instantiate ABMIL-specific versions of $f$ and $\sigma$.
% For example, Set Transformer \citep{lee2019set} uses an attention mechanism in both $f$ and $\sigma$. ABMIL is simpler, with attention only used in pooling.

 %Our method builds upon ABMIL~\citep{ilse2018attention}, so when needed we instantiate 
%Our method builds up
%ABMIL \citep{ilse2018attention} uses an attention mechanism in pooling function $\sigma$. Set Transformer \citep{lee2019set} uses the attention mechanism in both $f$ and $\sigma$. 

\paragraph{Instance representation layer $f$.} 
Let $f$ be a row-wise feedforward layer that processes each instance $x_k \in \mathcal{X}$
independently and identically, producing an instance-specific embedding $h_k=f(x_k)$, where $h_k \in \mathbb{R}^{M}$. Following ~\citet{ilse2018attention}'s ABMIL, we use a stack of convolution layers and a MLP layer to extract and project each instance's feature representation to low-dimensional embedding. More details in App.~\ref{app:Architecture}.
%producing a bag of embeddings
%$H = \{h_1, \ldots, h_K\}$ \in \mathbb{R}^{L \times 1}$. 

\paragraph{Pooling layer $\sigma$.}
Following ABMIL, our pooling layer produces a bag-level representation $z \in \mathbb{R}^{M}$ via an attention-weighted average of the $K$ instance embeddings $\{h_1, \ldots h_K\}$:
\begin{equation}
    \label{eq:patient_embedding_abmil}
    z = \sum_{k=1}^K a_k h_k, \quad 
    a_k = \frac{\exp(w^\top \tanh(U h_k))}{\sum_{j=1}^K \exp(w^\top \tanh(U h_j))},
\end{equation} 
where vector $w \in \mathbb{R}^{L}$ and matrix $U \in \mathbb{R}^{L \times M}$ are trainable parameters of layer $\sigma$. Gated attention modules are also possible~\citep{ilse2018attention}, but we find accuracy gains are marginal.

\paragraph{Output layer $g$.} Given a bag-level feature vector $z = \sigma(f(X))$, the output layer performs probabilistic classification for the 3 levels of AS severity (0=none, 1=early, 2=significant) via a standard linear-softmax transformation of $z$:
\begin{align}
    p( Y = r | X) = g(z)_r ~\text{for}~ r \in \{0,1,2\}, \qquad g(z) = 
    \left[
    \frac{\exp( \eta_0^{\top} z)}{S(\eta,z)}, \frac{\exp( \eta_1^{\top} z)}{S(\eta,z)}, \frac{\exp( \eta_2^{\top} z)}{S(\eta,z)}
    \right].
\end{align}
Here, $\eta_0, \eta_1, \eta_2$ represent weights for each of the 3 severity levels of AS, and denominator $S = \sum_{r=0}^2 \exp( \eta_r^{\top} z)$ ensures the probabilities sum to one. We do include an intercept term for each class, but omit from notation for clarity.

\paragraph{Training.}
This 3-component deep MIL architecture has parameters $\eta$ for the output layer as well as $\theta$ for the pooling and representation layers ($\theta$ includes $w, U$ from Eq.~\eqref{eq:patient_embedding_abmil}).
We train these parameters by minimizing the cross-entropy loss between each study's observed AS diagnosis $Y$ and the MIL-predicted probabilities given each bag of images $X$ 
\begin{align}
    \theta^*, \eta^* = 
    \arg\!\min_{\theta, \eta} 
    \sum_{X, Y} \mathcal{L}_{\text{CE}}\left( Y, g_{\eta}( \sigma_{\theta}( f_{\theta}(X) ) \right) 
    \label{eq:L_CE}
\end{align}
In practice, weight decay is often used to regularize the model and improve generalization.
% Subscripts here remind us which layers depend on which parameters. In practical, additional regularization such as a weight decay loss may be needed to improve generalization.

% {@MCH I found prior works ususally don't include the WD term when talking about loss function. Shall we follow this convention, so that the notation is cleaner here, also when writing the total loss in Sec 4.5}

% \todo{with standard weight decay}.

% \begin{align}
%     \theta^*, \eta^* = \arg\!\min_{\theta, \eta} ~\sum_{i=1}^N \mathcal{L}_{\text{CE}}\left( Y_i, g_{\eta}( \sigma_{\theta}( f_{\theta}(X_i) ) \right) + \lambda \theta^T \theta + \lambda \eta^T \eta
% \end{align}

% {@MCH I found prior works ususally don't include the WD term when talking about loss function. Shall we follow this convention, so that the notation is cleaner here, also when writing the total loss in Sec 4.5}

\subsection{Contribution 1: Attention supervised by a view classifier}
\label{sec:methods_SA}

We find the attention-based architecture described above yields unsatisfactory performance in our diagnostic task (see entry labeled ABMIL in Tab.~\ref{tab:TMED2_BACC}). Furthermore, the learned attention values used in Eq.~\eqref{eq:patient_embedding_abmil} do not pass a clinical sanity check: attention should be paid only to PLAX and PSAX AoV view types, as only these show the aortic valve (see Fig.~\ref{fig:Attention_View_Alignment}).
This last observation suggests a path forward: supervising the attention mechanism. Suppose we have access to a trustworthy \emph{view-type-relevance} classifier $v : \mathcal{X} \rightarrow [0.0,1.0]$, which maps an image to the probability that it shows a relevant view depicting the aortic valve (either a PLAX or PSAX AoV view), rather than another view type (such as A2C, A4C, A5C, etc.).
This classifier could be used to guide the attention to focus on relevant images.
Classifying the view-type of a 2D TTE image has been demonstrated with high accuracy by several groups ~\citep{madani2018fast, zhang2018fully, long2018identification, huang2021new}. 

\paragraph{Supervised attention.}
To implement this idea, we introduce a new loss term, which we call supervised attention (SA), that steers the attention weights $A = \{a_1, \ldots a_K\}$ produced by Eq.~\eqref{eq:patient_embedding_abmil} to match relevance scores $R = \{r_1, \ldots r_K\}$ from a view-relevance classifier $v$: 
\begin{equation}
    \label{eq:L_SA}
    \mathcal{L}_{SA}(w, U) = \text{KL}(R || A) = \sum_{k=1}^K r_k \log \frac{r_k}{a_k}, \qquad  r_k = \frac{\exp(v(x_k)/\tau_{v})}{\sum_{k=1}^K \exp( v(x_k) / \tau_{v} ) } 
\end{equation}
%\mch{@HZ, please confirm direction of KL is correct. KL is not symmetric. I would have thought we should do $KL(R || A)$.}
Here, $\text{KL}$ means the KL-divergence between two discrete distributions over the same $K$ choices, and $R \in \Delta^K$ is a non-negative vector that sums to one obtained via a softmax transform of the view relevance probabilities with temperature scaling $\tau_{v} > 0$. We define view relevance probability as the sum of probability that the image is PLAX or PSAX. 

%$KL$ is the KL-divergence. $VR = {vr_1, \ldots, vr_K}$ is a bag of predicted view relevance where $vr_i = p(PLAX|x_i; \theta) + p(PSAX|x_i; \theta)$ and $\theta$ is a pretrained view classifier.
%\begin{equation}
%\phi(vr_i) = \frac{\exp(vr_i/\tau)}{\sum_{i=K}^P \exp(vr_i/\tau)}
%\end{equation} 
%normalize the vector $VR$ to a valid distribution with temperature scaling $\tau$. 

This supervision ensures the MIL diagnostic model attends to instances that are clinically plausible for the disease in question. That is, attention to PLAX or PSAX views that show the aortic valve is encouraged, and attention to irrelevant view types like A4C or A2C is discouraged. 
We emphasize that our approach is classifier-guided because reliable human-annotated labels are not always available. Only 40\% percent of images in TMED-2 training set have view labels. If expert-derived labels were more readily available, we could have supervised directly on those. Using classifier-provided probabilistic labels $R$ allows us to train easily on ``as-is'' data without expensive annotation effort.

Our supervised attention module can be seen as an example of \emph{knowledge distillation}~\citep{hinton2015distilling}, because the MIL model is ``taught'' to output attentions weight similar to the relevant view predictions from the pretrained view classifier. In a sense, the knowledge from the view classifier is distilled directly into the MIL model.

% regularization guides the MIL model to attend to where it suppose be attend as if it is a cardiologist (e.g., attend to PLAX or PSAX that show the aortic valve, instead of irrelevant background images or irrelevant views like A4C or A2C etc).

\paragraph{View classifier.}
We trained the view classifier $v$ on TMED-2's labeled \emph{and} unlabeled sets via a recently proposed semi-supervised learning method~\citep{huang2023fix} designed to be robust to realistic medical image datasets.
%The classifier is trained on images with view labels from the train set and all images in the unlabeled set. 
The classifier is trained to recognize the view type of an image, classifying it as either PLAX, PSAX or Other. 
To prevent data leakage, separate view classifiers are trained for each data split. 
See App ~\ref{app:ViewClassifier} for details.

\paragraph{Flexible attention.}
A potential drawback of enforcing a strict alignment between attention weights and predicted view relevance is reduced flexibility. 
Ideally, even after identifying images of \emph{relevant views}, we would like freedom to focus on some images more than others.
To achieve this, we introduce another set of attention weights $\mathrm{B} = \{b_1, \dots, b_K\}$.
%\begin{equation}
%    b_i = \frac{\exp(w_{b}^\top \tanh(U_b h_i))}{\sum_{i=1}^K \exp(w_{b}^\top \tanh(U_b h_i))},
%\end{equation} 
Together, the view-classifier-supervised attention $A$ and the flexible attention $B$ are combined to produce the final study-level represention $z \in \mathbb{R}^{M}$ by a simple construction,
\begin{equation}
\label{eq:patient_embedding_samil}
    z = \sum_{k=1}^K c_k h_k, \quad c_k(A,B) = \frac{a_k b_k}{\sum_{j=1}^K a_j b_j }, 
    \quad b_k = \frac{\exp(w_{b}^\top \tanh(U_b h_k))}{\sum_{j=1}^K \exp(w_{b}^\top \tanh(U_b h_j))}.
\end{equation}
In this way, the ultimate attention $c_k$ paid to an image can span the full range of 0.0 to 1.0 if that image is a relevant view, but is likely to be near 0.0 if the classifier deems that image's view irrelevant.
The trainable parameters that determine $B$ -- $w_b \in \mathbb{R}^{L}$ and $U_b \in \mathbb{R}^{L \times M}$ -- are not supervised by view-relevance, unlike their counterparts $w, U$ that determine $A$.
%from Eq.~\eqref{eq:patient_embedding_abmil}.

%The final attention weights for our model are based on the view regularized attention weights and the free attention weights 
%\begin{equation}
%    \alpha^f_i = \frac{\alpha_i \cdot \beta_i }{\sum_{i=1}^K %\alpha_i \cdot \beta_i} 
%\end{equation}
%And the final representation of the study is 

%move to Related work section
% \paragraph{Related work.}
% C2C~\citep{sharma2021cluster} also use KL-divergence to guide the attention weights, but different from ours the method is proposed for WSI classification. Further, the KL-divergence is measured between the attention weights against a predefined distribution (uniform distribution) with the purpose of minimize intra-cluster variance for patches coming from the same cluster. 

\subsection{Contribution \#2: Contrastive learning of entire study representations}
\label{sec:methods_CL}

Self-supervised learning (SSL) is an effective way to pre-train models that can be later fine-tuned to downstream tasks. 
As reviewed earlier, most previous methods \citep{holste2022self,saillard2021self,dehaene2020self} applying SSL to MIL tasks focus on pretraining the instance-level feature extractor $f$  (or part of $f$) aiming to learn better instance-level representations. % processing each image independently.
In contrast, we propose to pretrain the 2-component network $\sigma \circ f$, thus
refining the study-level representation vector $z$ summarizing all $K$ images in a routine echocardiogram.
In the vocabulary of MIL, we call this pretraining the ``bag-level'' representation.
Later results in Tab.~\ref{tab:Pretraining strategy ablation} show that our study-level pretraining leads to substantial accuracy gains at AS diagnosis compared to image-level pretraining.

%\todo{In contrast to most previous methods that pretrain only the feature extractor of the MIL model using each \textbf{image} independently, we propose to pretrain on the whole MIL network with each \textbf{study}, empirical results ~\ref{tab:Pretraining strategy ablation} shows that our pretraining strategy is better suited for the problem of diagnosing Aortic Stenosis using multi-view ultrasound images, leading to substantial performance gain compared to common approaches.}

\paragraph{MoCo(v2) for representations of individual images.}
Our pretraining strategy builds upon MoCo~\citep{he2020momentum,chen2020improved}, a recent method for self-supervised image-level contrastive learning (img-CL) that yields state-of-the-art representations via an instance discrimination task~\citep{wu2018unsupervised, ye2019unsupervised, bachman2019learning}. 
The learned embedding for a training image is encouraged to be similar to embeddings of slight transformations of itself, while being different from the embeddings of other images.

To obtain embeddings that should be similar, %given a training set of $J$ images $x_j \in \mathcal{X}$, 
each image $x_j$ in training goes through different transformations (e.g., random augmentation) to yield two versions of itself,
$x'_j$ and $x^+_j$, referred to as the ``query'' and the ``positive key''. These images are then \emph{encoded} into an $L$-dimensional feature space by composing a projection layer $\psi$ (a feed-forward network with $l_2$ normalization) onto the output of the instance-level representation layer $f$.
% : $\phi = \psi \circ f$.

To obtain embeddings that should be \emph{dissimilar} to a given query, MoCo retrieves $P$ previous embeddings from a first-in-first-out queue data structure.
For each new query, these are treated as $P$ ``negative keys''. In practice, this queue is updated throughout training at each new batch: 
the oldest elements are dequeued and all key embeddings from the current batch are enqueued. $P$ is usually set to the size of the queue \citep{he2020momentum}.  
% \todo{(usually nearly the size of the entire dataset)}.
%Let $N^{-}$ denote the number of elements in the queue, following convention in ~, we set $P$ to $N^{-}$. 

%will be sampled to serve as the embeddings of the ``negative key''. This queue contains the embeddings from previous mini-batches, and is updated each iteration by enqueuing the current minibatch's key embeddings and dequeuing its oldest elements. Let $N^{-}$ denote the number of elements in the queue, following convention in ~\citep{he2020momentum}, we set $P$ to $N^{-}$. 

To train image-level encoder $\phi = \psi \circ f$ that composes projection head $\psi$ with feature layer $f$ given a training set of $J$ images, we minimize InfoNCE loss~\citep{oord2018representation}:
\begin{equation}
    \label{eq:regular InfoNCE}
    \mathcal{L}_{\text{img-CL}}(\phi_q) = \sum_{j=1}^J -\log \frac{\exp(q_j^{\top} k^+_j/t)}{\exp(q_j^{\top} k^+_j/t)+\sum_{p=0}^P \exp(q_j^{\top} k^{-}_{jp}/t)}, 
    ~ 
    \begin{array}{cc}
    q_j = \phi_{q}(x'_j)
    \\
    k^+_j = \phi_{k}(x^+_j)
    %\\
    %k^-_{jp} = \phi_{k}(x_{jp}^{-})
    \end{array}
\end{equation}
\noindent Here, $q_j \in \mathbb{R}^L$ is an embedding of the ``query'' image, $k_j^+ \in \mathbb{R}^L$ is an embedding of the ``positive key'', and $k^-_{j1}, \ldots k^-_{jP} \in \mathbb{R}^L$ are $P$ embeddings of ``negative keys'' retrieved from the queue.
Scalar temperature $t > 0$ is a  hyperparameter~\citep{he2020momentum}. 

To improve representation quality, in MoCo queries and keys are encoded by separate networks: a query encoder $\phi_q$ with parameters $\theta_q$ and a key encoder $\phi_k$ with parameters $\theta_k$. 
%MoCo uses a query encoder $\phi_{q}$ with parameters $\theta_q$ and a key encoder $\phi_{k}$ with parameters $\theta_k$ to encode the queries and keys.
The query encoder $\phi_q$ is trained via standard backpropagation to minimize the loss above. The key encoder $\phi_k$ is only updated via momentum-based moving average of the query encoder: $\theta_k = m \theta_k + (1 - m)\theta_q$.
Momentum $m \in [0, 1)$ is often set to a relatively large value such as 0.999 to make the key embeddings more consistent over time:

\paragraph{Adapting MoCo to bag-level representations.}
%\todo{(@MCH we say 'many prior studies', do we need to cite more here?)} 

Most prior studies in the MIL literature, such as  ~\citet{li2021dual},  use an ``off-the-shelf'' version of image-level contrastive learning algorithm (e.g., SimCLR ~\citep{chen2020simple} or MoCo~\citep{he2020momentum,chen2020improved}) to pretrain feature extractor $f$ as described above.
However, we find that naively applying MoCo in this way does not yield useful results for our AS diagnosis problem.

Reasoning that what ultimately matters is the quality of the study-level representation $z$ produced by our MIL architecture, we adapted MoCo to produce solid representations of entire echocardiogram studies.
Correspondingly, we modified the InfoNCE loss to operate on the bag-level representations $z$. Given a training set of $N$ bags $X_1, \ldots X_N$, our approach to ``bag-level'' contrastive learning tries to pull together positive pairs of \emph{studies} and push away (make dissimilar) negative pairs of studies, via the bag-level contrastive-learning loss
\begin{equation}
    \label{eq:our InfoNCE}
    \mathcal{L}_{\text{bag-CL}}(\phi_q) = \sum_{i=1}^N -\log \frac{\exp(\tilde{z}_i^{\top} \tilde{z}_i^{+}/t)}{\exp(\tilde{z}_i^{\top} \tilde{z}_i^{+}/t) + \sum_{p=0}^P \exp(\tilde{z}_i^{\top} \tilde{z}_{ip}^{-})/t)}, 
    ~
    \begin{array}{cc}
         \tilde{z}_i = \phi_q(X'_i),
         \\
         \tilde{z}_i^{+} = \phi_k(X^{+}_i).
    \end{array}
\end{equation}
Here, encoder $\phi = \psi \circ \sigma \circ f$ now operates on \emph{all images in a study}, composing the same feature extractor $f$ with pooling layer $\sigma$ and projection head $\psi$ (note that pooling $\sigma$ is not used in Eq.~\eqref{eq:regular InfoNCE}).
 %$f$ and $\psi$ are the same feature extractor and projection head as in image-level case. 
 %However, a pooling layer $\sigma$ is used and the input is now . 
 $\tilde{z}=\psi(z)$ is the projection of $z$, $z_{i} = \sigma_q(f_q(X_i'))$ is the bag-level representation of the ``query'' study, and $z_{i}^{+} = \sigma_k(f_k(X_i^{+}))$ is the bag-level representation of the ``positive key'' study.
$X'$ and $X^+$ are obtained from the given study $X$ by applying different random augmentation to each of its images. 
$\tilde{z}_{ip}^{-}$ are again sampled from MoCo's queue. 
The enqueue and dequeue mechanisms and update rules of $\phi_q$ and $\phi_k$ are the same as the image level case.

\subsection{SAMIL Pipeline}
% \paragraph{self-supervised pretraining.} We pretrain the MIL network using studies in the labeled train set (both the view labeled images and the view unlabeled images in the study are used), as well as the large unlabeled set with our proposed bag-level pretraining strategy (see ~\ref{sec:methods_CL}). The projection head $f_{p}$ is discarded following convention ~\citep{chen2020improved, chen2020simple}.

\paragraph{Stage 1: Self-Supervised Pretraining.} 
We pretrain our SAMIL network on TMED-2 data utilizing our proposed bag-level pretraining strategy (Sec.~\ref{sec:methods_CL}).  This method can learn from all available studies, including both the labeled train set as well as the much larger unlabeled set (over 350,000 images).
After pretraining finishes, following convention~\citep{chen2020improved, chen2020simple}, the projection head $\psi_q$ is discarded, and parameters of $\sigma_q$ and $f_q$ are retained to warm-start the supervised fine-tuning. More details in App~\ref{app:SSL_Pretraining}.

% \paragraph{supervised fine-tuning of MIL using diagnosis label}

% We initialize the MIL network with the self-supervised pretrained weights. Next, we fine-tune the MIL network using studies in the labeled train set (both the view labeled images and the view unlabeled images in the study are used), supervised by the diagnosis label of each study, minimizing the loss
% \begin{equation}
%     \mathcal{L} =  \mathcal{L}_{CE} + \lambda \mathcal{L}_{SA}
%     % \mathcal{L}_{SA}(w, U) = \text{KL}(R || A) = \sum_{k=1}^K r_k \log \frac{r_k}{a_k}, \qquad  r_k = \frac{\exp(v(x_k)/\tau_{v})}{\sum_{k=1}^K \exp( v(x_k) / \tau_{v} ) }    
% \end{equation}
% Where $L_{CE}$ is the standard cross-entropy loss and $L_{SA}$ is the supervised attention loss in ~\ref{eq:L_SA}. Overall workflow can be seen in Fig~\ref{fig:workflow_diagram}.

\paragraph{Stage 2: Fine-Tuning to Diagnose Aortic Stenosis (AS).}
After initializing $f$ and $\sigma$ via stage 1, we fine-tune $f, \sigma$ and $g$ using complete studies (all available 2D images regardless of view label availability) from TMED-2's train set by minimizing the overall loss 
\begin{equation}
\label{eq:total_loss}
\mathcal{L} = \mathcal{L}_{CE} + \lambda_{SA} \mathcal{L}_{SA},
\end{equation}
Here, the primary supervision comes from each study's diagnosis label $Y$ (via cross entropy loss $\mathcal{L}_{CE}$ defined in Eq.~\eqref{eq:L_CE}), while the predicted view relevance of each image provides additional supervision to the attention module (via supervised attention loss $\mathcal{L}_{SA}$ in Eq.~\eqref{eq:L_SA}). 
Hyperparameter $\lambda_{SA} > 0$ sets the relative weight of the SA loss term.

\section{Results} 
\label{sec:Results}
\paragraph{Performance metrics.}
We use \emph{balanced accuracy} 
as our primary performance metric.
The class imbalance in TMED-2 means standard accuracy is less suitable~\citep{huang2021new}.  
Given a dataset of $N$ true labels $Y_{1:N}$ and $N$ predicted labels $\hat{Y}_{1:N}$, with each AS diagnosis label in $\{0, 1, 2\}$, we compute balanced accuracy as $\sum_{c=0}^{2} \frac{\text{TP}_{c}(Y_{1:N}, \hat{Y}_{1:N})}{N_{c}(Y_{1:N})}$, where $\text{TP}_c(\cdot)$ counts \emph{true positives} for class $c$ and $N_c(\cdot)$ counts all examples with class label $c$.
Later evaluations of screening potential assess discrimination between two classes via \emph{area under the ROC curve}.

%\begin{align}
%\text{balanced-accuracy}(y_{1:N}, \hat{y}_{1:N}) &= 
%\frac{1}{C} \sum_{c=1}^{C} \frac{\text{TP}_{c}(y_{1:N}, \hat{y}_{1:N})}{N_{c}(y_{1:N})}.
%\label{eq:balanced_accuracy}
%\end{align}
%Let $\text{TP}_c(\cdot)$ count \emph{true positives} for class $c$ (that is, the number of correctly classified examples whose true label is $c$), and $N_c(\cdot)$ gives the total number of examples with true label $c$. 

\paragraph{Comparisons.}
We compared our methods with a set of strong baseline including general-purpose multiple-instance learning algorithms ~\citep{ilse2018attention, lee2019set, li2021dual} and prior methods for Aortic Stenosis diagnosis using deep neural networks~\citep{wessler2023automated, holste2022automated, holste2022self}. %Results are shown in Table \ref{tab:TMED2_BACC}. 
We also tried DeepSet~\citep{zaheer2017deep}, but omit those results as we could not get DeepSet to perform better than random chance on this challenging diagnostic task despite substantial hyperparameter tuning (details in App.~\ref{App:DeepSet}).

%We did not show DeepSet in Tabel \ref{tab:TMED2_BACC}, because  despite substantial effort on hyperparameter tunning, we wouldn't able train DeepSet to obtain meaningful result. More details can be found in App ~ . 
%This again demonstrates how challenging the problem of diagnosing AS from ultrasound image is. 

% \todo{Another general-purpose \footnote{Although DSMIL is purposed mainly for WSI classification, the author claimed that the method also work well on general MIL problems} state-of-the-art deep MIL method is DSMIL~\citep{li2021dual}. We didnt' compare with DSMIL since we found that two of the three key components: multi-scale fusion and identifying key instances from max-pooling then calculate attention weight of each instance based on the distance to the key instance, are inappropriate for our problem if apply directly without substantial modification.}    

\subsection{Accuracy vs. Model Size Evaluation on TMED-2}
Table \ref{tab:TMED2_BACC} compares methods on test-set balanced accuracy for 3 diagnostic classes of AS across the 3 splits of TMED-2.
%In the results shown in Table \ref{tab:TMED2_BACC},
Our proposed method, SAMIL, scores 76\%, significantly better than 4 other state-of-the-art attention-based MIL architectures we tested (which span 60-67\%). 
SAMIL consistently improves over its predecessor ABMIL by a remarkable 13-17\% gain across all 3 splits.
%When compared to ABMIL, upon which SAMIL is built, our method achieves a remarkable performance gain of over \textbf{16\%}, with consistent gains across all three splits.
SAMIL also outperforms more recent MIL architectures like Set Transformer, which employs self-attention for both feature extraction and pooling, and the state-of-the-art DSMIL~\citep{li2021dual}, which leverages a two-stream architecture. 
%Despite SAMIL's conceptual simplicity, it performs notably better: approximately \textbf{14\%} better than Set Transformer and \textbf{9\%} better than DSMIL.

\begin{table}[!t]
    \begin{tabular}{l|rrr| r | r r}
	    & \multicolumn{4}{c}{Test Set Bal. Accuracy} & & \\
     Method & split 1 & split 2 & split 3 & average & \# params & view clf.?\\
    \hline
    Filter then Avg. [b] & 62.06 & 65.12 & 70.35 & 65.90 & 11.18 M & Yes
    \\
    W. Avg. by View Rel. [c]$*$
    & 74.46 & 72.61 & 76.24 & 74.43 &  5.93 M & Yes
    \\
    %JASE  & 11.87 M & 74.46 & 72.61 & 76.24 & 74.43 & Yes\\
    %Yale  & 11.18 M & 62.06 & 65.12 & 70.35 & 65.90 & Yes\\
    SAMIL (ours)            & 75.41 & 73.78 & 79.42 & \textbf{76.20}& 2.31 M & No
    \\
    \hline
    ABMIL [d]                 & 58.51 & 60.39 & 61.61 & 60.17 & 2.25 M & No\\
    ABMIL + Gate Attn. [d]  & 57.83 & 62.60 & 59.79 & 60.07 & 2.31 M & No \\
    Set Transformer [e]  & 60.95 & 62.61 & 62.64 & 62.06&  1.98 M & No \\
    DSMIL [f]  & 60.10 & 67.59 & 73.11 & 66.93&  2.02 M & No \\
    % VR ABMIL (ours)  & 2.31 M & 72.72 & 71.60 & 73.46 & 72.59& No \\
    \end{tabular}
    \\
    % {\small [b] \citet{holste2022automated}, [c] \citet{wessler2023automated} 
    %  [d] \citet{ilse2018attention} [e] \citet{lee2019set} [f] \citet{li2021dual} 
    % }%end citation list
     {\footnotesize [b] \citet{holste2022automated}, [c] \citet{wessler2023automated} 
     [d] \citet{ilse2018attention} [e] \citet{lee2019set} [f] \citet{li2021dual} 
    }%end citation list
    \caption{AS diagnosis results on TMED2. Showing balanced accuracy (percentage, higher is better) on the test set across three train/test splits. Methods b, c, d are diagrammed in corresponding panel in Fig.~\ref{fig:diagrams}.
    Methods above the line are approaches specialized to the AS task, others are generic MIL methods. Column ``\# params'' shows number of trainable parameters. Column ``view clf.?'' shows whether an additional view classifier is needed at deployment. $*$: value from the cited paper.
    }%endcaption
    \label{tab:TMED2_BACC}
\end{table}

Table \ref{tab:TMED2_BACC} also compares to two previous approaches dedicated to AS diagnosis: \emph{Filter then Average} and \emph{Weighted Average by View Relevance}.
Results suggest that our SAMIL method achieves better accuracy at substantially \emph{smaller model size}. Moreover, once trained, SAMIL can process the entire TTE study (dozens of images of different views) without the need to deploy an additional view classifier to filter \citep{holste2022automated,holste2022self} or downweight \citep{wessler2023automated} images. 
We thus find SAMIL to be an effective and portable alternative.
%This highlights the efficiency and effectiveness of SAMIL in comparison to other approaches.

To understand the source of SAMIL's gains, in the appendix we provide confusion matrices in Fig.~\ref{fig:confusion_matrix}.
%comparing it to W. Avg. by View Rel ~\citep{wessler2023automated} DSMIL ~\citep{li2021dual} and ABMIL~\citep{ilse2018attention}. 
SAMIL outperforms W. Avg. by View Rel. in early AS recall, while maintaining similar or slightly lower no AS and significant AS recall. Compared to DSMIL, SAMIL improves no AS and early AS recall, with similar significant AS recall. Compared to ABMIL, SAMIL performs better in all three categories.
Further results in Fig~\ref{fig:TMED2_roc} show ROC curves indicating discriminative performance of three clinical use cases for binary screening (no vs some AS, early vs significant, and significant AS vs not). SAMIL outperforms ABMIL and DSMIL across all tasks. 
In comparison to W. Avg. by View Rel, SAMIL reaches similar performance in screening No AS vs. Some AS, while doing better in the other two tasks.

% Compared to W. Avg. by View Rel, SAMIL achieves better recall for early AS while being on par or slightly worse on recall for no AS and Significant AS. Compared to DSMIL, SAMIL achieve better recalls on no AS and early AS, while being on par on significant AS. Compared to ABMIL, SAMIL achieves better recall for all the three categories.  

%\subsection{Using Automatic Study-Level Diagnosis as a Preliminary Screening Tool for AS}

% Discriminatory performance for various clinical use cases are shown in Fig~\ref{fig:TMED2_roc}. SAMIL performs significantly better than ABMIL and DSMIL on all three tasks. Compared to W. Avg. by View Rel, despite being smaller in model size, SAMIL achieves comparable performance on No AS vs Some AS, while clearly better at Early vs Significant and No Significant vs Significant tasks.

\subsection{Screening Evaluation on 2022-Validation Dataset}
We further validate methods on the separate 2022-Validation dataset described earlier, which contains 225/48/50 examples of no/early/significant AS. 
Results in Tab.~\ref{tab:AUC_analysis_323}
compare SAMIL to the best-performing baselines from previous section. 
SAMIL achieved competitive performance on two critical screening tasks: It seems best on Significant-vs-Not and equivalent to the best on No-vs-Some. On the more challenging Early-vs-Significant task, where both classes have 50 or fewer examples in this set, all methods have wide uncertainty intervals from bootstrap resamples and SAMIL scores slightly below DSMIL.

%Note that the early-vs-significant task has especially wide uncertainty intervals due to the small number of available cases (9) of early AS.

% \begin{table}[h]
% \centering
% \begin{tabular}{c|c|c|c}
%        & \multicolumn{3}{c}{AUROC for AS screening}
%        \\
% Method & No vs Some & Early vs Significant & Significant vs. Not \\
% \hline
% W. Avg. by View Rel.  & 0.948 (0.898, 0.980)  & 0.582 (0.312, 0.829) & 0.909 (0.859, 0.949)
%  \\
% DSMIL.  & 0.899 (0.839, 0.945)  & 0.881 (0.730, 0.991) & 0.941 (0.893, 0.978)\\
% SAMIL (ours) & 0.935 (0.867, 0.978)  & 0.805 (0.645, 0.931) & 0.958 (0.923, 0.983) \\

% \end{tabular}
% \caption{AUROC for AS screening on temporarily distinct cohort. Values in parenthesis show 2.5th and 97.5th percentiles of AUROC values computed from 5000 bootstrap resamples of 263 studies.}
% \label{tab:AUC_analysis_263}
% \end{table}

\begin{table}[h]
\centering
\begin{tabular}{c|c|c|c}
       & \multicolumn{3}{c}{AUROC for AS screening}
       \\
Method & No vs Some 
    & Significant vs. Not
    & Early vs Significant 
    \\
\hline
W. Avg. by View Rel.  & 0.934 (0.904, 0.959)  
    & 0.881 (0.837, 0.921)
    & 0.653 (0.539, 0.760) 
 \\
DSMIL.  & 0.897 (0.862, 0.929)  
    & 0.902 (0.857, 0.941)
    & 0.765 (0.664, 0.857) 
    \\
SAMIL (ours) & 0.923 (0.885, 0.955)  
    & 0.921 (0.886, 0.951) 
    & 0.717 (0.610, 0.813) 
    \\
\end{tabular}
\caption{AUROC for AS screening on temporarily distinct cohort. Values in parenthesis show 2.5th and 97.5th percentiles of AUROC values computed from 5000 bootstrap resamples of 323 studies.}
\label{tab:AUC_analysis_323}
\end{table}

\subsection{Attention Quality Assessment}

\begin{figure}
\begin{tabular}{c c c }
     Split1 & Split2 & Split3 
    \\
    \includegraphics[width=0.32\textwidth]{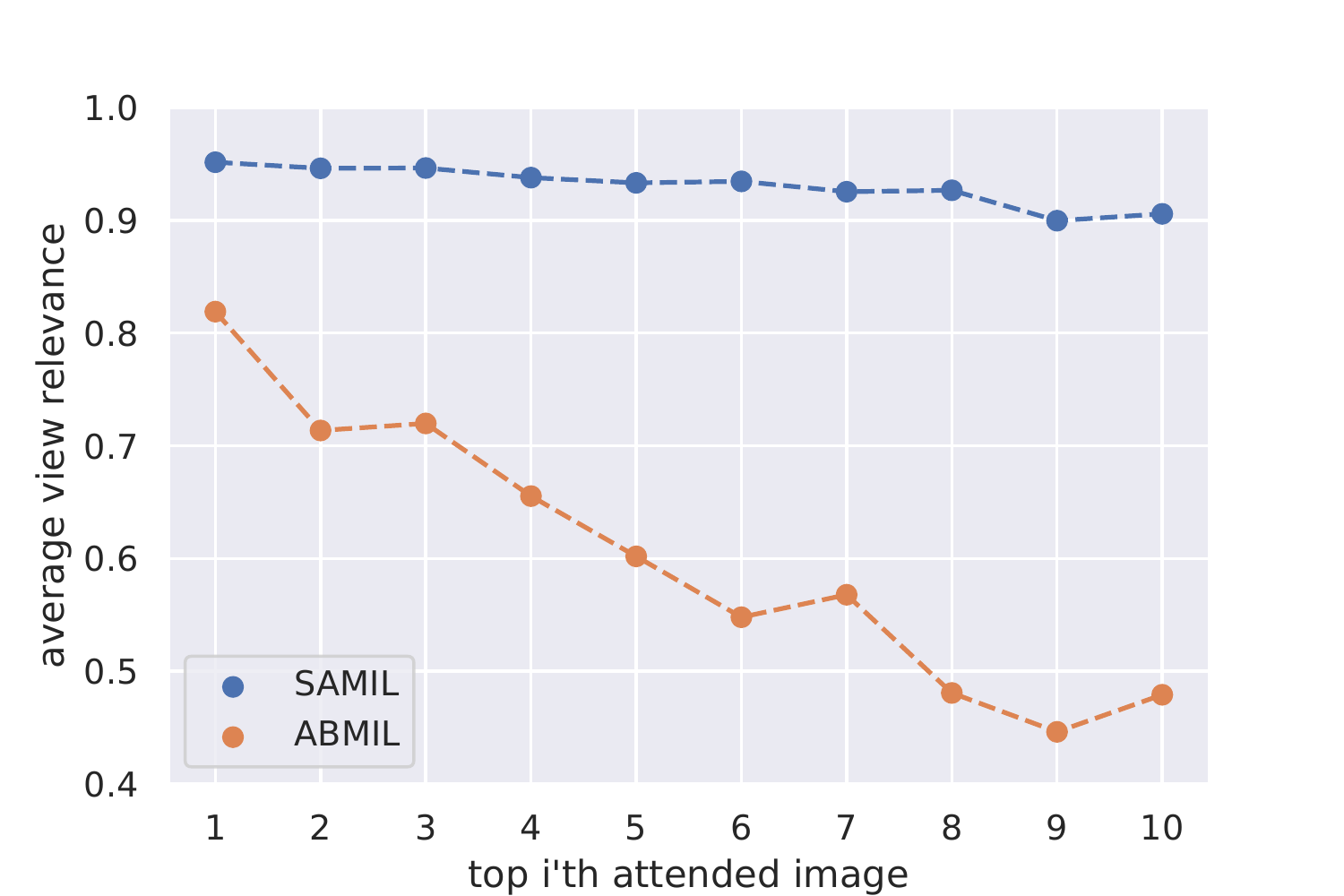}
    &
    \includegraphics[width=0.32\textwidth]{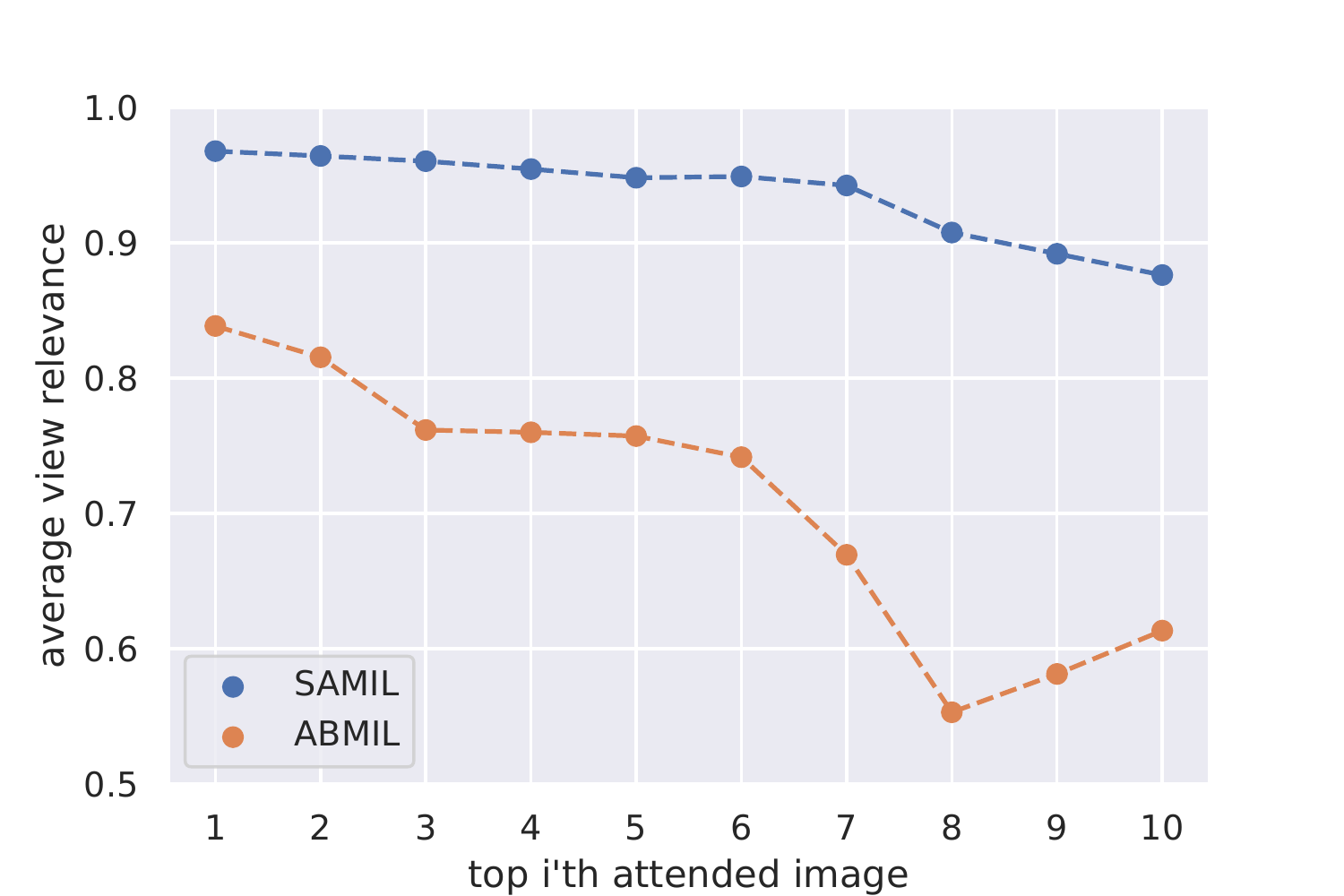}
    &
    \includegraphics[width=0.32\textwidth]{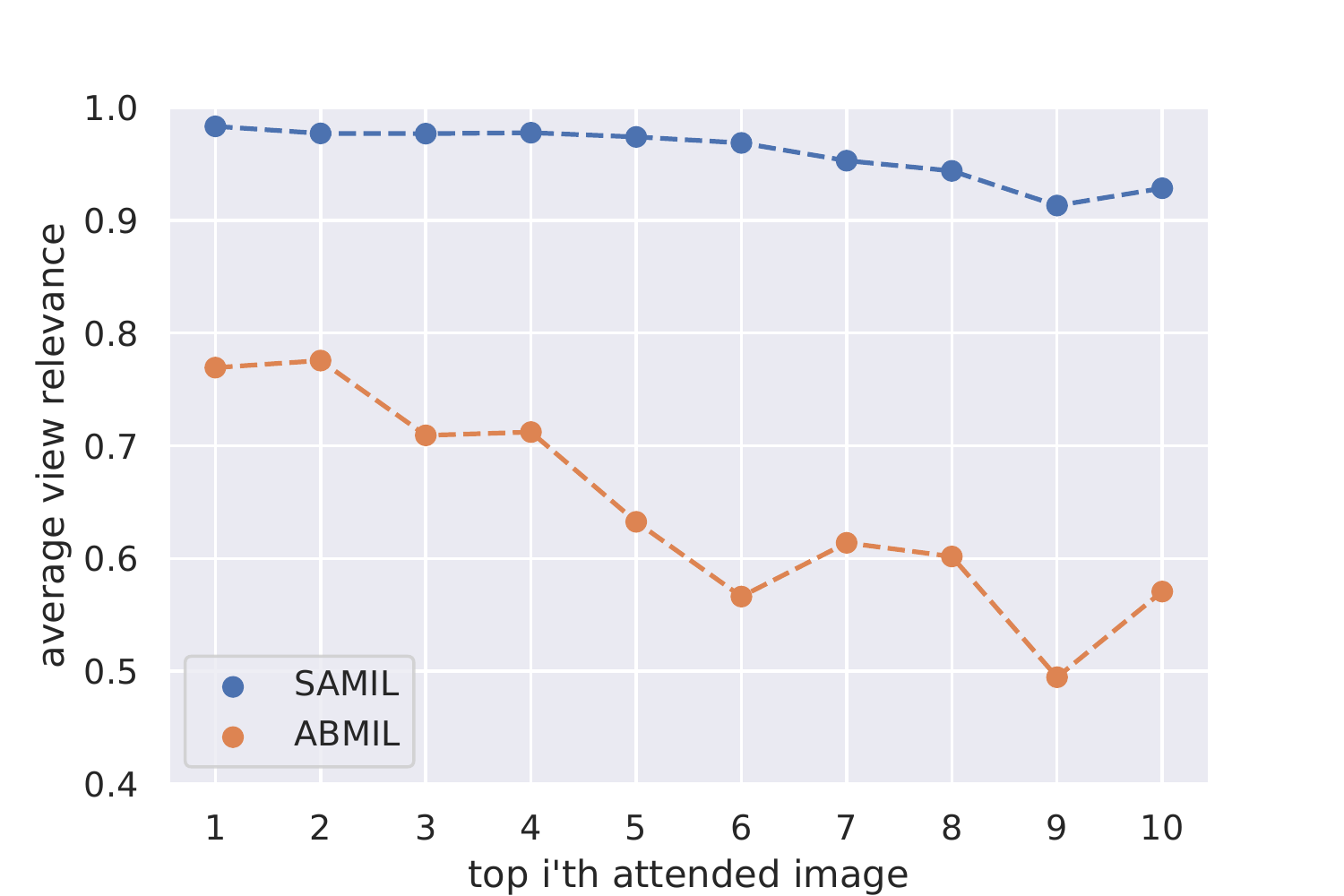}
    \end{tabular}	
    \caption{
    Predicted view relevance of top-ranked images by attention (higher is better).
    %Showing the attention weights and view relevance alignment across all studies in the test set for multiple train/test splits. 
    Supervised attention (SAMIL, ours) outperforms off-the-shelf ABMIL by wide margin across all 3 splits.
    The x-axis indicates a rank position of images within an echo study when sorted by attention (1 = largest $a_k$, 2 = second largest, etc.). The y-axis indicates the average view relevance (across studies in test set) assigned by view classifier $v(x)$ to image $x$ at rank $k$. 
     }
    \label{fig:Attention_View_Alignment}
\end{figure}

Our supervised attention module is intended to ensure that the model's decision-making process is consistent with human expert intuition, by only using relevant views to make diagnostic judgments.
Here, we evaluate how well the attention mechanisms of various models align with this goal. 
Fig \ref{fig:Attention_View_Alignment} compares the predicted view relevance of SAMIL's and ABMIL's attended images, aggregating across all studies in the test set. For instance, the first panel reveals that after ranking by attention, the 9th ranked image by ABMIL on average has less than 0.5 view relevance. This means that for many studies, some images in the top 9 (as ranked by attention) are likely from irrelevant views. In contrast, SAMIL's 9th ranked image has an average view relevance above 0.9.
Overall, the figure demonstrates that SAMIL bases decisions on clinically relevant views, while ABMIL fails this clinical sanity check. 
We hope these evaluations reveal how our SAMIL's improved attention module contributes to helping audit a model's overall interpretability, which is key to gaining trust from clinicians and successfully adopting an ML system in medical applications ~\citep{holzinger2017we, lundberg2017unified, tonekaboni2019clinicians}. 

%In addition to achieving superior performance, our proposed method enhances the interpretability and trustworthiness of the model compared to standard off-the-shelf MIL models. 

% Fig ~\ref{fig:Attention_View_Alignment} compares the predicted view relevance of each attended images of SAMIL with those of ABMIL. This visual shows an aggregated perspective across all studies in the test set. For example, in split 1, the top 9th attended images across all studies, on average only has a predicted view relevance of less than 0.5. This indicates that for many studies in this test split, the 9th attended images by ABMIL is likely from \textbf{irrelevant views}. On the other hand, for the same figure, the top 9th attended images by SAMIL has an average predicted view relevance of above 0.9, which substaintially higher than that of the ABMIL (0.5), indicating that these attended images by SAMIL are very likely to from relevant views. The figure as a whole, demonstrates that SAMIL makes decisions based on clinically relevant views, while ABMIL fails the clinical sanity check.

We provide two additional sanity checks for our supervised attention module. 
First, Fig ~\ref{fig:top_attended_images} illustrates the top 10 images ranked by attention from one study (the first in the test set to avoid cherry-picking).
Among the top 10 images attended by ABMIL, 5 are actually irrelevant views. In contrast, the top 10 images attended by SAMIL are all from relevant views.
Second, we assess the view classifier's performance on the view classification task in App~\ref{app:ViewClassifier}, supporting that its predicted view relevance serves as a reliable indicator for assessing whether an image comes from a relevant view or not.

\setlength{\tabcolsep}{0.1cm}

% Concretely, clinicians use clinically relevant views (PLAX and PSAX) to make AS grading. Similar our model  to make the predictions based on relevant views from each TTE studies. \todo{elaborate/cite paper how this help gain trust from clinicians}.

\subsection{Ablation Evaluations of Supervised Attention and Pretraining}

Tables~\ref{tab:View Regularization ablation} and \ref{tab:Pretraining strategy ablation} verify the impact of our attention (Sec.~\ref{sec:methods_SA}) and pretraining (Sec.~\ref{sec:methods_CL}) methods.

\begin{table}[!h]
\parbox{.45\linewidth}{
\centering
    \begin{tabular}{l|rrr|c}
	    & \multicolumn{4}{c}{Test Set Bal. Accuracy} \\
     Method & 1 & 2 & 3 & average \\
    \hline
    ABMIL             & 58.5 & 60.4 & 61.6 & 60.2 \\
    ABMIL Gate Attn.  & 57.8 & 62.6 & 59.8 & 60.1  \\
    SAMIL no pretrain & 72.7 & 71.6 & 73.5 &\textbf{72.6}
    \end{tabular}
    \caption{Ablation of \textbf{attention} strategies on TMED2. Showing balanced accuracy for AS severity (higher is better) on the test set across splits. Model sizes are matched to (roughly) 2.3M parameters.
    }%endcaption
    \label{tab:View Regularization ablation}
}%endparbox
\hfill 
\parbox{.45\linewidth}{
    \centering
    \begin{tabular}{l|rrr|c}
	          & \multicolumn{4}{c}{Test set Bal. Accuracy} \\
     Method & 1 & 2 & 3 & average \\
    \hline
    
    SAMIL no pretrain & 72.7 & 71.6 & 73.5 & 72.6 \\
    SAMIL w/ img-CL & 71.2 & 67.0 & 75.8 & 71.4 \\
    SAMIL           & 75.4 & 73.8 & 79.4 & \textbf{76.2}
    \end{tabular}
    \caption{Ablation of \textbf{pretraining} strategies on TMED-2. Reporting balanced accuracy for AS severity (higher is better) on the test set across splits.
    Model sizes are matched.
    % to 2.3M parameters.
	Last row uses recommended ``bag-CL'' pretraining.
    %Our proposed SAMIL uses study-level contrastive learning (CL), which we compare to commonly-used image-level CL and no pretraining at all.
    }%endcaption
    \label{tab:Pretraining strategy ablation}
}%endparbox
\end{table}

Our ablation of the attention module used within pooling layer $\sigma$ in Table~\ref{tab:View Regularization ablation} demonstrates the effectiveness of SAMIL's supervised attention (Eq.~\eqref{eq:L_SA}).
SAMIL achieves an improvement of over 12\% compared to ABMIL, the model it builds upon, even without any pretraining.

To understand what SAMIL's built-in study-level (aka bag-level) pretraining adds, we compare image-level contrastive learning (``w/~img-CL'') and without pretraining at all. Table~\ref{tab:Pretraining strategy ablation} shows that image-level pretraining does not improve AS diagnosis performance, while our proposed study-level pretraining strategy in SAMIL delivers gains.

% \begin{table}[!h]
%     \centering
%     \begin{tabular}{l|l|rrr|c|c}
% 	    & & \multicolumn{3}{c}{Split-specific Test Set} & & \\
%      Method & parameters & 1 & 2 & 3 & average \\
%     \hline
%     ABMIL             & 2.25 M & 58.51 & 60.39 & 61.61 & 60.17 \\
%     ABMIL Gate Attn.  & 2.31 M & 57.83 & 62.60 & 59.79 & 60.07  \\
%     SAMIL no pretrain & 2.31 M & 72.72 & 71.60 & 73.46 &\textbf{72.59} \\
%      \\
    
%     \end{tabular}
%     \caption{Ablation of attention strategies on TMED2 AS diagnosis. Showing balanced accuracy on the test set across multiple train/test splits.
%     }%endcaption
%     \label{tab:View Regularization ablation}
% \end{table}

% \begin{table}[!h]
%     \centering
%     \begin{tabular}{l|l|rrr|c|c}
% 	    & & \multicolumn{3}{c}{Split-specific Test Set} & & \\
%      Method & parameters & 1 & 2 & 3 & average \\
%     \hline
    
%     SAMIL no pretrain & 2.31 M & 72.72 & 71.60 & 73.46 & 72.59 \\
%     SAMIL w/ Image CL & 2.31 M & 71.24 & 67.04 & 75.84 & 71.37 \\
%     SAMIL            & 2.31 M & 75.41 & 73.78 & 79.42 & \textbf{76.20} \\
    
%     \end{tabular}
%     \caption{Ablation of SSL pretraining strategies on TMED2 AS diagnosis.  Showing balanced accuracy on the test set across multiple train/test splits.
%     Our proposed SAMIL uses study-level contrastive learning (CL), which we compare to commonly-used image-level CL and no pretraining at all.
%     }%endcaption
%     \label{tab:Pretraining strategy ablation}
% \end{table}

\section{Discussion} 
\label{sec:Discussion}
We have developed an approach to deep multiple instance learning for diagnosing a common heart valve disease (aortic stenosis) from the dozens of images collected in a routine echocardiogram. In our evaluations on the open-access TMED-2 dataset, we find our approach reaches better classifier accuracy than several alternatives, including two recent methods dedicated to AS screening. 
We suspect that gains come from two sources. First, SAMIL can use images of both PLAX and PSAX views, not just PLAX as in \citet{holste2022automated}. Second, SAMIL's flexible attention (Eq.~\eqref{eq:patient_embedding_samil}) does not weight each relevant view equally. 
\citeauthor{holste2022automated}'s \emph{Filter-then-Average} and \citeauthor{wessler2023automated}'s \emph{Weighted Average by View Relevance} essentially treat each high-confidence PLAX image \emph{equally} in diagnosis. Instead, we emphasize that our method can learn a study-specific subset of PLAX images to attend to, based on image quality, anatomic visibility, or other factors.

\paragraph{Limitations in diagnostic potential.}
Human experts assess AS using several additional factors not available to our method. These include patient demographics, clinical variables, and (most importantly) other imaging technologies such as doppler echocardiography as well as high-resolution cineloop videos from 2D TTE (not just lower-resolution single frame images used here). 
Adapting SAMIL to these modalities could provide improved accuracy.
% further gains.

\paragraph{Limitations in evaluation.}
As of this writing, TMED-2 is the only open-access dataset of echos known to us with diagnostic labels for AS or other valve disease.
However, it is limited in size and in covered demographics due to drawing from just one hospital site.
Further assessment is needed to understand how our proposed method generalizes, especially to populations underrepresented at the Boston-based hospital where this data was collected.

\paragraph{Advantages.}
Our SAMIL approach is designed to perform automatic screening of an echo study without requiring a first-stage manual or automatic prefiltering to relevant view types.
We further leverage large unlabeled data collections for pretraining. Another direction to utilize unlabeled data is semi-supervised learning~\citep{zhu2005semi}, which we leave for future work.

SAMIL could be applied to other structural heart diseases including cardiomyopathies and mitral and tricuspid disease if suitable labels were available for some studies. 
Similar multi-view image diagnostic problems occur in fetal ultrasound, lung ultrasound, and head CT applications; we hope translation to these other domains could bear fruit. 
Both key innovations -- supervised attention to steer toward clinically-relevant views for the diagnostic task and study-level representation learning -- are applicable to other prediction tasks.
Ultimately, we hope our study plays a part in transforming early screening for AS and other burdensome diseases to be more reproducible, effective, portable, and actionable.

% ACKNOWLEDGEMENTS ONLY GO IN THE CAMERA-READY, NOT THE SUBMISSION
\acks{
We acknowledge financial support from the Pilot Studies Program at the Tufts Clinical and Translational Science Institute (Tufts CTSI NIH CTSA UL1TR002544). 
We are also grateful for computing infrastructure support from the Tufts High-performance Computing cluster, partially funded by the National Science Foundation under grant OAC CC* 2018149.
Author B. W. was supported in part by K23AG055667 (NIH-NIA).
}

\begin{small}
\bibliography{refs_manual.bib,refs_from_zotero_multiview.bib}    
\end{small}

\clearpage
\appendix

%% Config Table-of-Contents to track the sections of the appendix
\startcontents[sections]

\counterwithin{table}{section}
\setcounter{table}{0}
\counterwithin{figure}{section}
\setcounter{figure}{0}
%\counterwithin{algorithm}{section}
%\setcounter{algorithm}{0}

%% Use ONE counter for all figs and tables to give unique identifiers in supplement
\makeatletter 
\let\c@table\c@figure
\let\c@lstlisting\c@figure
\let\c@algorithm\c@figure
\makeatother

% Print Table of Contents
\subsection*{Appendix Contents}
\printcontents[sections]{l}{1}{\setcounter{tocdepth}{2}}

\section{Further Results}
\subsection{Confusion matrix}
\newcommand{\BW}{0.29}
\setlength{\tabcolsep}{0.05cm}
% \begin{figure}[!h]
\begin{figure}[H]
\begin{tabular}{r c c c }
    & Split 1 & Split 2 & Split 3
    \\
    {\rotatebox{90}{~~~~~W. Avg. by View Rel.}}
    & 
    \includegraphics[width=\BW\textwidth]{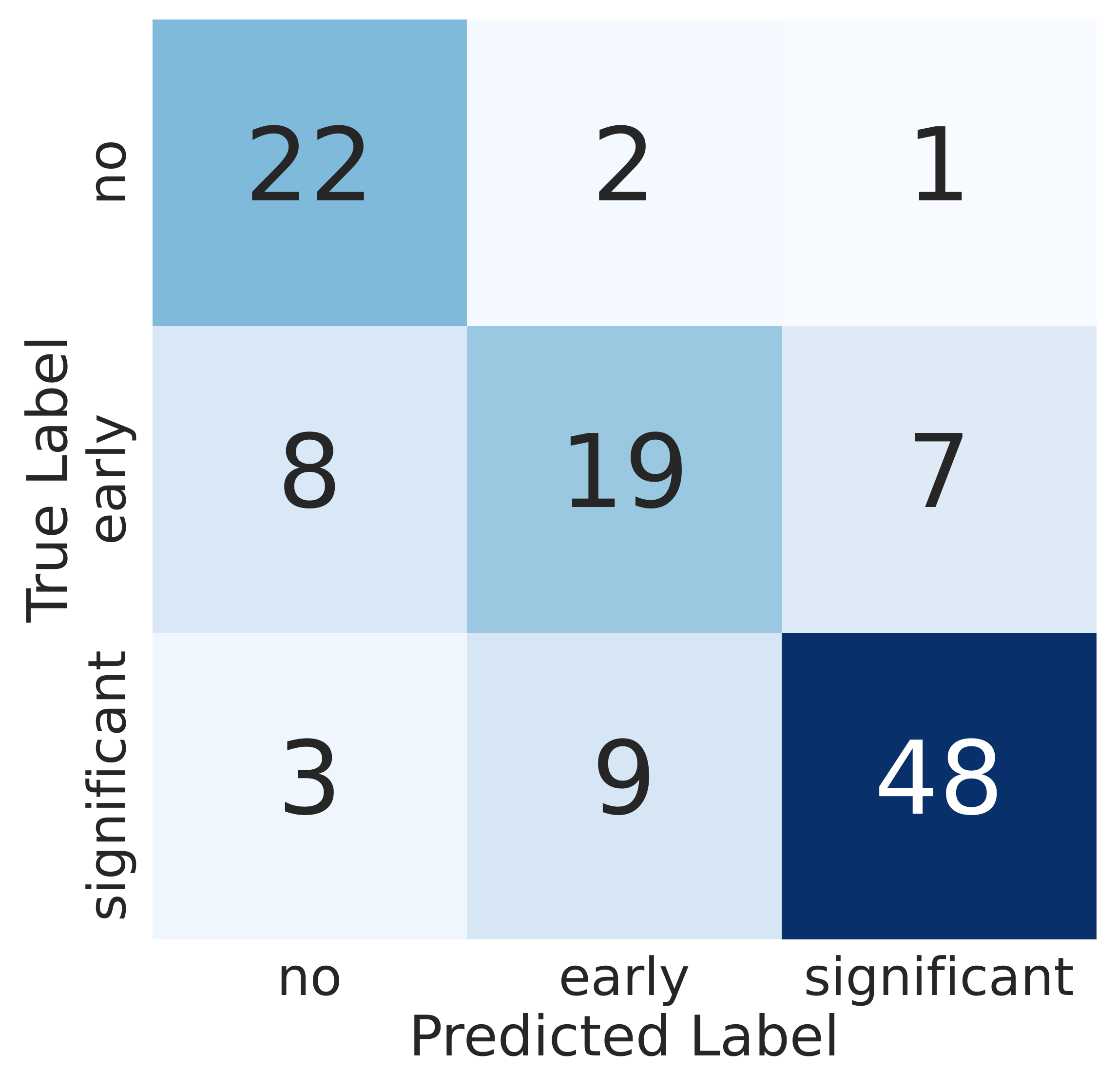}
    &
    \includegraphics[width=\BW\textwidth]{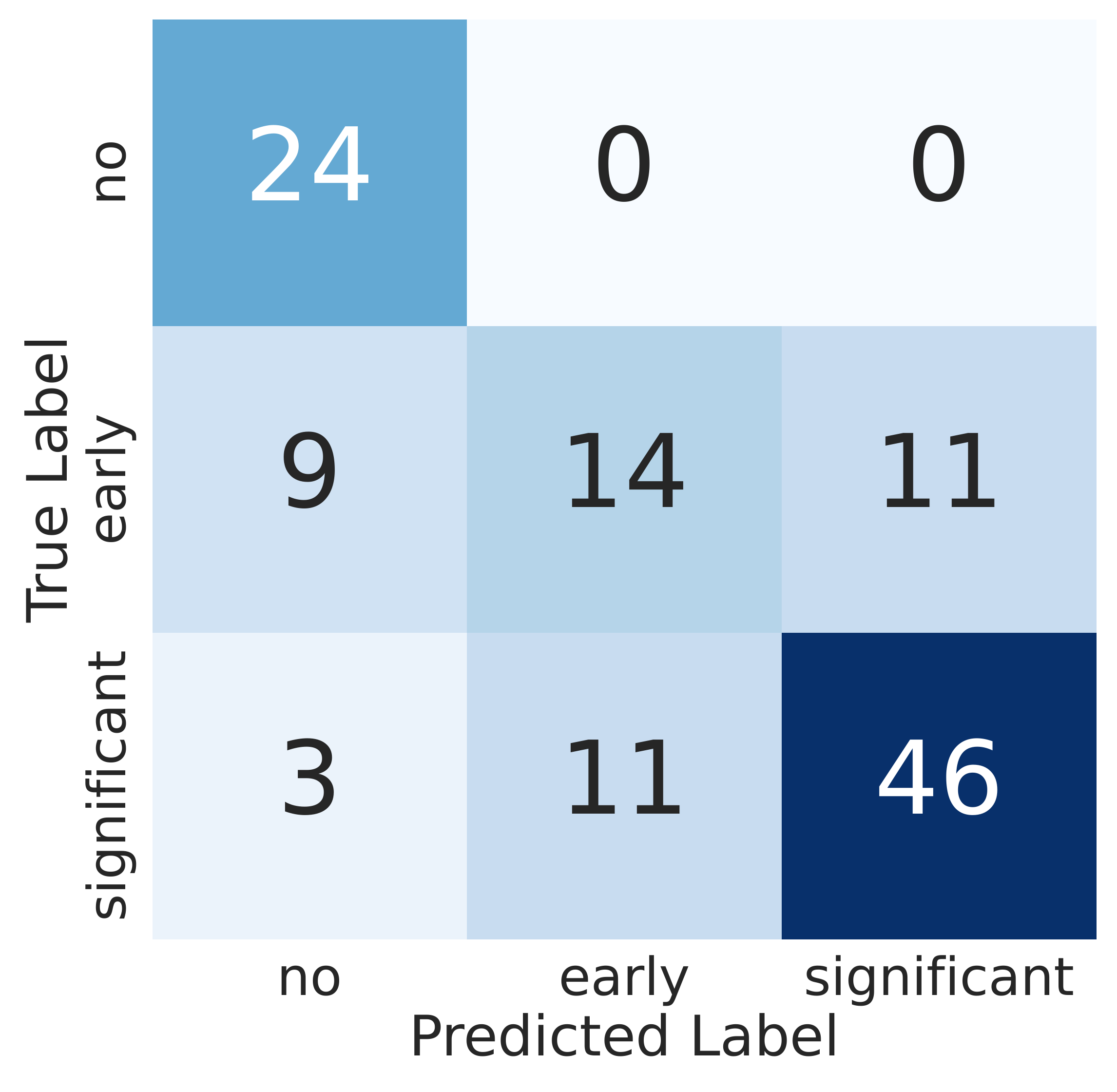}
    &
    \includegraphics[width=\BW\textwidth]{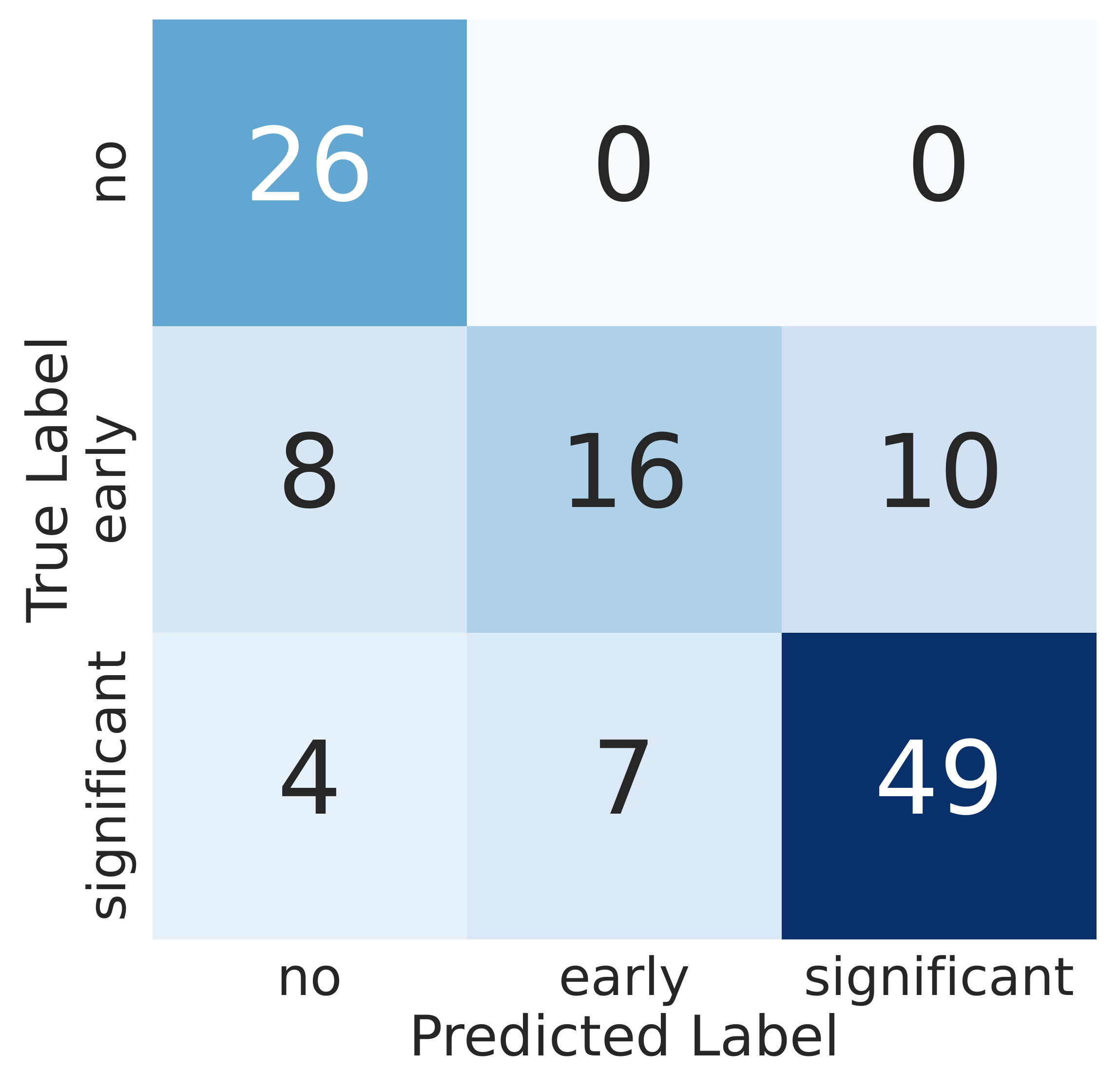}
    \\
    {\rotatebox{90}{~~~~~~~~~~~~ABMIL}}
    & 
    \includegraphics[width=\BW\textwidth]{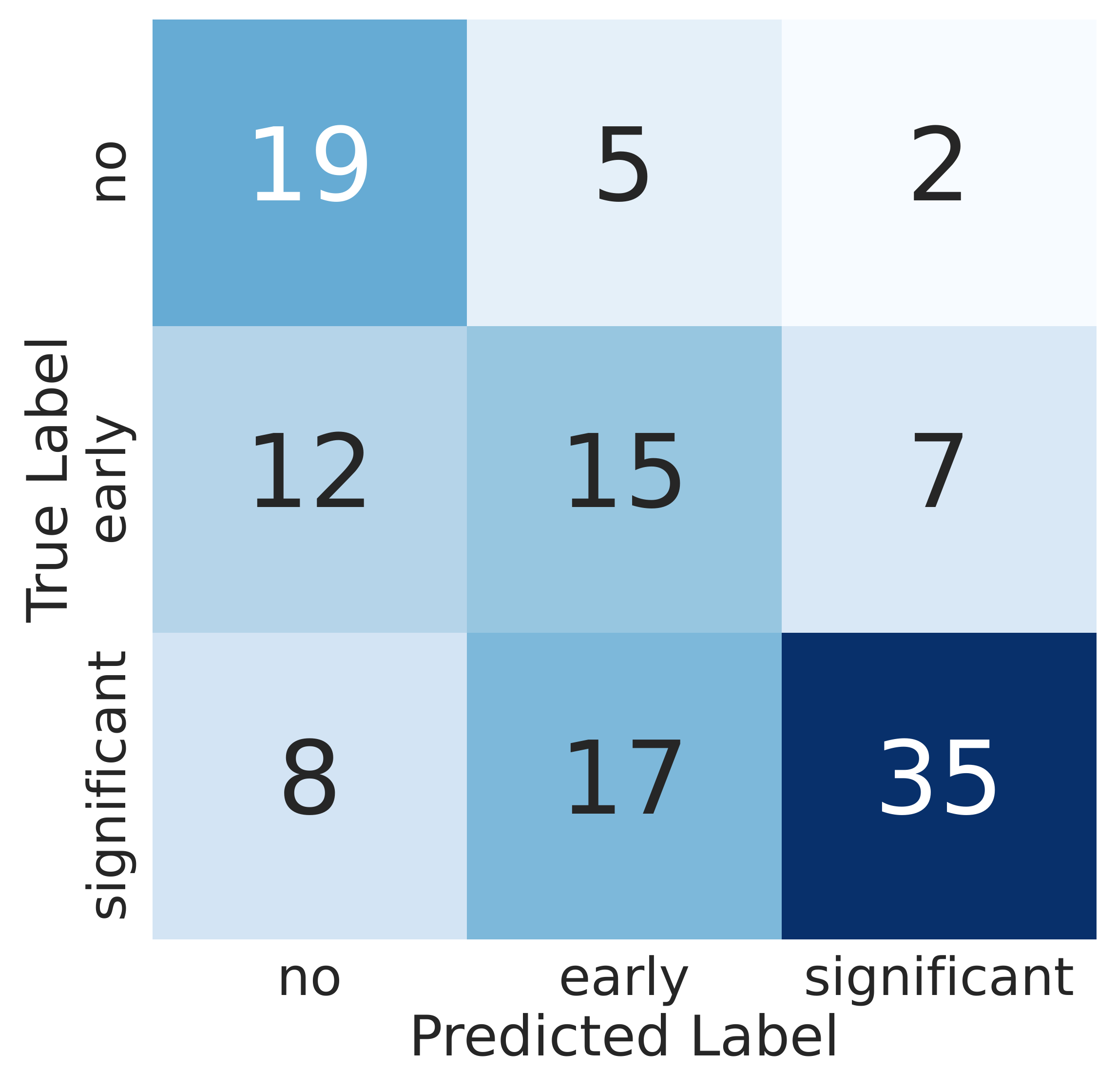}
    &
    \includegraphics[width=\BW\textwidth]{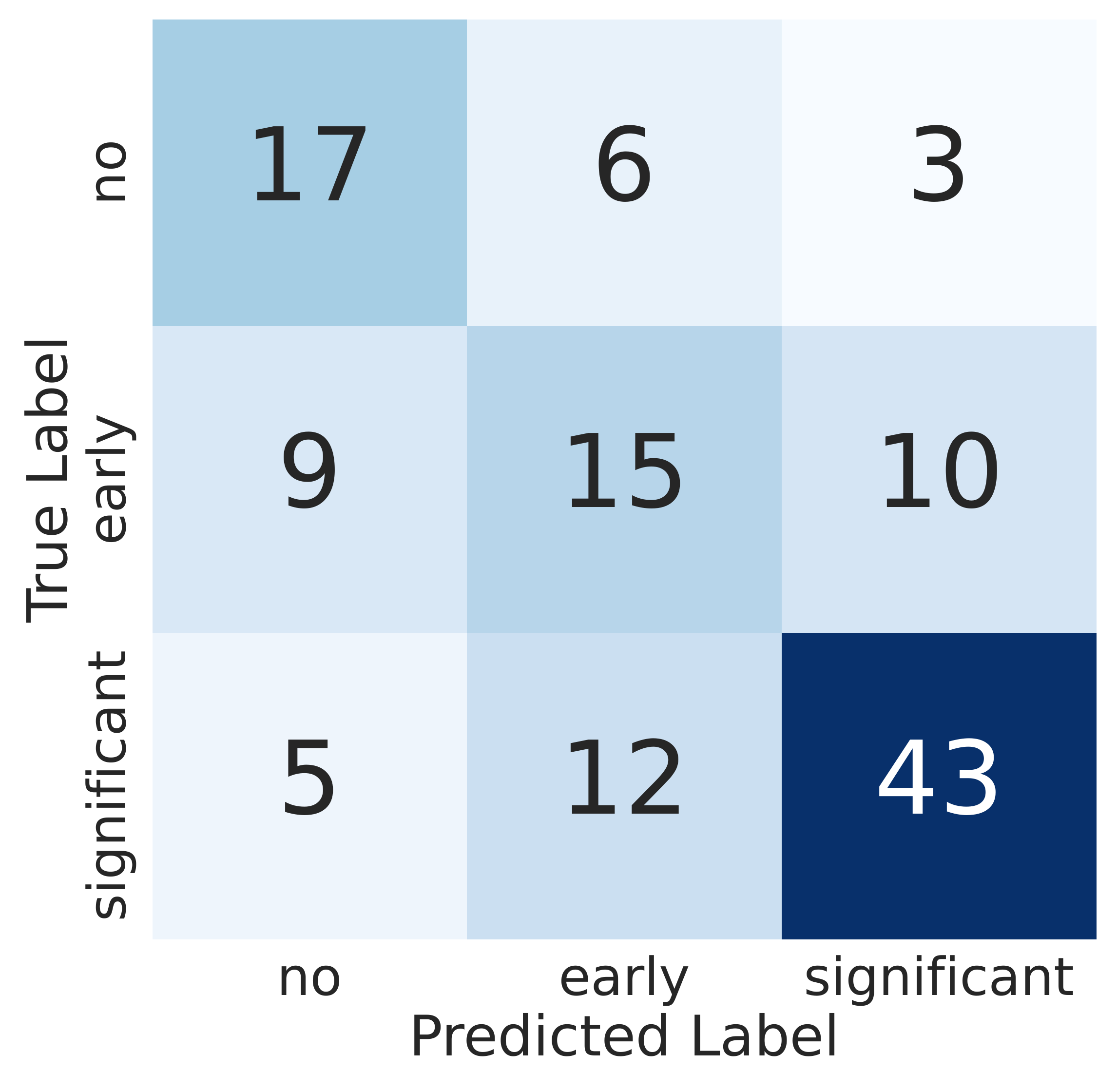}
    &
    \includegraphics[width=\BW\textwidth]{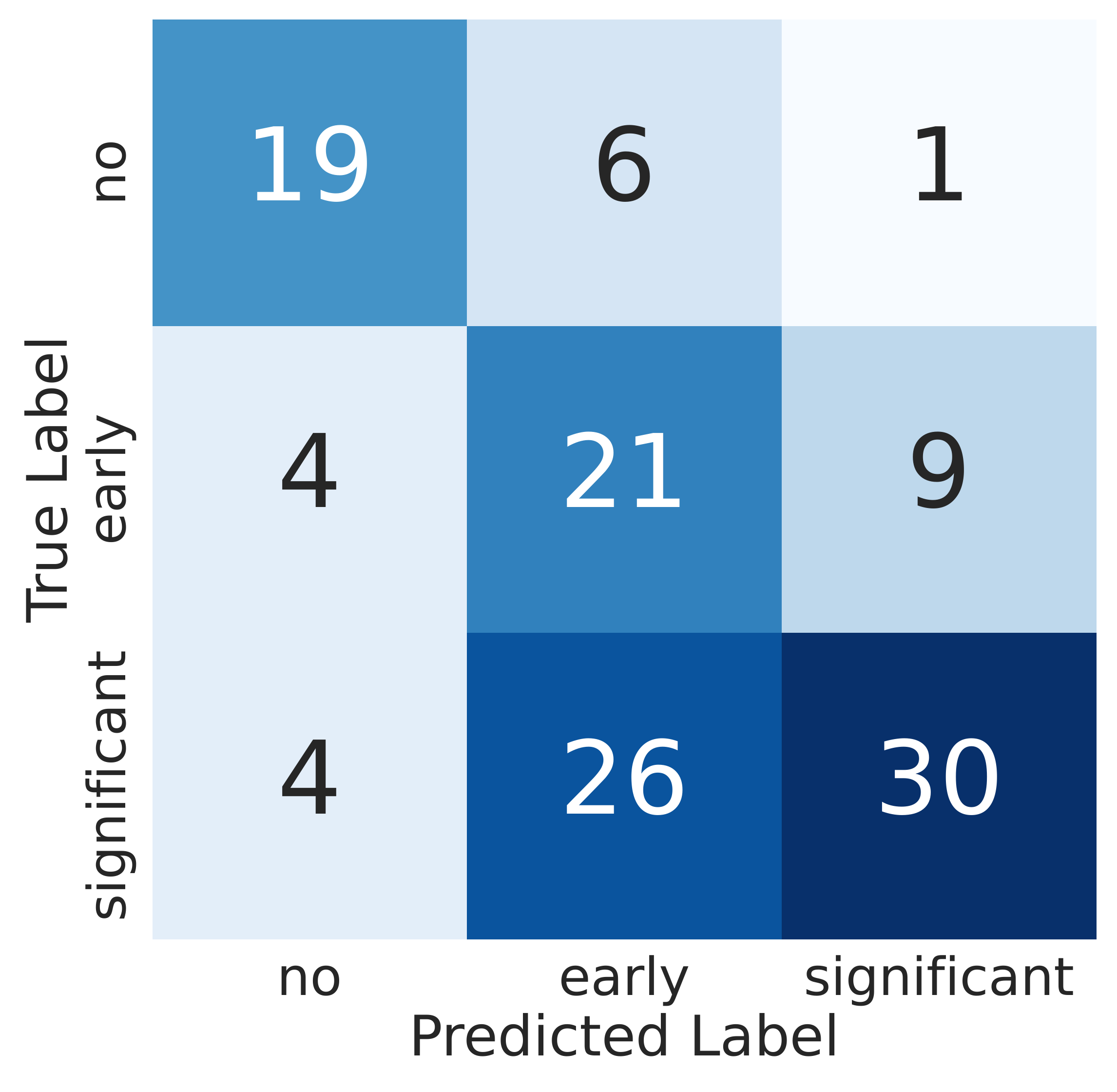}
    \\
    {\rotatebox{90}{~~~~~~~~~~~~DSMIL}}
    & 
    \includegraphics[width=\BW\textwidth]{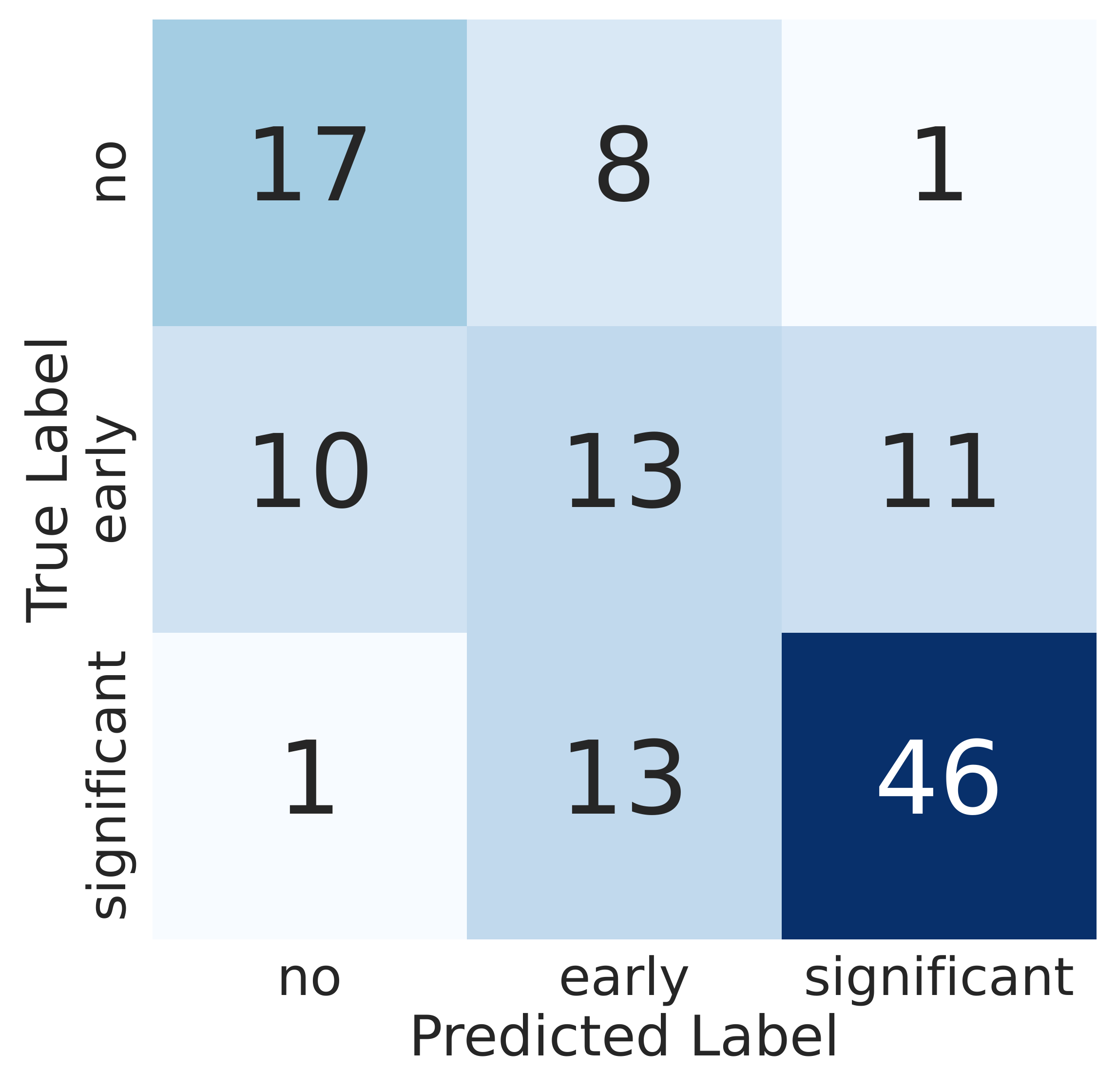}
    &
    \includegraphics[width=\BW\textwidth]{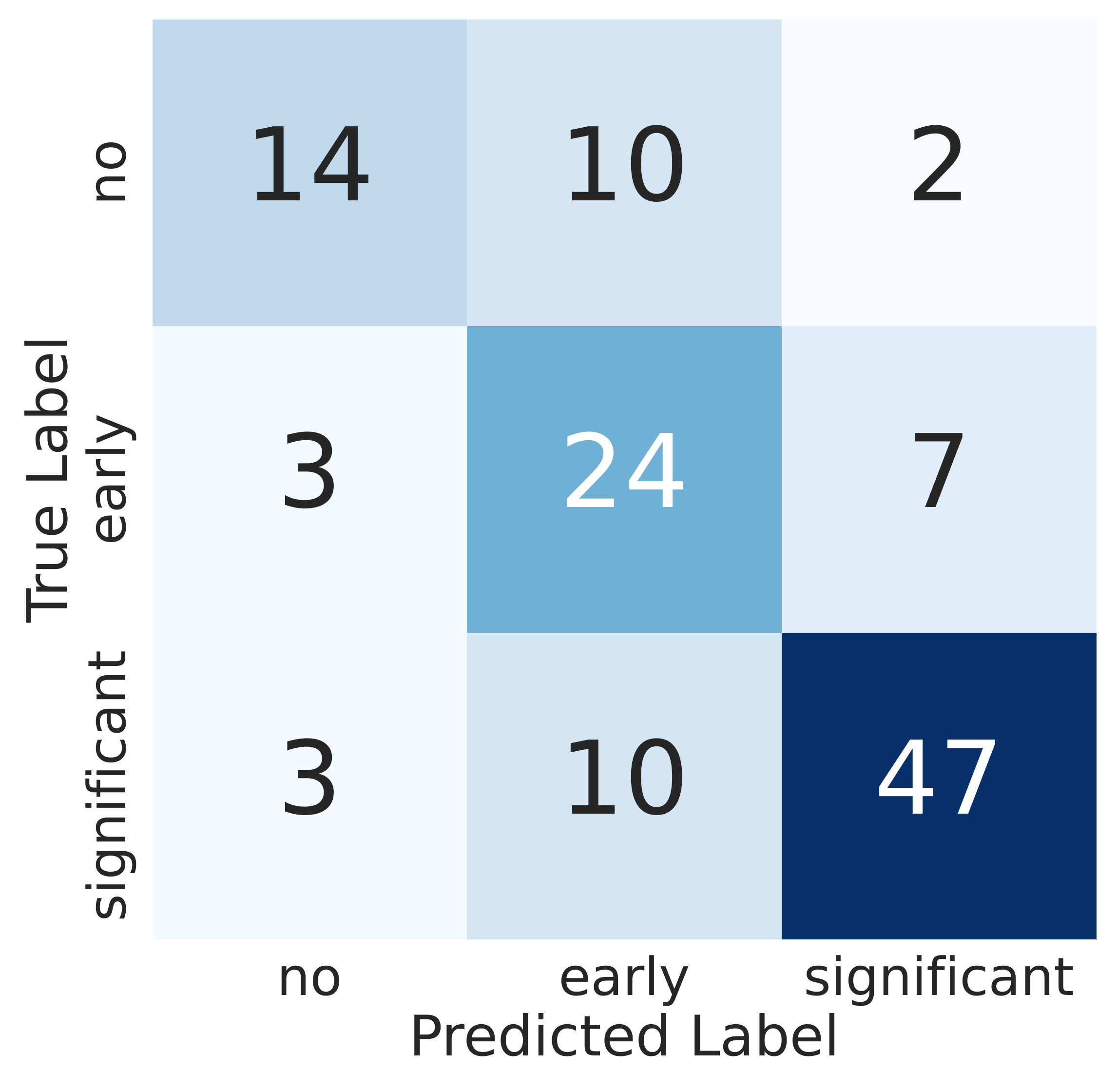}
    &
    \includegraphics[width=\BW\textwidth]{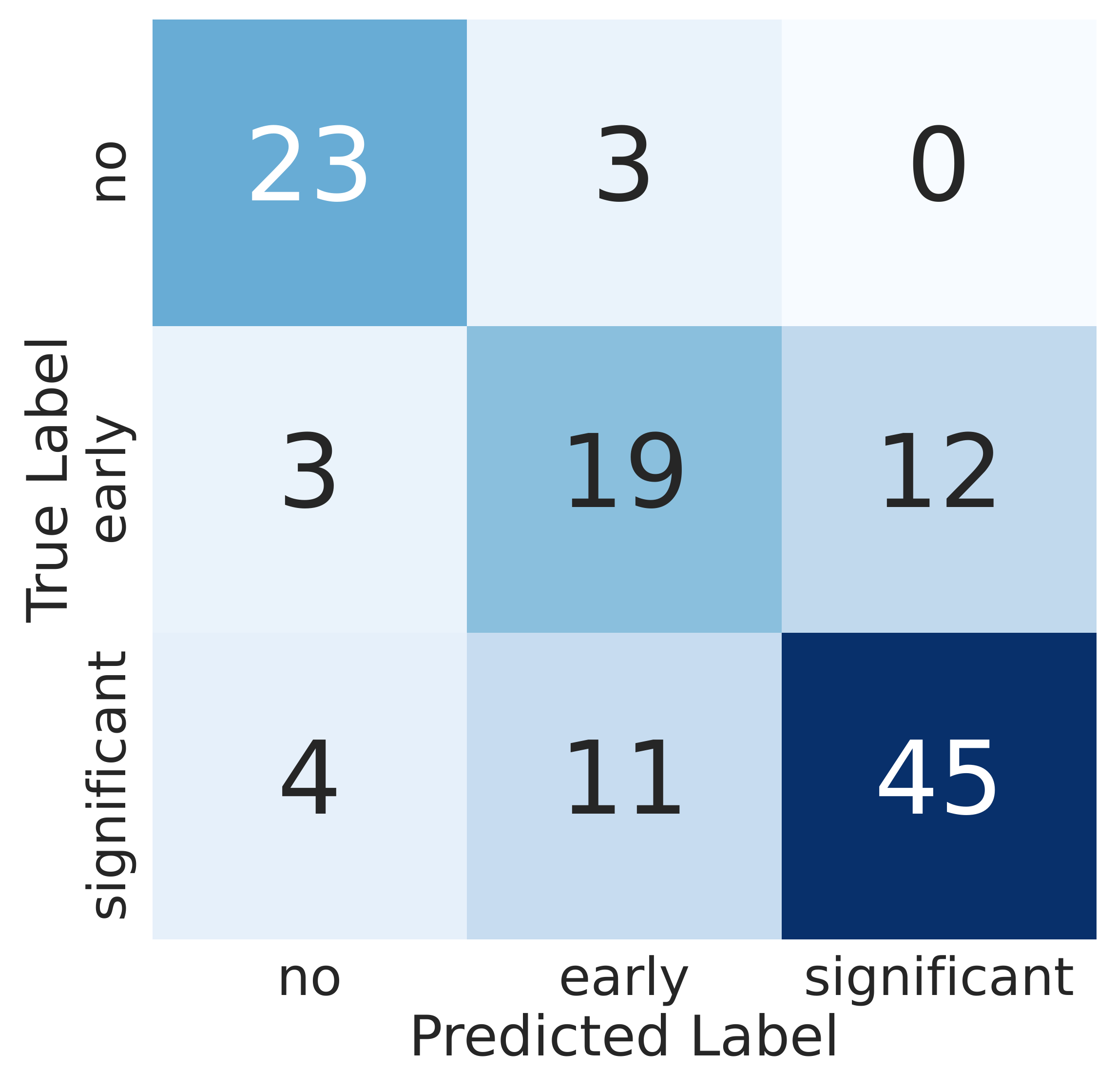}
    \\
    {\rotatebox{90}{~~~~~~~~ SAMIL (ours)}}
    & 
    \includegraphics[width=\BW\textwidth]{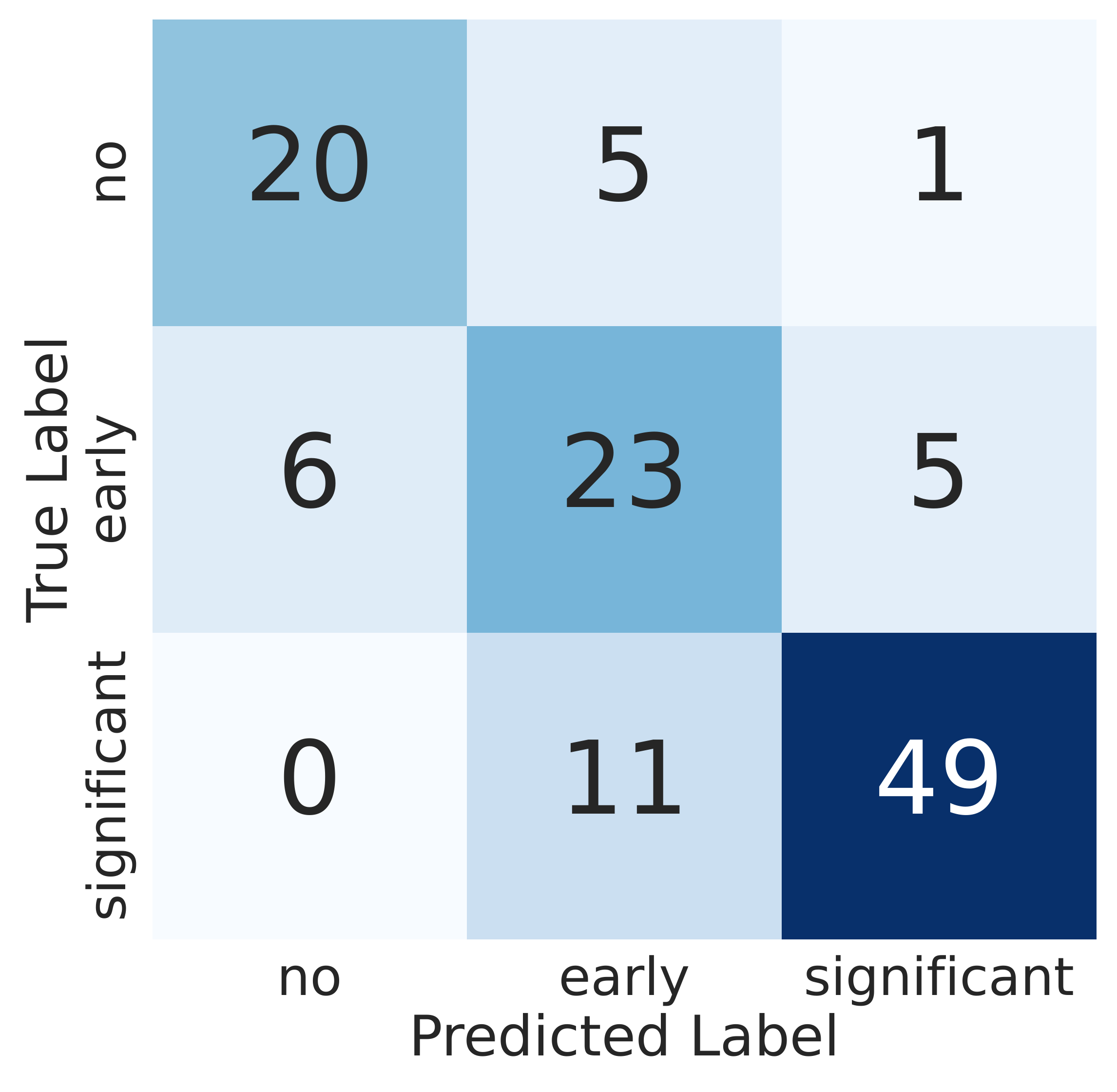}
    &
    \includegraphics[width=\BW\textwidth]{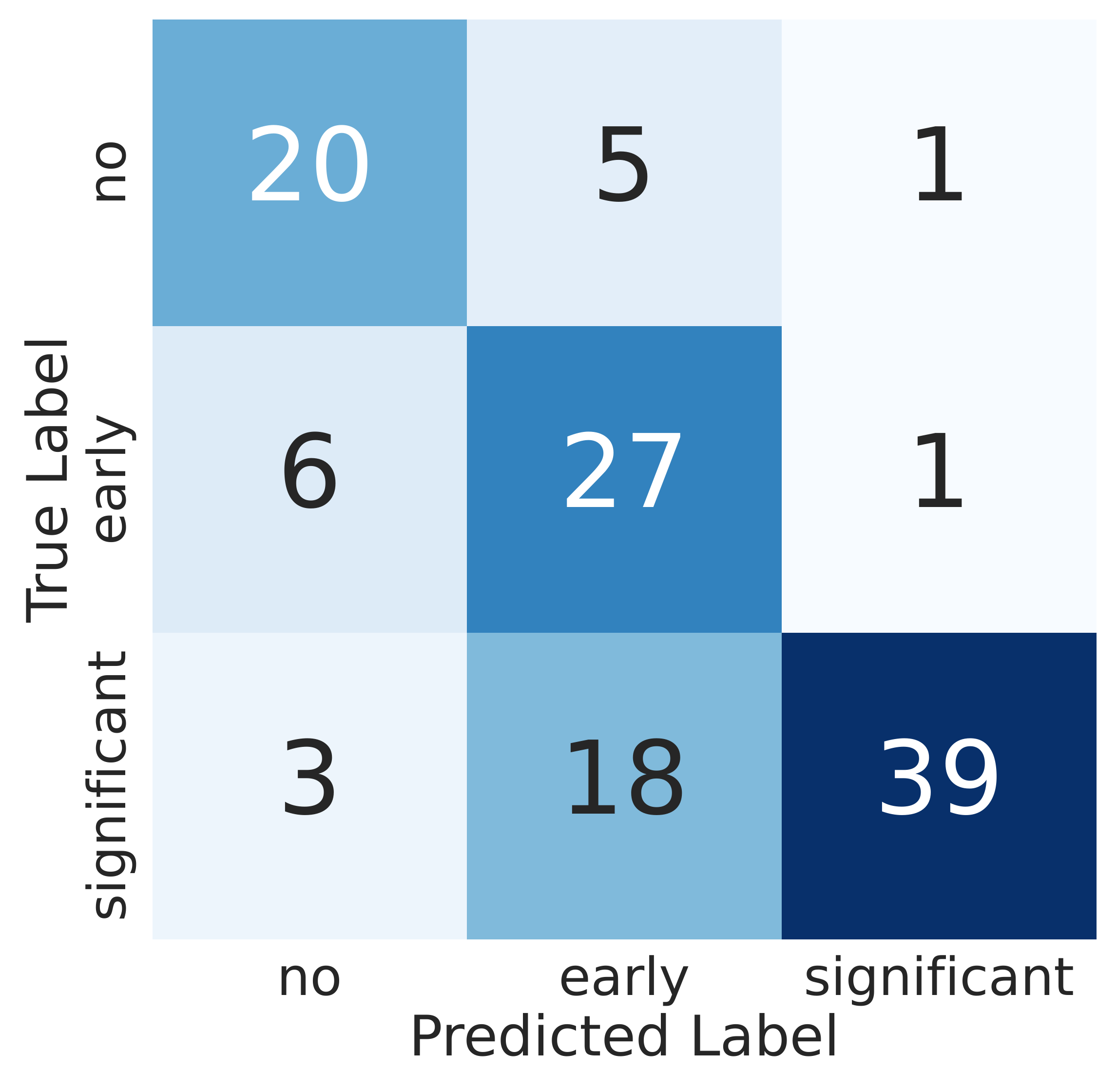}
    &
    \includegraphics[width=\BW\textwidth]{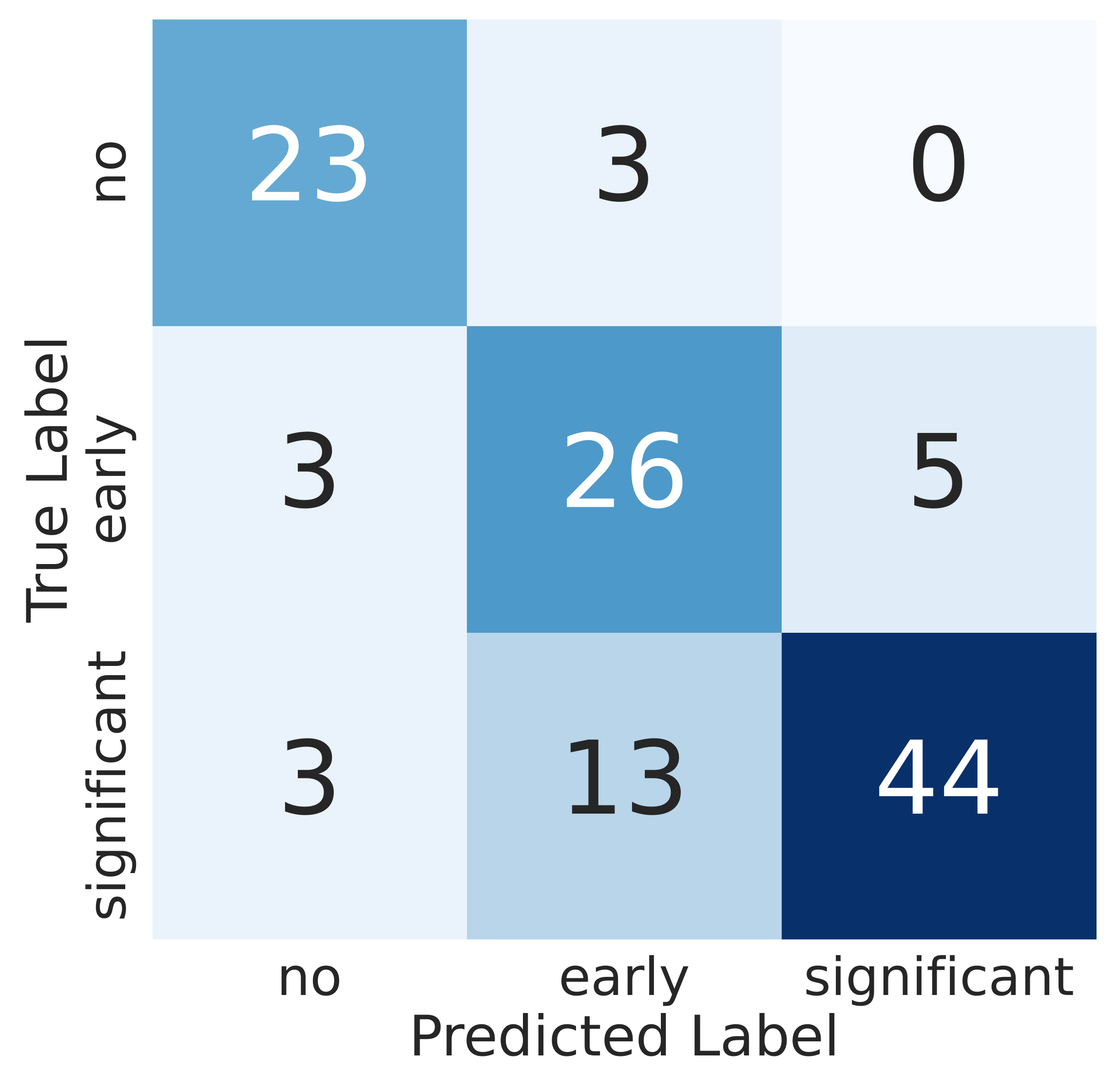}
    \end{tabular}
    \vspace{-.5cm} %% WHITESPACE HACK
    \caption{Confusion matrices for the patient-level AS diagnosis classification, across three predefined train/test splits of TMED2.
     }
    \label{fig:confusion_matrix}
\end{figure}

\subsection{ROC for AS Screening Tasks}
\newcommand{\BWWW}{0.36}
\setlength{\tabcolsep}{0.01cm}
\begin{figure}[H]
\begin{tabular}{r c c c }
    & No vs Some AS & Early vs Significant AS & NoSig vs Significant AS
    \\
    {\rotatebox{90}{~~~~Split1}}
    & 
    \includegraphics[width=\BWWW\textwidth]{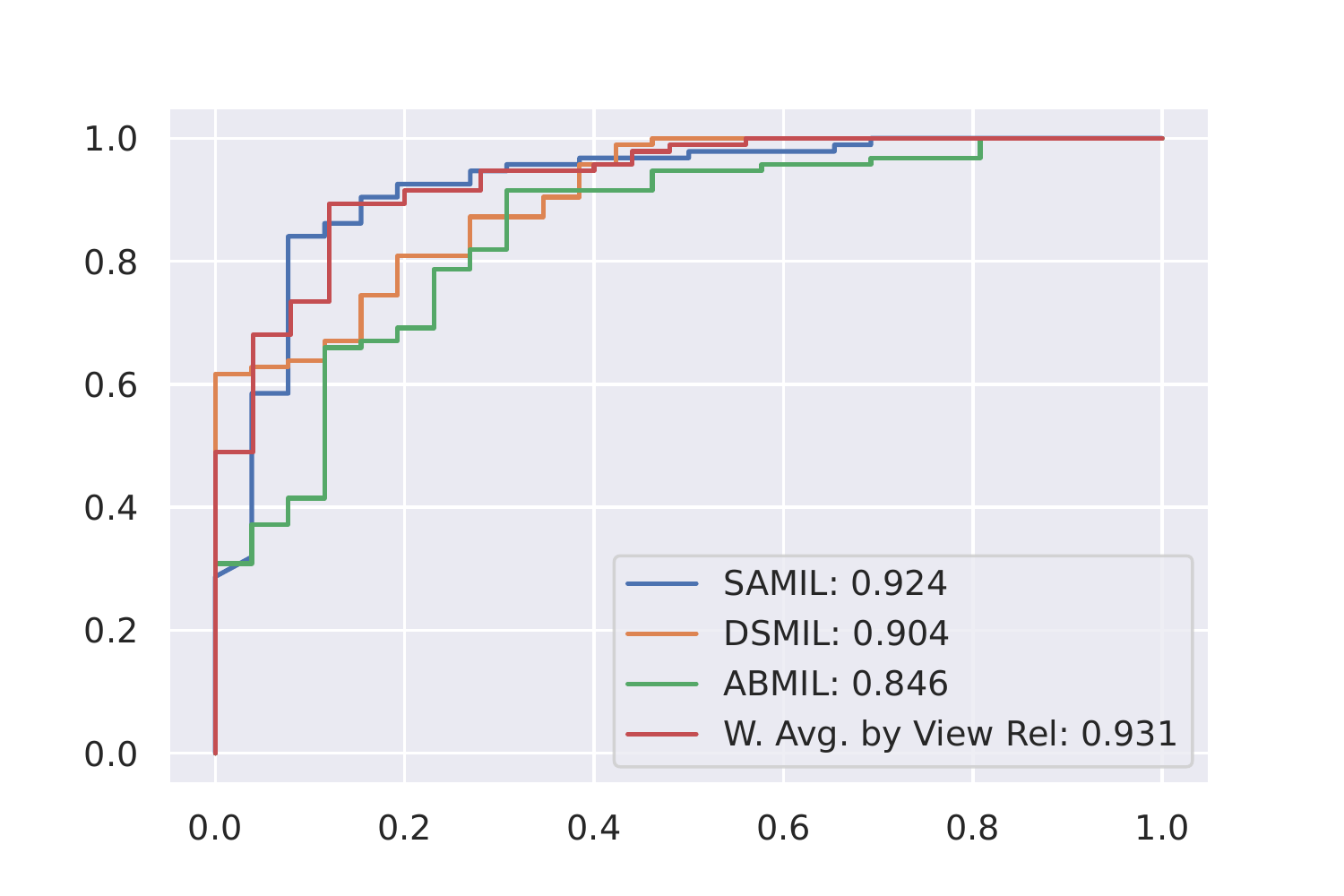}
    &
    \includegraphics[width=\BWWW\textwidth]{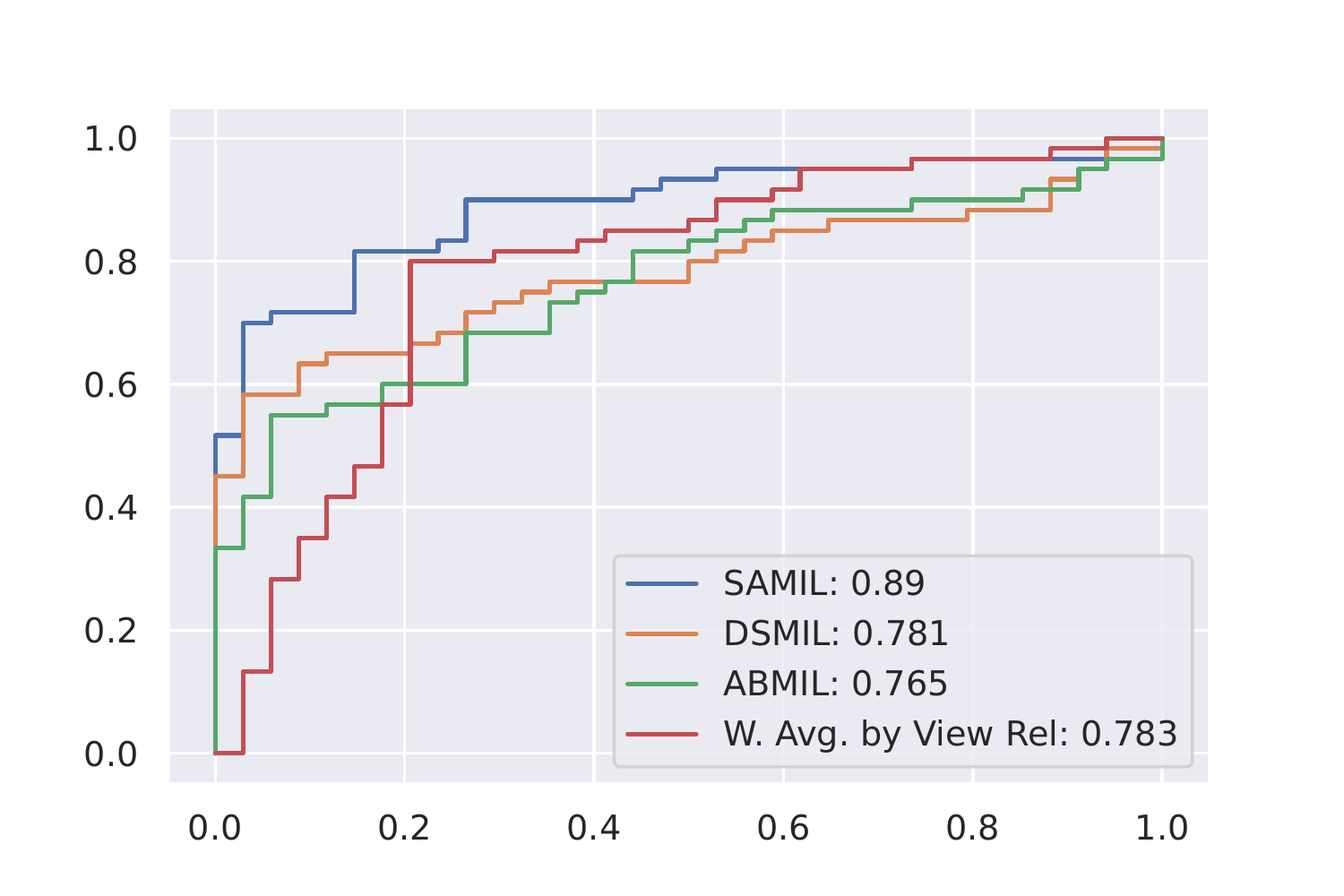}
    &
    \includegraphics[width=\BWWW\textwidth]{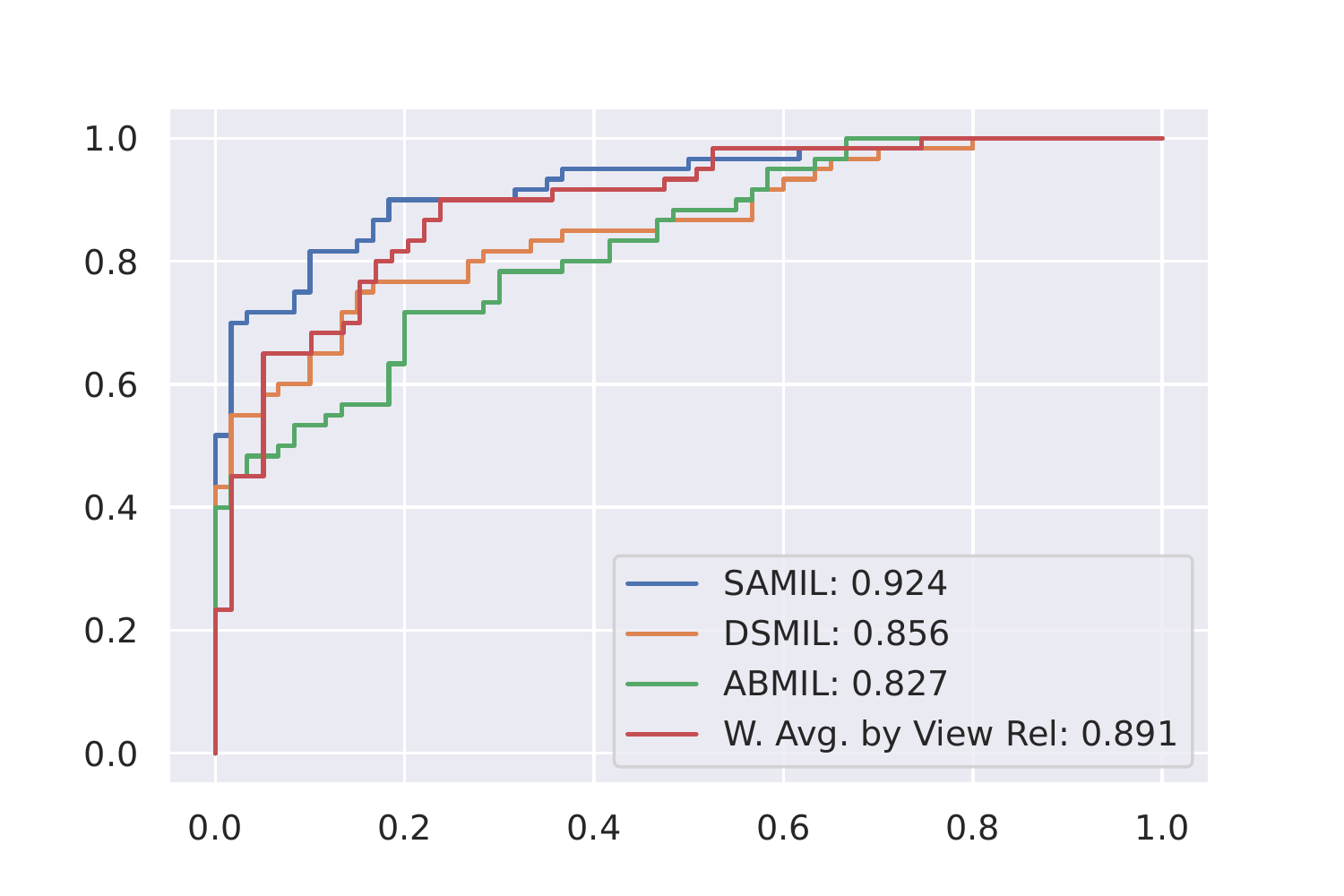}
    
    \\
    {\rotatebox{90}{~~~~Split2}}
    & 
    \includegraphics[width=\BWWW\textwidth]{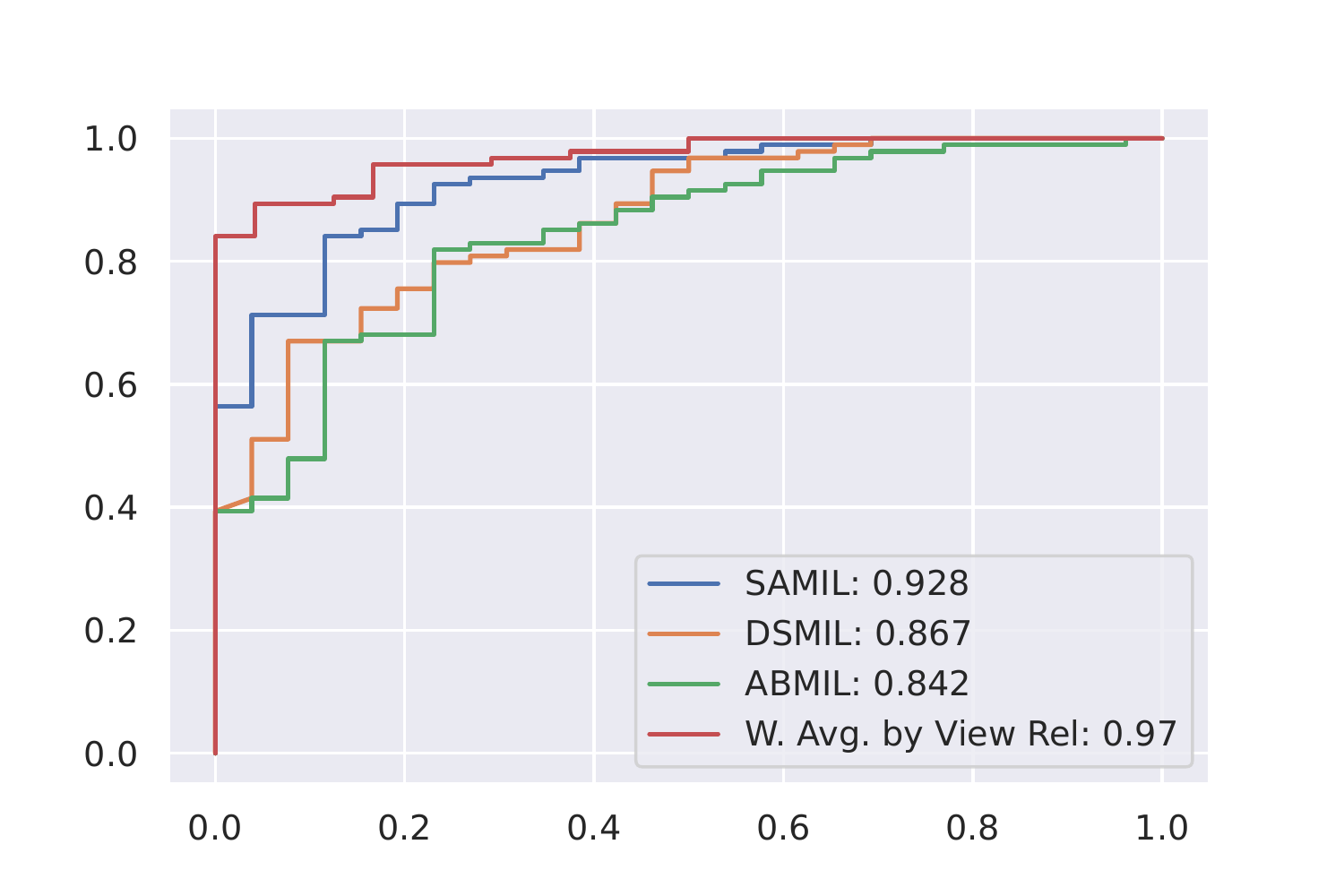}
    &
    \includegraphics[width=\BWWW\textwidth]{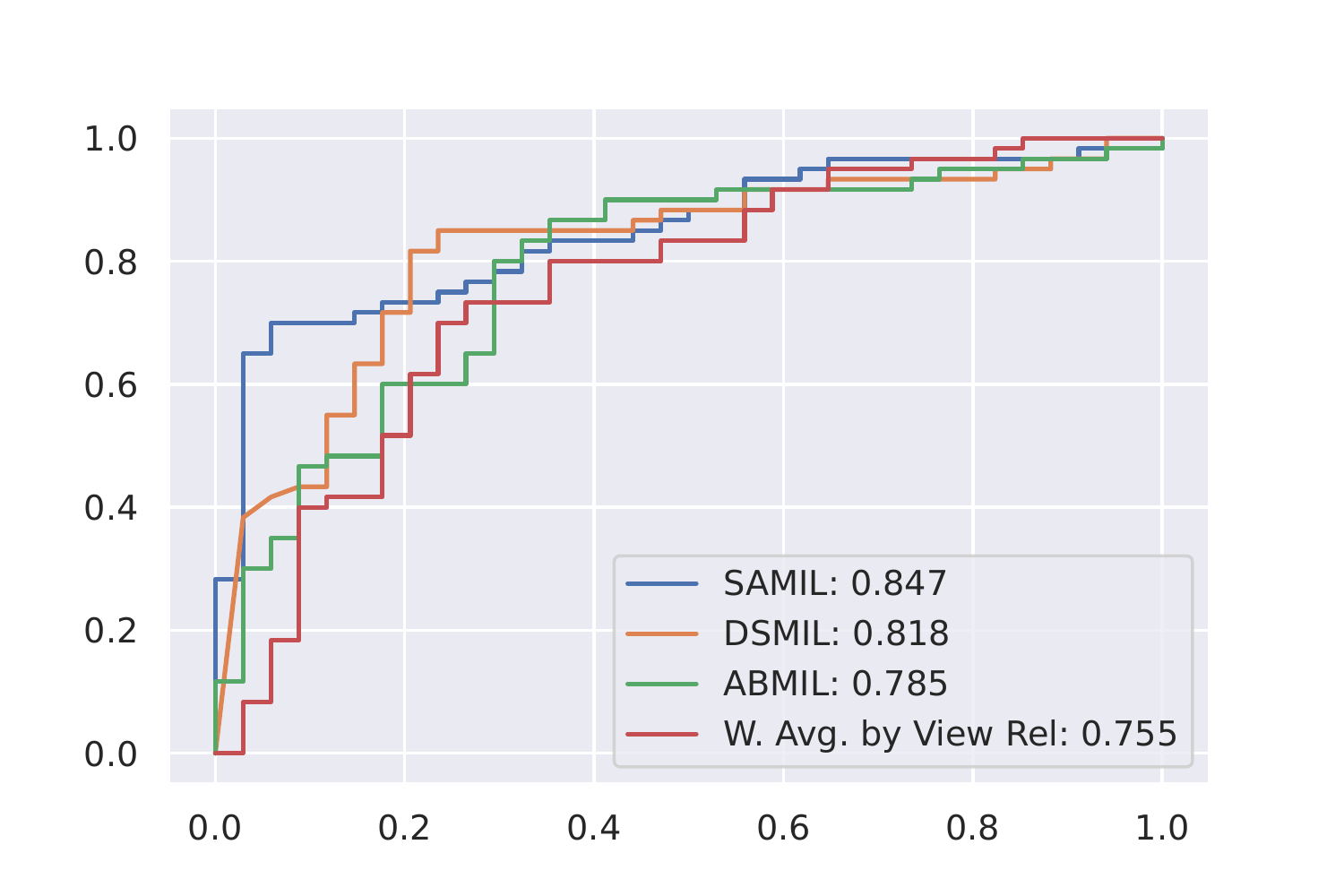}
    &
    \includegraphics[width=\BWWW\textwidth]{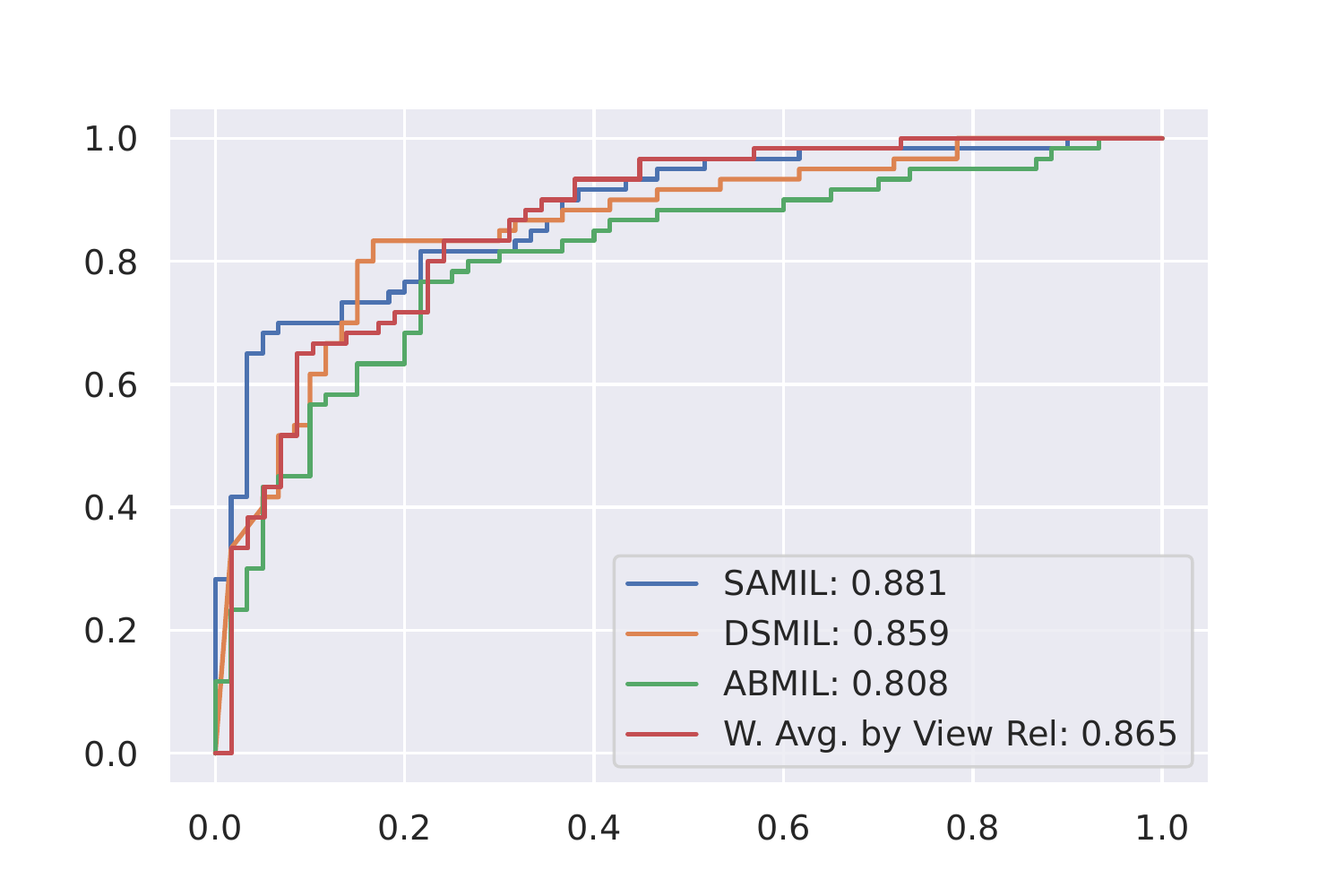}

    \\
    {\rotatebox{90}{~~~~Split3}}
    & 
    \includegraphics[width=\BWWW\textwidth]{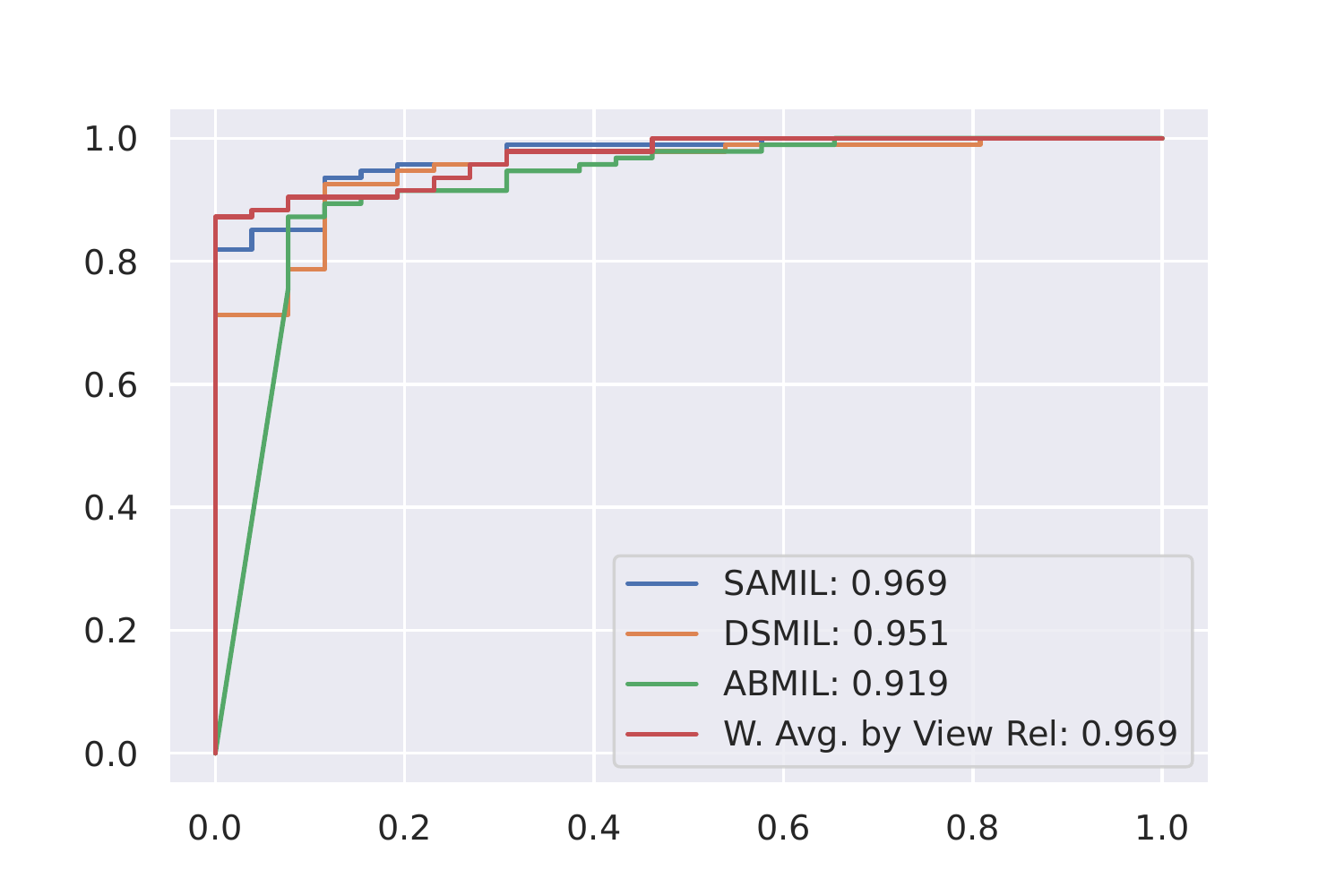}
    &
    \includegraphics[width=\BWWW\textwidth]{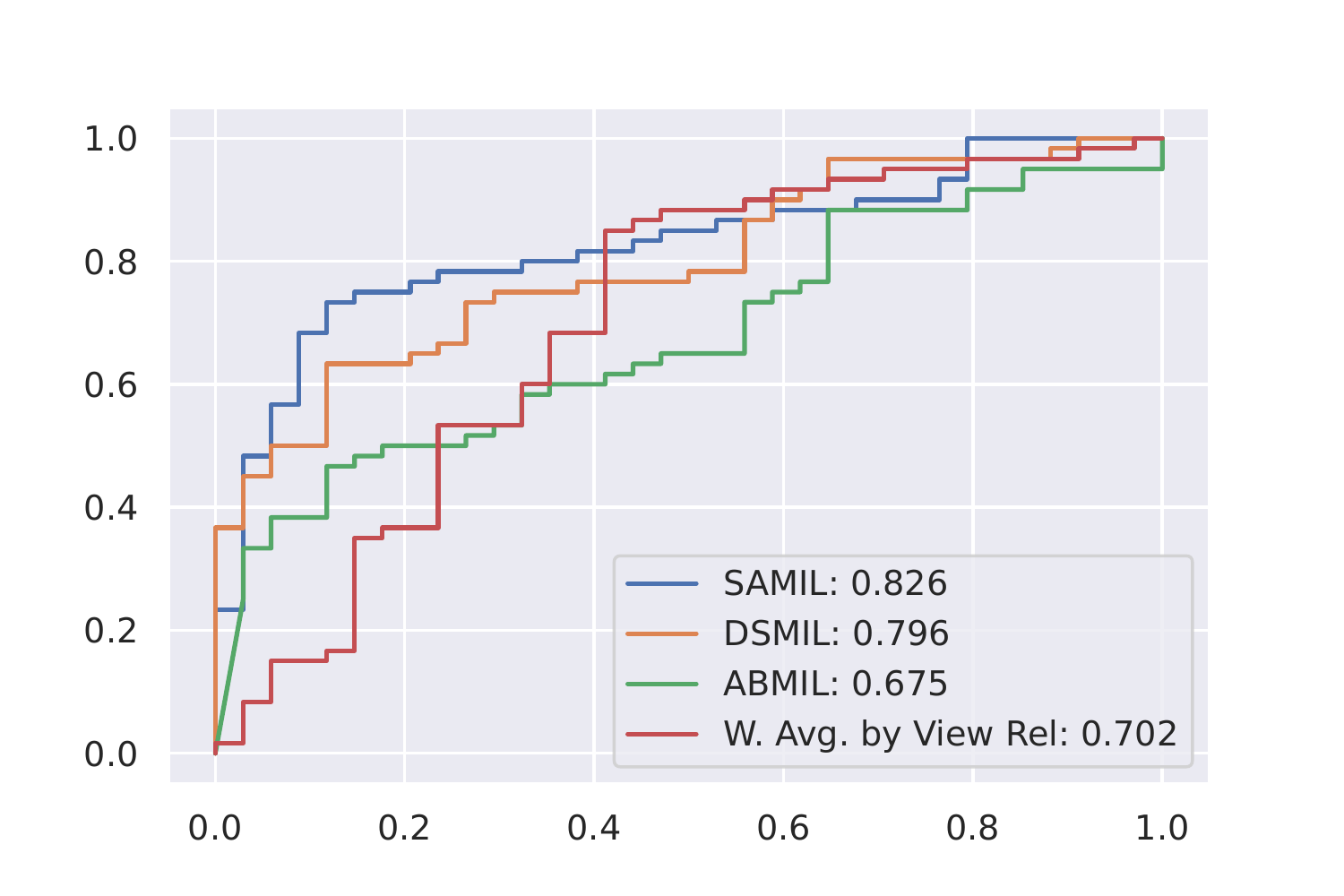}
    &
    \includegraphics[width=\BWWW\textwidth]{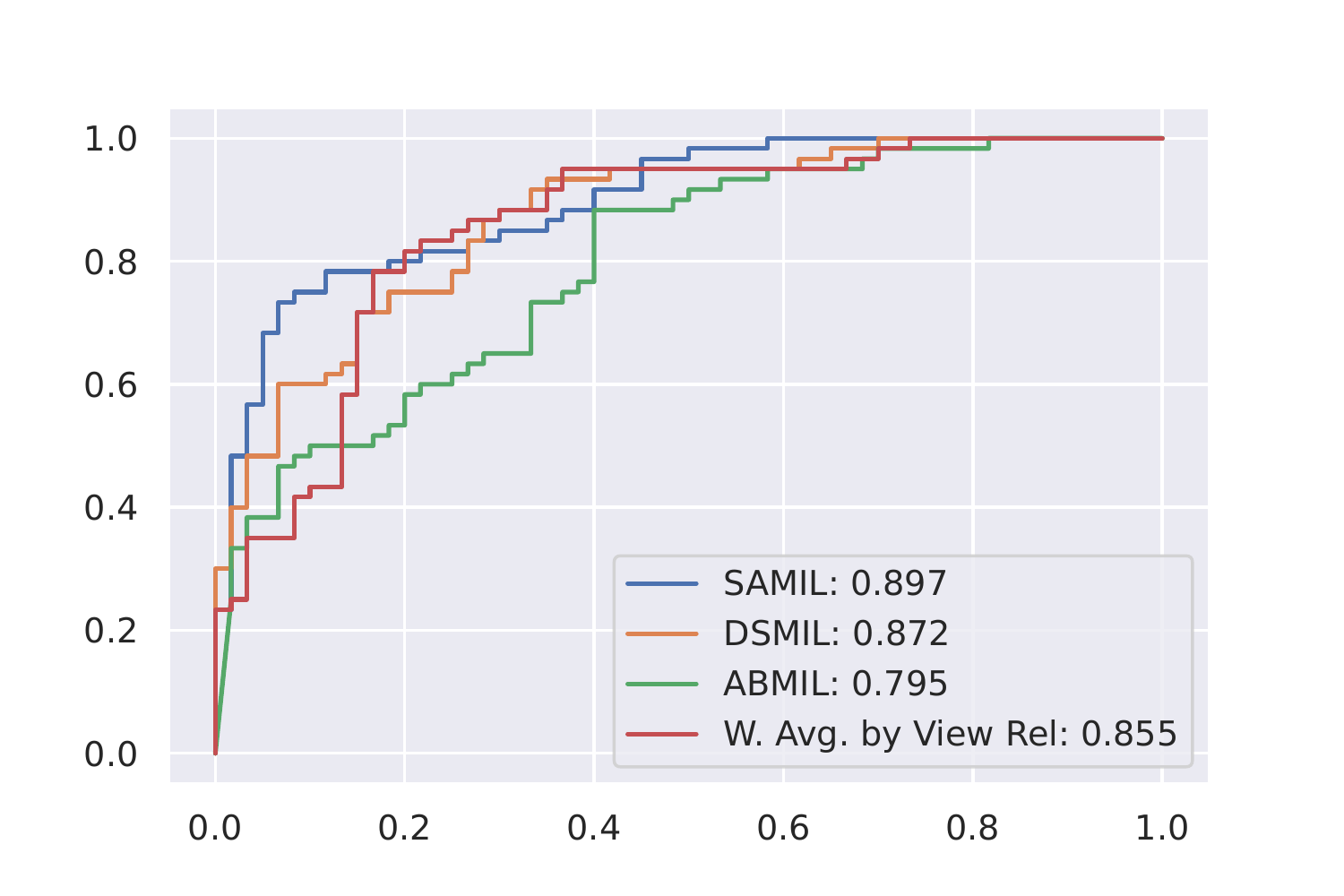}
    
    \end{tabular}	
    \caption{Diagnosis classification receiver operator curves. Showing results across three predefined train/test splits of TMED2 and three clinically relevant screening tasks.
     }
    \label{fig:TMED2_roc}
\end{figure}

\subsection{Attended Images by SAMIL and ABMIL}
\renewcommand{\BW}{0.18}
\setlength{\tabcolsep}{0.1cm}
% \begin{figure}[!h]
\begin{figure}[H]
\begin{tabular}{r c c c c c}
    \\
    {\rotatebox{90}{~~~~~~ABMIL}}
    & 
    \fcolorbox{green}{white}{\includegraphics[width=\BW\textwidth]{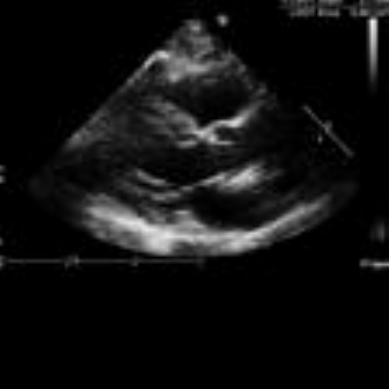}}
    &
    \fcolorbox{green}{white}{\includegraphics[width=\BW\textwidth]{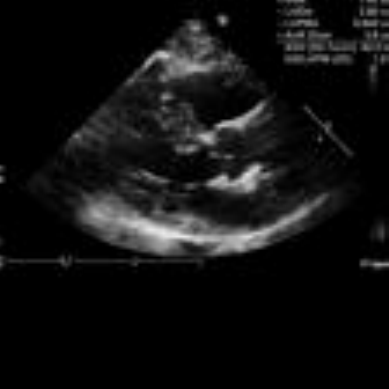}}
    &
    \fcolorbox{green}{white}{\includegraphics[width=\BW\textwidth]{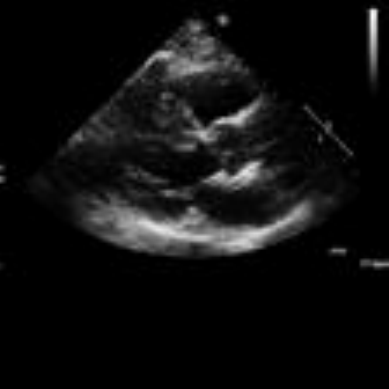}}
    &
    \fcolorbox{green}{white}{\includegraphics[width=\BW\textwidth]{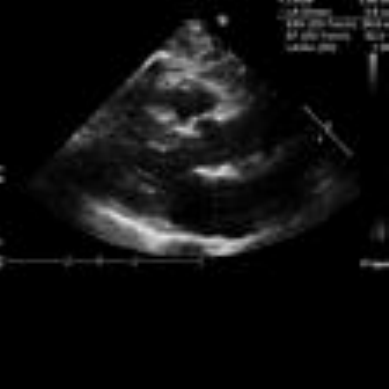}}
    &
    \fcolorbox{green}{white}{\includegraphics[width=\BW\textwidth]{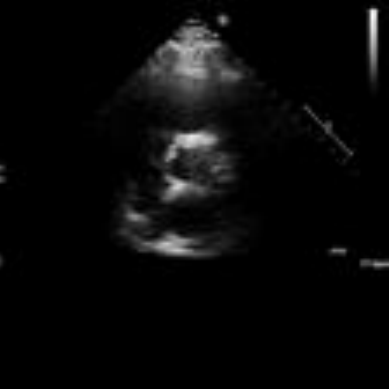}}
    
    \\
    {\rotatebox{90}{~~~~~~~~~~~~ABMIL}}
     & 
    \fcolorbox{red}{white}{\includegraphics[width=\BW\textwidth]{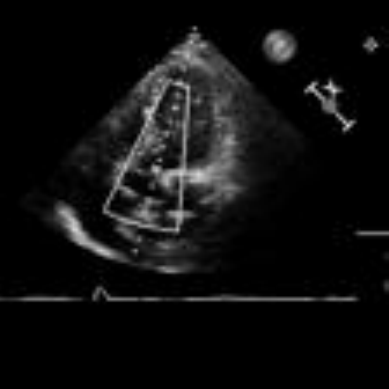}}
    &
    \fcolorbox{red}{white}{\includegraphics[width=\BW\textwidth]{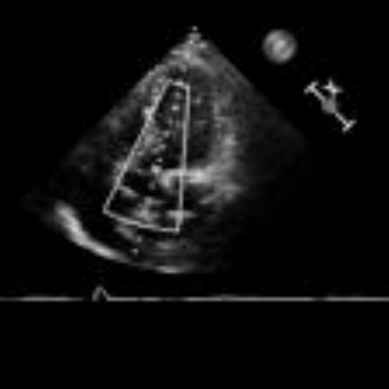}}
    &
    \fcolorbox{red}{white}{\includegraphics[width=\BW\textwidth]{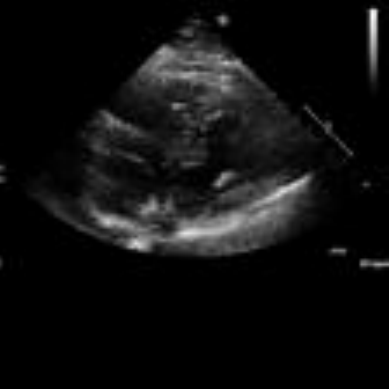}}
    &
    \fcolorbox{red}{white}{\includegraphics[width=\BW\textwidth]{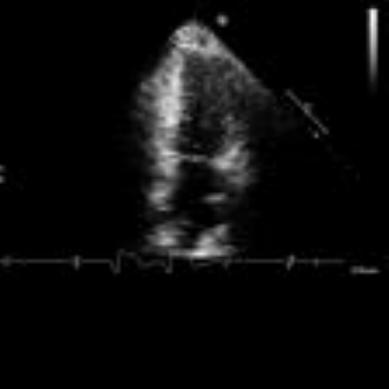}}
    &
    \fcolorbox{red}{white}{\includegraphics[width=\BW\textwidth]{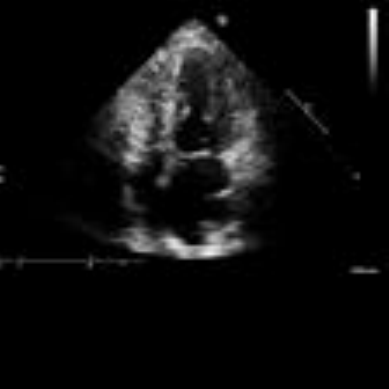}}
    
    \\
    {\rotatebox{90}{~~~~~~~~~~~~SAMIL}}
     & 
    \fcolorbox{green}{white}{\includegraphics[width=\BW\textwidth]{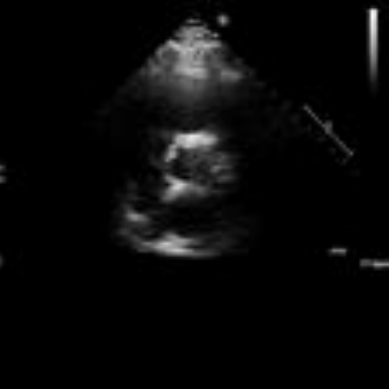}}
    &
    \fcolorbox{green}{white}{\includegraphics[width=\BW\textwidth]{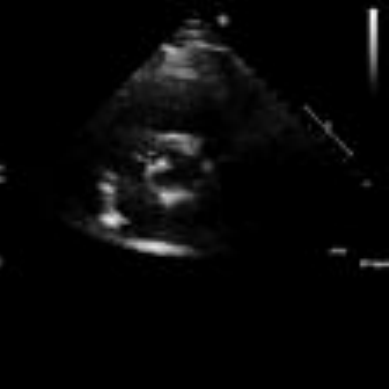}}
    &
    \fcolorbox{green}{white}{\includegraphics[width=\BW\textwidth]{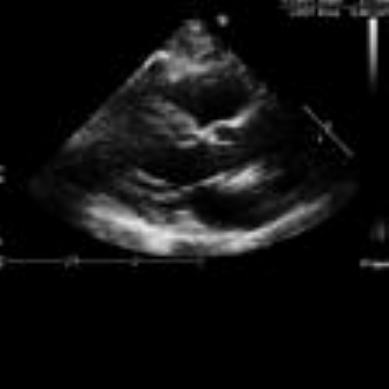}}
    &
    \fcolorbox{green}{white}{\includegraphics[width=\BW\textwidth]{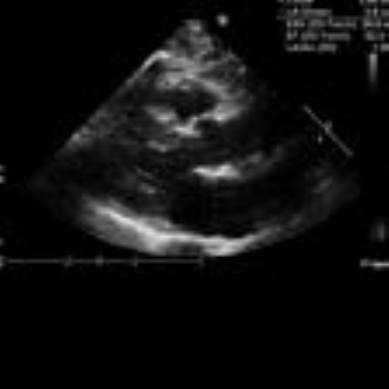}}
    &
    \fcolorbox{green}{white}{\includegraphics[width=\BW\textwidth]{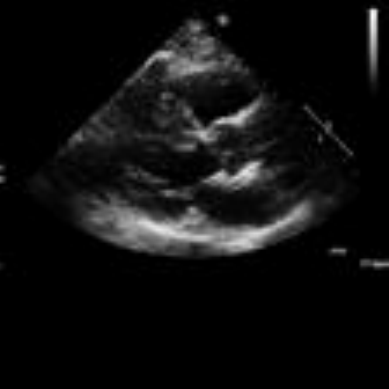}}
    
    \\
    {\rotatebox{90}{~~~~~~~~ SAMIL}}
     & 
    \fcolorbox{green}{white}{\includegraphics[width=\BW\textwidth]{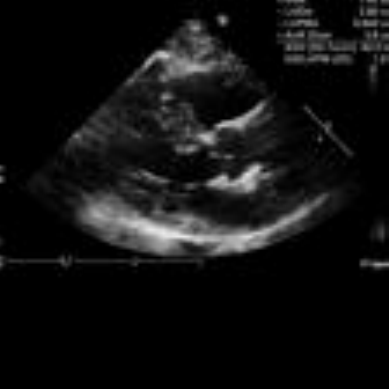}}
    &
    \fcolorbox{green}{white}{\includegraphics[width=\BW\textwidth]{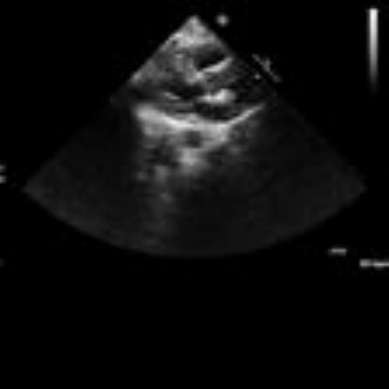}}
    &
   \fcolorbox{green}{white}{\includegraphics[width=\BW\textwidth]{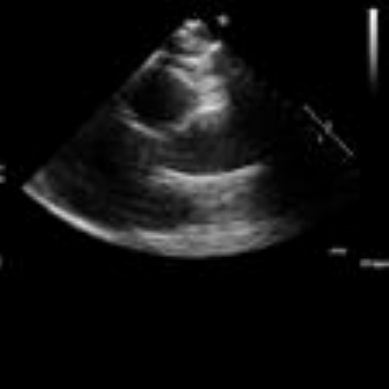}}
    &
    \fcolorbox{green}{white}{\includegraphics[width=\BW\textwidth]{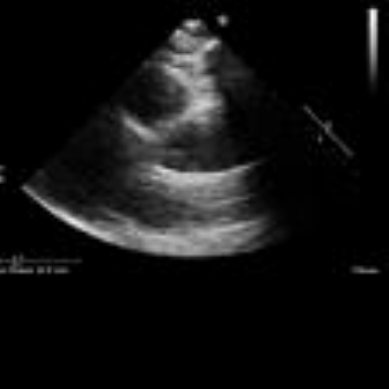}}
    &
    \fcolorbox{green}{white}{\includegraphics[width=\BW\textwidth]{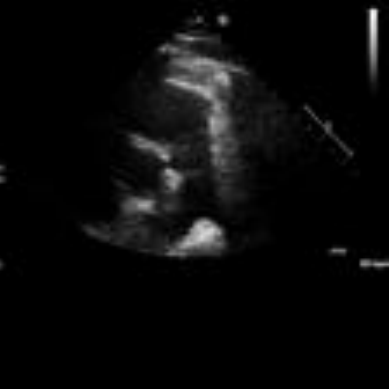}}
    
    \end{tabular}	
    \caption{Showing top attended images for the first study in the test set. The top 2 rows show the top 10 attended images by ABMIL, bottom 2 rows show the top 10 attended images by SAMIL. Red box indicates the image is not a clinically relevant view for AS diagnosis.
     }
    \label{fig:top_attended_images}
\end{figure}

% \section{Methods Supplement}
% \subsection{Architecture}
% \label{app:Architecture}
% \input{app_Architecture}

% \subsection{View Classifier}
% \label{app:ViewClassifier}
% \input{app_ViewClassifier}

% \section{MIL Experiment Details}
% \input{app_experiment_details}

% \section{Self-supervised Pretraining}
% \label{app:SSL_Pretraining}
% \input{app_SSL_Pretraining}

% \section{Additional Related Work}
% \label{app:relatedworks}
% \input{app_relatedworks}

\section{Methods Details}
\subsection{Mapping between the TMED-2 labels and finer-grained clinical scale}
\label{App:TMED2_label_mapping}

Here we show how the 3-level course diagnosis classes in TMED-2 (advocated by \citet{wessler2023automated}) map to the common 5-level fine-grained clinical scale used by clinicians.

\begin{table}[h]
\centering
\begin{tabular}{|c|c|}
\hline
\textbf{5-Level Scale} & \textbf{TMED2 Label} \\ 
\hline
no AS & no AS \\ 
\hline
mild AS & early AS \\ 
\hline
mild-to-moderate AS & early AS \\ 
\hline
moderate AS & significant AS \\ 
\hline
severe AS & significant AS \\ 
\hline
\end{tabular}
\caption{Mapping between the TMED-2 labels and finer-grained clinical scale}
\label{tab:TMED2_label_mapping}
\end{table}

%We chose this 3-level coarse label structure to focus on automated screening use cases.
%We hope to be able to use classifiers that can distinguish between these three levels of severity to identify individuals who should be referred for comprehensive imaging and AS-related care.

We chose this mapping because our study was designed with three overarching clinical considerations: (1) An AS screening framework should be designed primarily to be sensitive for identifying disease rather than for comprehensive phenotyping of AS given the complexity of this clinical syndrome; (2) Given the challenges with contemporary diagnosis (and the many subtypes of severe AS) and the concerns that severe AS might masquerade as moderate AS with certain low-flow subtypes, we designed our disease classifiers to identify ‘significant AS’, a category that includes moderate and severe AS. This was purposely done to maximize utility as a screening tool; and (3) The expected clinical application of our fully automated MIL models is that it will trigger referral for comprehensive echocardiography and heart team evaluation.

\subsection{MIL Architecture}
Below we report the architecture details for SAMIL. For feature extractor $f$, we use a simple stack of convolution layers as done in ABMIL~\citep{ilse2018attention}.
\begin{table}[!h] 
\centering 
\begin{tabular}{l}
~~~~~Feature Extractor $f$ \\
\midrule
Conv2d(3, 20, kernel=(5,5))\\
ReLU()\\
MaxPool2d(2, stride=2)\\
Conv2d(20, 50, kernel=(5,5))\\
ReLU()\\
MaxPool2d(2, stride=2)\\
Conv2d(50, 100, kernel=(5,5))\\
ReLU()\\
Conv2d(100, 200, kernel=(5,5))\\
ReLU()\\
MaxPool2d(2, stride=2)\\

\bottomrule
\end{tabular}
\caption{Details of Feature Extractor $f$}
\label{tab:Feature Extractor $f$}
\end{table} 
We used the same feature extractor $f$ shown in \ref{tab:Feature Extractor $f$} for SAMIL, ABMIL, Set Transformer and DSMIL.

The feature extractor $f$ maps each of the original images into 200 feature maps with smaller size. In practice, a MLP can be use (optional) to further process the flattened feature maps (also see Fig~\ref{fig:workflow_diagram}). We use the same MLP [Linear(32000, 500), ReLU(), Linear(500, 250), ReLU(), Linear(250, 500), ReLU()] for both SAMIL and ABMIL. For Set Transformer, we directly flattened the extracted feature maps and feed them to the Set Transformer's ISAB blocks. Please refer to original paper~\citep{lee2019set} for more details. For DSMIL, the extracted feature maps are flattened and projected to vectors of dimension 500 by a linear layer followed by ReLU, and then feed to its two streams. Please refer to original paper~\citep{li2021dual} for more details.

For the pooling layer $\sigma$, we use the same MLP architectures (shown in \ref{tab:attention_MLP}) for both the supervised attention branch and flexible attention branch in SAMIL. Note that this is also the same MLP architecture to learn attention weights in ABMIL.

\begin{table}[!h] 
\centering 
\begin{tabular}{l}
MLP learning attention weights \\
\midrule
Linear(500, 128)\\
Tanh()\\
Linear(128, 1)\\

\bottomrule
\end{tabular}
\caption{Details of MLP used to learn attention weights for SAMIL and ABMIL}
\label{tab:attention_MLP}
\end{table} 

For output layer $g$ both SAMIL and ABMIL use a simple linear layer (with softmax). Our experiments for DSMIL,  and Set Transformer are mainly based on the official open-source code from corresponding paper. Please refer to the original papers for more details on their $\sigma$ and $g$.

\label{app:Architecture}
\subsection{Details on Filter then Avg. Approach}
\label{App:Filter then Avg.}
To apply the Filter then Avg. approach proposed on TMED2, we follow closely the steps outlined in the paper ~\citet{holste2022automated,holste2022self}. We first use the same view classifiers that are used for SAMIL to prefilter images in the dataset, keeping only images that are predicted as PLAX. We then use a 2D ResNet18~\citep{he2016deep} to train the diagnosis classifier to classify each retained PLAX image as no AS, early AS or significant AS. In aggregation step, we average the AS predictions of all PLAX images in a study to obtain the study-level AS prediction. Note that author in ~\citep{holste2022self} uses a 3D ResNet18~\citep{tran2018closer} since their proprietary dataset consists of 3D videos while the open access TMED2 consists of 2D images. For the same reason, we are not able to directly use their self-superivsed training strategy that are proposed for 3D videos.

\subsection{Details on DeepSet}
\label{App:DeepSet}
DeepSet~\citep{zaheer2017deep} process each instance in the bag independently, and aggregate the processed feature embedding using simple pooling (mean or max). Fully connected layers are then used to map the aggregated feature embeddings into a bag prediction.

We perform the same hyperparameter search for DeepSet as shown in App~\ref{App:Hyper}. However, we won't able to obtain any meaningful results, which suggest that problem of using multiple ultrasound images for AS diagnosis is too challenging for simple architecture like DeepSet.

\section{MIL Training Details}
% \subsection{Training.} 
Our open source code (\url{https://github.com/tufts-ml/SAMIL/}) uses PyTorch ~\citep{paszke2019pytorch}. For all methods compared, we use SGD~\citep{robbins1951stochastic} as optimizer. Each method is set to train for 2000 epochs, and early stop if validation performance does not increase for 200 consecutive epochs. Each training run uses one NVIDIA A100 GPU.

% \subsection{Hyperparameter.} 
\label{App:Hyper}
We perform a grid search for each algorithm and each data split. From our preliminary experiments, we found that learning rate around 0.0005 and weight decay around 0.0001 is a good starting point. 

For DSMIL, ABMIL, Set Transformer, DeepSet and Filter then Avg, we search learning rate in [0.0003, 0.0005, 0.0008, 0.001, 0.003] and weight decay in [0.00001, 0.00003, 0.0001, 0.0003, 0.001]. SAMIL involves two additional hyperparameters, a temperature scaling term $\tau_{v}$ used in eq.~\ref{eq:L_SA}, and $\lambda_{SA}$ in eq.~\ref{eq:total_loss} that balance the supervised attention loss and the cross-entropy loss. For SAMIL, we search learning rate in [0.0005, 0.0008], weight decay in [0.0001, 0.001], $\tau_{v}$ in [0.1, 0.05, 0.03] and $\lambda_{SA}$ in [5, 15, 20]. Note that for ABMIL with gated attention, we did not search hyperparameters again, but directly use the corresponding best hyperparameter from its general attention version. Note that we perform same set of independent hyperparameter search for experiments on SAMIL with bag-level pretraining, image-level pretraining and without pretraining.

Final hyperparameter used are reported as follow:

% \begin{table}[!htb]
% \begin{table}[H]
% \centering
% \begin{tabular}{l|c|c|c}
% \multicolumn{4}{c}{SAMIL (with study-level SSL)} \\
% Hyperparameter & split1 & split2 & split3 \\
% \midrule
% Learning rate & 0.0005 & 0.0008 & 0.0005 \\
% Weight decay & 0.0001 & 0.001 & 0.001 \\
% Temperature T & 0.1 & 0.1 & 0.05 \\
% $\lambda_{SA}$ & 15.0 & 20.0 & 20.0 \\
% Learning rate schedule & cosine & cosine & cosine\\
% \end{tabular}
% \caption{Hyperparameter settings for SAMIL across different data splits.}
% \label{tab:SAMIL_hyper_nopretrain}
% \end{table}

\begin{table}[H]
\centering
\begin{tabular}{l|c|c|c}
\multicolumn{4}{c}{SAMIL (with study-level SSL)} \\
Hyperparameter & split1 & split2 & split3 \\
\midrule
Learning rate & 0.0008 & 0.0005 & 0.0005 \\
Weight decay & 0.0001 & 0.0001 & 0.001 \\
Temperature T & 0.1 & 0.05 & 0.1 \\
$\lambda_{SA}$ & 15.0 & 20.0 & 20.0 \\
Learning rate schedule & cosine & cosine & cosine\\
\end{tabular}
\caption{Hyperparameter settings for SAMIL across different data splits.}
\label{tab:SAMIL_hyper_nopretrain}
\end{table}

\begin{table}[!htb]
\centering
\begin{tabular}{l|c|c|c}
\multicolumn{4}{c}{DSMIL} \\
Hyperparameter & split1 & split2 & split3 \\
\midrule
Learning rate & 0.001 & 0.0008 & 0.0008 \\
Weight decay & 0.0001 & 0.00003 & 0.00001 \\
Learning rate schedule & cosine & cosine & cosine\\
\end{tabular}
\caption{Hyperparameter settings for DSMIL across different data splits.}
\label{tab:DSMIL_hyper}
\end{table}

\begin{table}[!htb]
\centering
\begin{tabular}{l|c|c|c}
\multicolumn{4}{c}{ABMIL} \\
Hyperparameter & split1 & split2 & split3 \\
\midrule
Learning rate & 0.0008 & 0.0005 & 0.0008 \\
Weight decay & 0.0001 & 0.00005 & 0.00005 \\
Learning rate schedule & cosine & cosine & cosine\\
\end{tabular}
\caption{Hyperparameter settings for ABMIL across different data splits.}
\label{tab:ABMIL_hyper}
\end{table}

\begin{table}[!htb]
\centering
\begin{tabular}{l|c|c|c}
\multicolumn{4}{c}{Set Transformer} \\
Hyperparameter & split1 & split2 & split3 \\
\midrule
Learning rate & 0.0010 & 0.0008 & 0.0008 \\
Weight decay & 0.00003 & 0.0001 & 0.00001 \\
Learning rate schedule & cosine & cosine & cosine\\
\end{tabular}
\caption{Hyperparameter settings for Set Transformer across different data splits.}
\label{tab:SetTransformer_hyper}
\end{table}

\begin{table}[!htb]
\centering
\begin{tabular}{l|c|c|c}
\multicolumn{4}{c}{Filter then Avg.} \\
Hyperparameter & split1 & split2 & split3 \\
\midrule
Learning rate & 0.003 & 0.001 & 0.003 \\
Weight decay & 0.00003 & 0.00001 & 0.00001 \\
Learning rate schedule & cosine & cosine & cosine\\
\end{tabular}
\caption{Hyperparameter settings for Filter then Avg. across different data splits.}
\label{tab:Filter then Avg._hyper}
\end{table}

\section{Self-supervised Pretraining Details}
Our implementation is based on the official code from MoCo~\citep{he2020momentum,chen2020improved}. For image-level contrastive learning, we set learning rate to 0.06, weight decay to 0.0005, batch size to 512, size of queue to 4096, momentum m to 0.99, softmax temperature to 0.1. For bag-level contrastive learning, we set learning rate to 0.00015 (following the linear Scaling Relu~\citep{goyal2017accurate}, which is also recommended by MoCo's author), weight decay to 0.0005, batch size to 1, size of queue to 4096, momentum m to 0.99, softmax temperature to 0.1. Note that we did not tune hyperparameters for the self-supervised pretraining. 

We train the model using the train set as well as the unlabeled set for both image-level and bag-level contrastive learning. The model is set to train for 200 epochs, with early stopping monitored by knn protocol on the validation set. The early stopping patience is set to 20.

\paragraph{projection head $\psi$.} The projection head is a two-layer MLP with the structure [Linear(500, 500), ReLU(), Linear(500, 128)]. The projection head is used to project the image or bag representation to a latent space where the contrastive loss is applied. The projection head is discarded after training following the convention from ~\citep{chen2020improved,chen2020simple}.
\label{app:SSL_Pretraining}

\section{View Classifier Details}
We train a view classifier for each of the three splits independently. We train the classifiers using a recently proposed semi-supervised learning method~\citep{huang2022fix} with Pi-model~\citep{laine2016temporal}. We used the view labeled images in each split's train set (as the labeled data) as well as the unlabeled set (as the unlabeled data). 

The view classifiers are trained to output probabilities of three category: PLAX, PSAX and Other. The view classifiers' performance is shown in \ref{tab:viewclassifier_performance}

\begin{table}[h]
\centering
\begin{tabular}{c|c|c|c}
Method & split1 & split2 & split3 \\
\hline
Fix-A-Step + Pi  & 97.20  & 98.14 & 98.00

\end{tabular}
\caption{Balanced accuracy on view classification. Showing view classification on TMED2 test set's view labeled images.}
\label{tab:viewclassifier_performance}
\end{table}

\paragraph{Backbone.} The view classifiers use Wide ResNet~\citep{zagoruyko2016wide} as backbone,  specifically, the ``WRN-28-2'' that has a depth 28 and width 2.

\paragraph{Training and Hyperparameters.} We train the view classifiers using SGD~\citep{robbins1951stochastic} as optimizer. We train the classifiers for 500 epochs, and retain the checkpoint that has maximum validation accuracy on the validation set. Hyperparameters used are reported below~\ref{tab:ViewClassifier hyperparameters}

\begin{table}[!htb]
\centering
\begin{tabular}{l|c|c|c}
Hyperparameter & split1 & split2 & split3 \\
\midrule
Labeled batch size & 64 & 64 & 64 \\
Unlabeled batch size & 64 & 64 & 64 \\
Learning rate & 0.0003 & 0.009 & 0.009 \\
Weight decay & 0.05 & 0.0005 & 0.0005 \\
Max consistency coefficient & 0.3 & 0.3 & 0.3 \\
Beta shape $\alpha$ & 0.5 & 0.5 & 0.5 \\
Unlabeled loss warmup schedule & linear & linear & linear \\
Learning rate schedule & cosine & cosine & cosine\\
\end{tabular}
\caption{Hyperparameters used for the view classifiers in each split.}
\label{tab:ViewClassifier hyperparameters}
\end{table}

\label{app:ViewClassifier}

\section{Additional Related Work}
\label{app:relatedworks}
\paragraph{Classic approaches.}
% Numerous studies have been conducted on MIL even before the  ``deep learning era'' \footnote[1]{It is usually refered to 2012 when AlexNet achieved a significant improvement in image recognition accuracy over previous methods in the ImageNet Large Scale Visual Recognition Challenge}. 
% Numerous studies on MIL have been conducted even prior to the advent of deep learning. 
Examples of classic MIL methods includes iARP~\citep{dietterich1997solving}, 
Diverse Density~\citep{maron1997framework}, 
Citation-kNN~\citep{wang2000solving}, MI-Kernels~\citep{zhang2001dd}, MI/mi-SVM~\citep{andrews2002support}, mi-Graph ~\citep{zhou2009multi}, MILBoost~\citep{zhang2005multiple}, GPMIL~\citep{kim2010gaussian}, among others. 

\paragraph{Additional examples of medical applications of MIL.}
Other medical applications of MIL include diabetic retinopathy screening ~\citep{quellec2012multiple, li2021deep, li2021multi, kandemir2015computer}, bacteria clone analysis~\citep{borowa2020classifying}, drug activity prediction~\citep{dietterich1997solving, zhao2013drug}, and cancer diagnosis ~\citep{campanella2019clinical, chikontwe2020multiple, hou2016patch, ding2012breast, xu2014weakly}. 
%\citep{cosatto2013automated, xu2014weakly, kandemir2015computer, zhu2017deep, campanella2019clinical, borowa2020classifying, sharma2021cluster, shao2021transmil, lu2021data, li2021dual, rymarczyk2023protomil}. 

\end{document}